\documentclass[aps,twocolumn,showpacs,preprintnumbers,floatfix,amssymb,amsmath,amsfonts]{revtex4}

\usepackage{graphicx}
\usepackage{dcolumn}
\usepackage{bm}
\usepackage{epsfig}
\usepackage{amssymb}

\begin{document}
    
\preprint{APS/revtex4}

\title{$^{63}$Cu NQR Study of the Inhomogeneous Electronic State 
in La$_{2-x}$Sr$_{x}$CuO$_{4}$}

\author{P.M. Singer and A.W. Hunt}
\affiliation{Department of Physics and Center for Materials Science and 
Engineering, Massachusetts Institute of Technology, Cambridge, MA 02139, USA}

\author{T. Imai}
\affiliation{Department of Physics and Astronomy, McMaster University, 
Hamilton, ON L8S-4M1, Canada $^{*}$}
\affiliation{Department of Physics and Center for Materials Science and 
Engineering, Massachusetts Institute of Technology, Cambridge, MA 02139, USA}

\date{\today}

\begin{abstract}
We report detailed systematic measurements of the spatial variation 
in electronic states in the high $T_{c}$
superconductor La$_{2-x}$Sr$_{x}$CuO$_{4}$ ($0.04\leq x \leq 0.16$)
using $^{63}$Cu NQR for $^{63}$Cu isotope 
enriched poly-crystalline samples. 
We demonstrate that the spatial variation in local
hole concentration $^{63}x_{local}(\neq x$) given by $^{63}x_{local} = x \pm ^{63}\Delta x_{local}$,
where $x$ is the nominal hole concentration and $^{63}\Delta x_{local}$ is
defined as the amplitude (or extent) of the spatial variation,
is reflected in the frequency dependence of the spin-lattice relaxation rate $^{63}1/T_{1}$
across the inhomogeneous linebroadening of the $^{63}$Cu NQR
spectrum [P.M. Singer {\it et al.,} Phys. Rev. Lett. {\bf 88}, 47602 
(2002)]. We show that compared to nominal $x$, the electronic state in certain regions of 
the CuO$_{2}$ plane are locally more metallic ($^{63}x_{local} =x + ^{63}\Delta x_{local}$) while others are 
more insulating ($^{63}x_{local} =x - ^{63}\Delta x_{local}$).
By using high precision measurements of 
the temperature dependence of $^{63}1/T_{1}$ at various
positions across the $^{63}$Cu NQR lineshape, we demonstrate that
$^{63}\Delta x_{local} (\neq 0)$ increases below 500 - 600 K and reaches 
values as large as $^{63}\Delta x_{local}/x \simeq 0.5$ in the 
temperature region $\gtrsim 
150$ K. We find a substantial overlap
between the $^{63}$Cu NQR spectrum 
of samples with different $x$, and find 
that the extent of the overlap increases with decreasing temperature.
By incorporating the random positioning of $^{+2}$Sr donor ions in the lattice in a novel approach, a lower bound to the length scale of the spatial variation $^{63}R_{patch}$ is deduced by fitting the entire $^{63}$Cu NQR spectrum (including the ``B'' -line 
originating from $^{63}$Cu sites with $^{+2}$Sr ions directly above 
or below) using a patch-by-patch distribution of the spatial variation $^{63}x_{local}$ with the patch radius $^{63}R_{patch}\gtrsim 3.0$ nm (= 8-10 $a$, where $a$ is the lattice spacing) as the only free parameter.
A corresponding upper bound to the amplitude of the spatial variation
$^{63}\Delta x_{patch} (\propto 1/^{63}R_{patch})$ is deduced within the patch-by-patch model,
and consistent results are found with $^{63}\Delta x_{local}$ 
determined from the frequency dependence in $^{63}1/T_{1}$.
Using our pool of $^{63}$Cu NQR data, we also deduce the onset temperature 
$T_{Q} (\gtrsim 400$ K) of local orthorhombic lattice distortions which, in the region $x \gtrsim 0.04$, is found to be larger than the onset temperature of long range structural order.
\end{abstract}
\pacs{74.80.-g, 74.72.-h, 76.60.-k}

\maketitle
(*) Present and permanent address.

\section{Introduction}

NQR (nuclear quadrupole resonance) and NMR (nuclear magnetic 
resonance) have proven to give unique information regarding 
the inhomogeneous electronic state in the CuO$_{2}$ plane of various high $T_{c}$ cuprates 
\cite{yoshimura,tou,hammel,stat,fujiyama,hunt,singer,curro,haase,julien,hunt2,singerprl,tobe} including
La$_{2-x}$[Sr,Ba]$_{x}$CuO$_{4}$,
[La,Nd,Eu]$_{2-x}$[Sr,Ba]$_{x}$CuO$_{4}$ and 
La$_{2}$CuO$_{4+\delta}$. 
In these materials, the high $T_{c}$ superconductivity is achieved by doping 
holes into the CuO$_{2}$ plane.
In the case of La$_{2-x}$[Sr,Ba]$_{x}$CuO$_{4}$ and its Nd or Eu co-doped compound 
[La,Nd,Eu]$_{2-x}$[Sr,Ba]$_{x}$CuO$_{4}$, hole doping is 
achieved by substituting ions of different 
ionicity, thereby creating an alloy with intrinsic 
inhomogeneities caused by chemical substitution \cite{burgy}.
Over the past several years, studies using $^{63}$Cu NQR and NMR wipeout 
\cite{hunt,singer,curro,julien,hunt2} have characterised the glassy nature of the 
slowing down of the stripe inhomogeneity \cite{tranquada} in these 
materials, where the Coulomb potential from the distorted lattice slows down spin 
and charge density waves. In the case of 
La$_{2}$CuO$_{4+\delta}$, hole doping is achieved by
super-oxygenation, and
it has been shown that the high mobility of the excess oxygen atoms
results in electronic phase separation \cite{sokol,kivelson,castro} between the superconducting and 
antiferromagnetic phase, as evidenced by $^{139}$La NMR \cite{hammel,stat}.

No clear picture has emerged which discerns and relates the 
effects of genuine electronic phase separation, stripe modulation, and the
random substitution of donor ions. 
On the other hand, recent STM (scanning tunnelling 
microscopy) studies on the surface state 
of Bi$_{2}$Sr$_{2}$CaCu$_{2}$O$_{8+\delta}$ cleaved at low 
temperature \cite{pan} reveal spatial variations of the electronic 
state on a short length scale $\sim$nm in the surface plane. Whether such nm modulations are universally 
observable in the bulk and other high $T_{c}$ cuprates remains 
to be seen, but the STM results have enhanced the interest and potential impact of 
the spatial inhomogeniety of the electronic properties in cuprates. 
One emerging counter example is YBa$_{2}$Cu$_{3}$O$_{y}$ 
where $^{89}$Y NMR \cite{alloulpre} measurements indicate that the spatial inhomogeneity in YBa$_{2}$Cu$_{3}$O$_{y}$ is {\it less} 
than in Bi$_{2}$Sr$_{2}$CaCu$_{2}$O$_{8+\delta}$ or La$_{2-x}$Sr$_{x}$CuO$_{4}$.

Unlike scattering techniques which probe coherent phenomena over 
length scales larger than tens of nm's, NMR and NQR are strictly $local$ probes
which make them an ideal tool for revealing the short length scale physics of the
CuO$_{2}$ plane. Among 
earlier reports of the inhomogeneous state at short length scales 
in La$_{2-x}$Sr$_{x}$CuO$_{4}$ are:
the splitting of the 
$^{63}$Cu NQR and NMR lines due to inequivalent Cu sites known as the `A'' 
and ``B''-sites resulting from 
different EFG (electric field gradient) tensors 
\cite{yoshimura,stat} due to the presence of nearby $^{+2}$Sr ions, the drastic 
broadening of the zero field $^{63}$Cu line at 
temperatures below $\sim$ 4 K \cite{tou,hunt}, the $^{63}$Cu NMR line broadening from 
short length scale modulations in orbital shifts \cite{haase}, and
also the $^{63}$Cu NQR and NMR wipeout 
\cite{hunt,curro,julien} as a result of the glassy 
slowing down of the stripe inhomogeneity in the temperature region $\lesssim$ 100 K. 

These evidences clearly support the existence of some sort of short 
length scale inhomogeneity in the CuO$_{2}$ planes of La$_{2-x}$Sr$_{x}$CuO$_{4}$   
which raises questions regarding recent theoretical debates of a ``universal electronic phase diagram'', 
including La$_{2-x}$Sr$_{x}$CuO$_{4}$ \cite{hwang}, which are based on 
the assumption that hole doping is homogeneous.
More recently, we reported evidence of an inhomogeneous electronic 
state in La$_{2-x}$Sr$_{x}$CuO$_{4}$ using $^{63}$Cu NQR in $^{63}$Cu isotope enriched 
samples in the range $0.04\leq x \leq 0.16$ \cite{singerprl}.
As discussed in Ref. \cite{singerprl}, the two essential
ingredients for the spatial variation in local hole 
concentration $^{63}x_{local} (\neq 0)$ 
are (a) the fact that hole doping is achieved by random substitution
of donor ions in the lattice with different ionicity, and (b) the presence of a short electronic length scale $^{63}R_{patch} 
\gtrsim 3.0-4.0$ nm for the spatial variation.
To the best of our knowledge, our results reported in this paper and 
in Ref. \cite{singerprl} are the 
first of its kind to detect the temperature dependence of the 
inhomogeneous electronic state in La$_{2-x}$Sr$_{x}$CuO$_{4}$,
or any other high $T_{c}$ materials with quenched disorder.

Most recently, $^{17}$O NMR in high-quality La$_{2-x}$Sr$_{x}$CuO$_{4}$ 
single crystals in the range $0.035\leq x \leq 0.15$ also reveal a 
substantial inhomogeneity in the local electronic state \cite{tobe}. 
The spatial variation in local hole concentration 
$^{17}x_{local}$ determined using $^{17}$O NMR \cite{tobe} is found to reveal a 
consistent value with $^{63}x_{local}$ reported here and in Ref. 
\cite{singerprl}. The spatial variation $^{17}x_{local}$ \cite{tobe} is determined 
through the observed spatial variation in the $spin$ susceptibility across the CuO$_{2}$ 
plane, whereas $^{63}x_{local}$ is determined through the 
spatial variation in the EFG which is 
purely a charge effect. The fact that $^{63}x_{local} \simeq ^{\;17\!}x_{local}$ 
therefore implies that the inhomogeneity in the spin and charge 
channels are highly correlated. 

As presented in Ref. 
\cite{tobe}, we also compare the 
extent of the inhomogeneity in high-quality single crystals \cite{tobe} and our poly-crystalline 
samples (reported here and in Ref. \cite{singerprl}) 
by comparing the extent of the frequency dependence of $^{63}1/T_{1}$ across the $^{63}$Cu 
NQR spectrum. We find consistent results between 
single crystal and poly-crystalline samples \cite{tobe}, which immediately establishes that the inhomogeneity  
in the electronic state is an $intrinsic$ phenomenon in 
La$_{2-x}$Sr$_{x}$CuO$_{4}$. 
It also rules out certain claims that our poly-crystalline 
samples are somehow more inhomogeneous than those reported elsewhere. 
Such claims on the quality of our poly-crystalline samples originated 
as an excuse to negate the fact that hole doping is inhomogeneous 
in La$_{2-x}$Sr$_{x}$CuO$_{4}$ \cite{singerprl}. 
In this paper we report the details of the $^{63}$Cu NQR results and analysis
originally outlined in Ref. \cite{singerprl}, and while 
the $^{17}$O NMR results are left to Ref. \cite{tobe},
we mention the conclusions of the $^{17}$O 
NMR results wherever appropriate.


The dynamics of the electronic state are probed using the spin-lattice relaxation rate
$^{63}1/T_{1}$. $^{63}1/T_{1}$ is a measure of the local spin fluctuations in the
CuO$_{2}$ plane at the
$^{63}$Cu NQR frequency $^{63}\nu_{Q} $ (ranging between $32 - 40$ MHz).
Due to the experimental ease of 
the measurement, the temperature dependence of $^{63}1/T_{1}$ is generally 
only measured at the peak of the spectrum.
The most striking feature 
of our $^{63}$Cu NQR data, however, is that $^{63}1/T_{1}$ shows 
qualitatively different {\it temperature dependence} depending on where 
along the NQR line it is measured \cite{singerprl}. This implies, without 
using any kind of model,
that certain regions of the CuO$_{2}$ plane are more metallic while others are more insulating \cite{singerprl}.

In order to measure the frequency dependence in $^{63}1/T_{1}$ and obtain any quantitative 
information over a wide temperature range,
it is $essential$ to separate the signal from $^{63}$Cu and $^{65}$Cu
isotopes. Previous work by S. Fujiyama {\it et al.}
\cite{fujiyama} measured $^{(63,65)}1/T_{1}$ for naturally abundant Cu 
in which they successfully deduced that a substantial frequency 
dependence exists across 
the $^{(63,65)}$Cu NQR spectrum, however, no statement beyond that could be 
inferred. Through $^{63}$Cu isotope enrichment, together with a 
systematic study as a function of $x$, we can use the extent of 
the frequency dependence in $^{63}1/T_{1}$ across the $^{63}$Cu NQR line
to determine the local hole concentration $^{63}x_{local}$ given by 
\begin{equation}
    ^{63}x_{local} = x \pm ^{63}\Delta x_{local},
\label{eqone}
\end{equation}
where $^{63}\Delta x_{local}$ is the characteristic amplitude or extent of the 
spatial variation $^{63}x_{local}$.

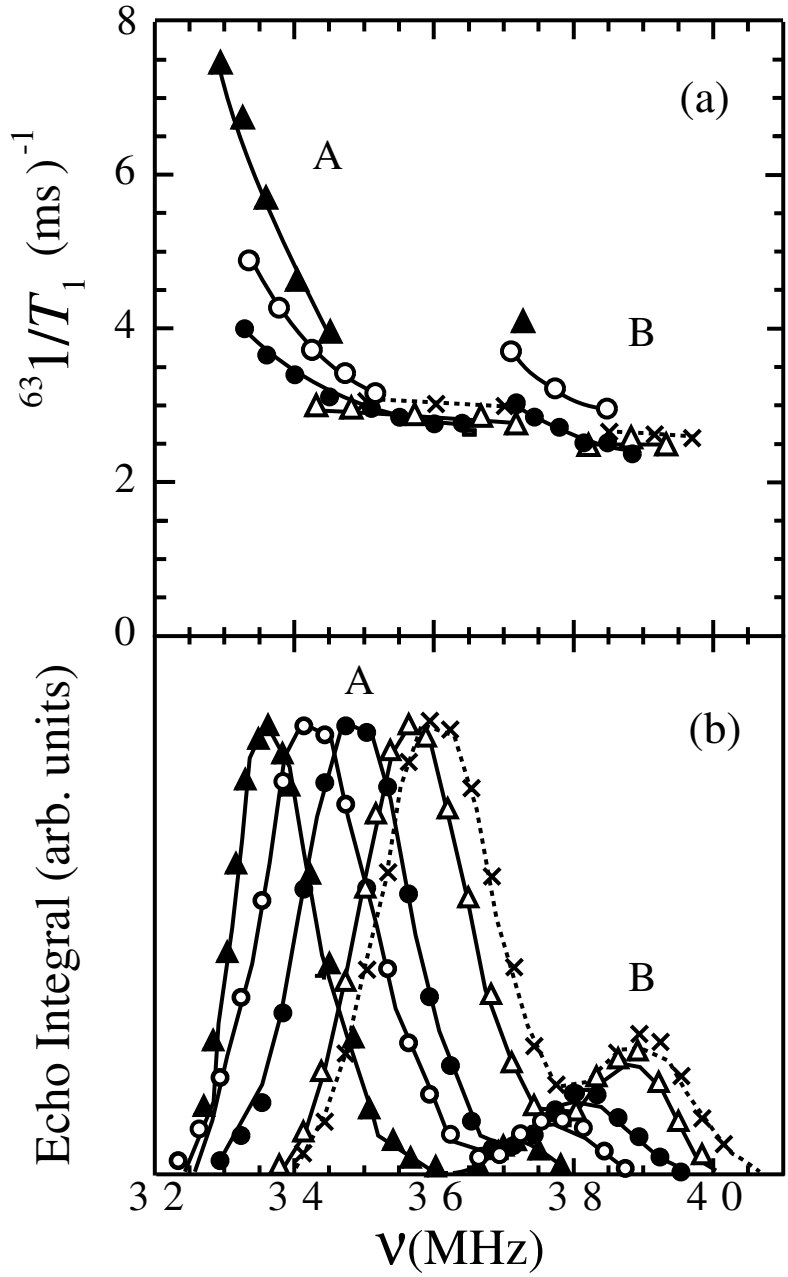
\begin{figure}
\caption{(a) Frequency dependence of $^{63}1/T_{1,A}$ and 
$^{63}1/T_{1,B}$ across the A and B-lines of the $^{63}$Cu NQR spectrum 
of $^{63}$Cu isotope enriched La$_{2-x}$Sr$_{x}$CuO$_{4}$ for $x=0.20$ 
($\times$), $x=0.16$ ($\triangle$),
$x=0.115$ ($\bullet$), $x=0.07$ ($\circ$) and 
$x=0.04$ ($\blacktriangle$). All data is taken at 300 K. Solid curves in (a) 
are a guide for the eye. (b) $^{63}$Cu NQR spectrum of the A and B-lines 
where the same symbols as part (a) are used. Solid curves in (b) are 
calculated fits using patch-by-patch model of spatial variation in
local hole concentration $^{63}x_{local}$. The only free parameter used in the fits is 
the lower bound for the patch radius $^{63}R_{patch} = 
2.1-3.1$ nm (using $d_{1}$ dopants, see later) in order of increasing $x$, respectively.}
\label{300T1f}
\end{figure}

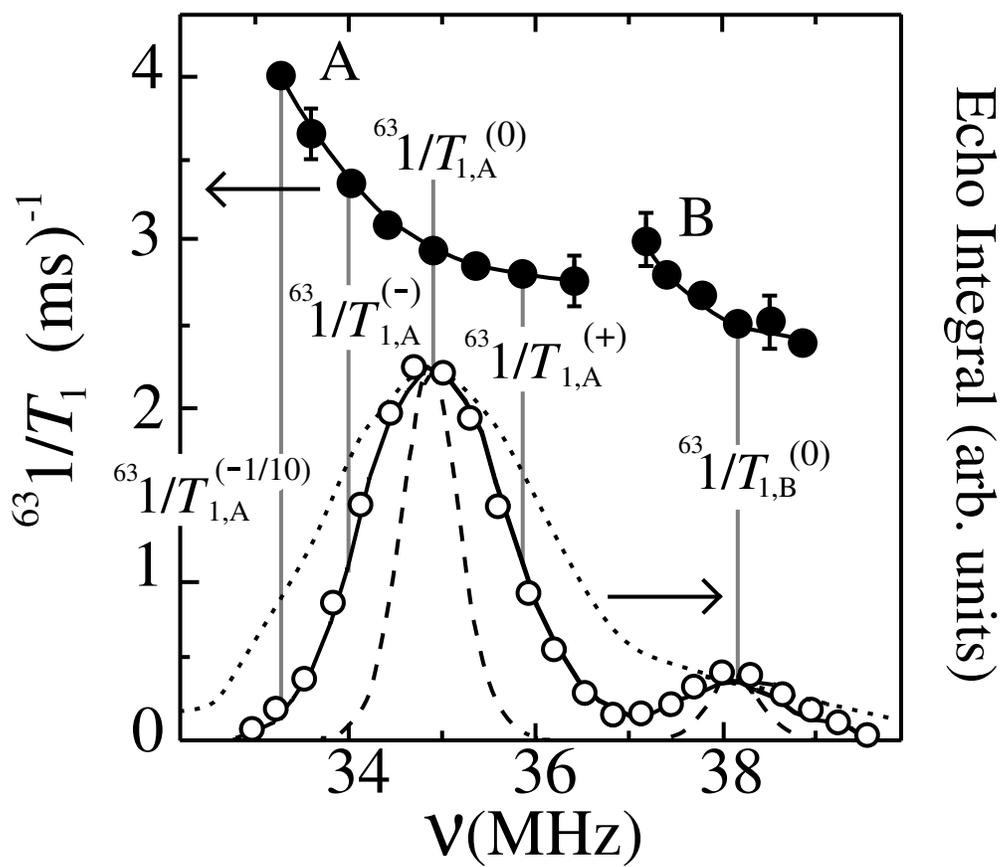
\begin{figure}
\caption{Frequency dependence of $^{63}1/T_{1,A}$ and $^{63}1/T_{1,B}$ 
($\bullet$) across $^{63}$Cu NQR lineshape ($\circ$)
for $x=0.115$ at 300 K taken from Fig. \ref{300T1f}.
We define $^{63}1/T_{1,A}^{(-1/10)}$, $^{63}1/T_{1,A}^{(-)}$,
$^{63}1/T_{1,A}^{(0)}$, and $^{63}1/T_{1,A}^{(+)}$  
at the lower one-tenth, the lower half, the CG (center of gravity), and upper half 
intensity position of the A-line, along with $^{63}1/T_{1,B}^{(0)}$ 
at the CG of the B-line.
Curves show fit using patch-by-patch model
with a lower bound for the patch radius of $^{63}R_{patch} = 1.6$ nm (dotted curve), 
$^{63}R_{patch} = 2.6$ nm (solid curve), and $^{63}R_{patch} = \infty$ (dashed curve, whose {\it HWHM} is given by $\Delta \nu_{latt}^{k}$).}
\label{300T1f2}
\end{figure}

In Fig. \ref{300T1f} we show an example of the frequency dependence in 
$^{63}1/T_{1}$ across the $^{63}$Cu 
NQR line at 300 K. We first remark on the overlap between
samples with different nominal hole concentration $x$, shown 
in Fig. \ref{300T1f}(b). 
Taking $x=0.115$ as an example, we see that the upper (lower) half intensity 
point of the $x=0.115$ A-line (or B-line) roughly coincides with the peaks of the 
$x=0.16(0.07)$, which implies that the characteristic local hole  
concentration $^{63}x_{local}$ is given by $^{63}x_{local} \leq 
0.16$ in the more metallic regions, and by $^{63}x_{local} \geq 
0.07$ in the more insulating regions. Using these limits and Eq. 
(\ref{eqone}), we estimate 
the upper bound to the amplitude $^{63}\Delta x_{local} \leq 0.045$ for $x=0.115$ at 300 K.
The reason that only an upper boundary to $^{63}x_{local}$ (and an 
upper bound to $^{63}\Delta x_{local}$) can be deduced
is that we have assumed that the linewidths in Fig. \ref{300T1f}(b) 
are dominated by $^{63} x_{local} (\neq x)$ alone. In fact, as we 
calculate in section IV,
there is a substantial intrinsic lattice 
linebroadening (defined as $\Delta 
\nu_{latt}$) which is independent of the broadening arising from $^{63} x_{local}$.
$\Delta \nu_{latt}$ originates from the distribution in local 
EFG values due to the random positioning of 
$^{+2}$Sr ions in the lattice, and we calculate its size by 
using a point charge lattice summation.
In Fig. \ref{300T1f2} we illustrate the extent of $\Delta 
\nu_{latt}$ as the dashed curve, which is comparable to the 
experimentally observed linebroadening.  
In order to determine a better estimate of $^{63} 
x_{local}$ in a model independent way, we must therefore measure $^{63}1/T_{1}$ at various 
positions across the line as shown in Fig. \ref{300T1f}(a) and Fig. \ref{300T1f2}.

In order to determine $^{63}x_{local}$ from $^{63}1/T_{1}$, 
we first define the values $^{63}1/T_{1,A}$ 
and $^{63}1/T_{1,B}$ taken at various positions across the NQR 
lineshape (shown in Fig. \ref{300T1f2}). $^{63}1/T_{1,A}^{(0)}$ 
defines $^{63}1/T_{1}$ measured at the CG (center of gravity) of the A-line, 
$^{63}1/T_{1,A}^{(+)}$ defines $^{63}1/T_{1}$ measured at the half 
intensity of the upper (+) frequency side of the A-line,
$^{63}1/T_{1,A}^{(-)}$ at the lower half intensity side, and 
$^{63}1/T_{1,A}^{(-1/10)}$ at the lower one-tenth intensity of the 
A-line. We also define the corresponding quantities for the B-line. 
We shall use these definitions throughout this paper to deduce $^{63} 
x_{local}$ as a function of temperature and $x$. 
Next, we assume that
$^{63}1/T_{1,A}^{(0)}$ and $^{63}1/T_{1,B}^{(0)}$ taken at the CG of each spectrum represents 
the $^{63}1/T_{1}$ value for the nominal hole concentration $x$, regardless of the 
underlying linebroadening.
We can then compare $^{63}1/T_{1}$ at the upper and lower half
intensity points of the A-line (or B-line) to nominal values
$^{63}1/T_{1,A}^{(0)}$ from different $x$. As we shall show in detail in 
section III, 
$^{63}1/T_{1,A}^{(-)}$ for $x=0.115$ at 300 K
is the same as 
$^{63}1/T_{1,A}^{(0)}$ for $x=0.08$, while $^{63}1/T_{1,A}^{(+)}$ for $x=0.115$
is the same as $^{63}1/T_{1,A}^{(0)}$ for $x=0.15$. Using Eq. (\ref{eqone}) we therefore deduce 
that $^{63}\Delta x_{local} = 0.035$ for $x=0.115$
at 300 K.
Since we do not have the $continuous$ function $^{63}1/T_{1,A}^{(0)}$ 
as a function of nominal $x$ (i.e. we have a limited number of samples), we interpolate 
$^{63}1/T_{1,A}^{(0)}$ between the discrete set of $x$ we do have.
The same argument applies to the B-line, and we shall
demonstrate that $^{63}\Delta x_{local}$ deduced using the B-line is consistent 
with that from the A-line.
In order to be consistent with Ref. \cite{singerprl}, we define the characteristic amplitude $^{63}\Delta x_{local}$ with respect to 
the half intensity points, however, 
following the same proceedure using $1/T_{1}^{(-1/10)}$ instead
results in an overall 20 - 40 \% increase in $^{63}\Delta x_{local}$ and does not effect our 
conclusions.
Using the frequency dependence of $^{63}1/T_{1}$, we therefore deduce that within each sample with 
nominal $x$ there exist certain regions in the CuO$_{2}$ plane with 
higher and lower local hole concentrations ($^{63}x_{local} \neq x$). Systematic studies as a function of temperature then allow us to follow the temperature dependence of $^{63}x_{local}$.

The next task in this paper is the model analysis of the NQR spectrum.
For this task, we develop a static real space picture consisting of a patch-by-patch spatial variation 
for $^{63}x_{local}$ \cite{singerprl}. Using a point charge lattice summation of the EFG together with a patch-by-patch spatial variation, we successfully fit the NQR spectrum (including the spectral overlap for different $x$ and the B-line) with one free parameter ($^{63}R_{patch}$). 
As detailed in section IV, the simulation of the EFG consists of two 
essential ingredients: (a) the random 
positioning of $^{+2}$Sr donor ions in the lattice, and (b) the presence of a short length scale 
$^{63}R_{patch} \gtrsim 
3$ nm which describes the inhomogeneous electronic state.
In our model, the CuO$_{2}$ plane is sectioned into 
patches (i.e. circles) of equal radius $^{63}R_{patch} (\simeq 3$ nm), 
and each patch $i$ has a different local hole concentration $x_{local}^{i} (\neq x)$.
The value of $x_{local}^{i}$ is determined by the random number of 
$^{+2}$Sr donor ions in the vicinity of patch $i$. The
holes are placed uniformily within patch $i$, and 
the resonance frequency of the $^{63}$Cu nuclei within patch $i$
is calculated in a self-consistent way using our pool of NQR data. By sampling a large number 
of patches ($\sim 10^{4}$) and calculating the resonance frequency for each patch, we build up a 
histogram spectrum of resonance frequencies which constitutes the simulated NQR
lineshape. $^{63}R_{patch}$ is the only free parameter in the simulation, and 
we optimize $^{63}R_{patch}$ to fit the NQR
spectrum, as shown in Fig. \ref{300T1f2}.

Using $^{63}R_{patch} = \infty$ is equivalent to uniform homogeneous doping 
where $^{63}x_{local} = x$ throughout the plane (i.e. $^{63}\Delta 
x_{local} = 0$). This is also equivalent to excluding ingredient (b) 
listed above and calculating the broadening solely from the random 
positioning of $^{+2}$Sr donor ions in the lattice (i.e. $\Delta \nu _{latt}$). Clearly, however, in the case $^{63}R_{patch} = \infty$ the observed linewidth is 
underestimated by $\Delta \nu _{latt}$ alone, likewise for the B-line. Using a 
very short length scale $^{63}R_{patch} = 1.6$ nm
overestimates the NQR linebroadening, whereas the optimum value $^{63}R_{patch} = 2.6$ 
nm results in the best fit to the data, including the B-line. In Ref. \cite{tobe} we 
also show that $^{63}R_{patch}$ thus deduced is consistent with $^{17}R_{patch}$ 
determined in a similar way using the inhomogeneous linebroadening of the $^{17}$O NMR spectra. 

Once $^{63}R_{patch}$ is optimized to fit the NQR spectra,
we deduce the spatial variation $^{63}x_{patch}$ given by 
\begin{equation}
    ^{63}x_{patch} = x \pm ^{63}\Delta x_{patch},
\label{equan2}
\end{equation}
where $^{63}\Delta x_{patch}$ is the calculated amplitude for
$^{63}x_{patch}$. $^{63}\Delta x_{patch}$ is found to follow the simple relation $^{63}\Delta x_{patch}
\propto 1/^{63}R_{patch}$.
The results of the EFG simulation ($^{63}x_{patch}$) are found to be 
consistent with $^{63}x_{local}$ deduced from the frequency dependence 
in $^{63}1/T_{1}$. Since $^{63}x_{local}$ is determined in a 
model independent way, the fact that $^{63}x_{patch} \simeq ^{63\!\!}x_{local}$
justifies the use of our patch-by-patch model. We note, however, that 
apart from the linebroadening mechanisms (a) and (b) discussed above, we ignore all other possible NQR linebroadening
mechanisms in our model. Other possible mechanisms include 
distributions in local lattice distortions (section V). Our calculation in section IV ignores 
such possibilities which implies that $^{63}R_{patch}$ is a lower bound,
that $^{63}\Delta 
x_{patch} (\propto 1/^{63}R_{patch})$ is correspondingly an upper 
bound, and therefore that $^{63}x_{patch}$ is an upper 
boundary. 
The fact that 
$^{63}x_{patch} \sim ^{\;\;63\!\!}x_{local}$, however, indicates 
that our simulation of the NQR spectra and the patch-by-patch model is appropriate. 

In section V we introduce the effects of the local orthorhombic 
distortions into the simulation. We find that the NQR data separates into two 
temperature regions, above and below $T_{Q} (\gtrsim 400)$ K. For $T > T_{Q}$, our data 
indicates that the lattice is locally in the tetragonal phase and we 
use this region to estimate $^{63}R_{patch}$ and $^{63}x_{patch}$ as discussed above.
For $T \leq T_{Q}$, however, we find evidence for local orthorhombic 
distortions which, in the case of $x\gtrsim 0.04$, start somewhat above the 
structural phase transition temperature $T_{st} (\leq $ 515 K) 
according to neutron diffraction studies \cite{yamada}. 
In the doping region $x\gtrsim 0.04$, we show the 
systematic differences between the local and
long range orthorhombic order.

The rest of this paper is sectioned as follows: in section II we 
describe the experimental characteristics including the sample characteristics and NQR background.
In section III we present the experimental 
results which include the frequency dependence of $^{63}1/T_{1}$ and the determination of $^{63} 
x_{local}$. We also present the temperature dependence of the NQR spectra.
In section IV we describe the analysis of the NQR spectra using a patch-by-patch model for the spatial variation in $^{63}x_{local}$. We present the details of the EFG simulation incorporating 
randomness effects, and we optimize the length scale $^{63}R_{patch}$ 
(and $^{63}x_{patch}$) to fit the experimental data. In section V we incorporate the 
local orthorhombic distortions into the EFG simulation. In section VI we then present our conclusions.

\section{EXPERIMENTAL Characteristics}

\subsection{Samples}

All of the single phase, poly-crystalline samples of
La$_{2-x}$Sr$_{x}$CuO$_4$ used in this study were prepared
using conventional solid state reactions \cite{takagisusc,radaelli}.
We mix pre-dried La$_{2}$O$_{3}$ (99.995$\%$), 
SrCO$_{3}$ (99.995$\%$), and $^{63}$CuO (99.995$\%$) with correct nominal 
compositions using an agate mortar and pestle until an intimate mixture is 
obtained.
A pre-reaction is carried out for 20\,h in a box furnace at $850^{\circ}$C followed by repeated
grindings and sinterings (also 20\,h) at temperatures between $950^{\circ}$C and
$1000^{\circ}$C. During these initial firings the
materials are made into low density pellets with a hand press. The samples are then
pelletized with the pneumatic press and high temperature annealed in flowing O$_2$ gas at
$1100^{\circ}$C to $1150^{\circ}$C for 24\,h to 48\,h before a slow and controlled cooling cycle
that includes low temperature annealing at $800^{\circ}$C (24\,h) and $500^{\circ}$C (24\,h). 
The long annealing in O$_2$ insures that the oxygen content is uniform 
and stoichiometric and high annealing temperatures insure fast 
reaction kinetics. By
using a large number of grindings (typically 5-8), we achieve high quality
poly-crystalline samples. 

The most important test for the quality of our samples was discussed earlier 
(and presented in Ref. \cite{tobe}) where we confirmed that the 
frequency dependence of $^{63}1/T_{1}$ across the  
 $^{63}$Cu NQR spectrum in our poly-crystalline 
samples is the $same$ as in high-quality single crystals grown 
in Tokyo \cite{tobe}. This confirms that the inhomogeneity we report in this paper and in 
Ref.'s \cite{singerprl,tobe} are intrinsic to La$_{2-x}$Sr$_{x}$CuO$_{4}$.
Apart from $^{63}$Cu NQR itself, we characterise our poly-crystalline samples by measuring the superconducting 
transition $T_{c}$ and the room temperature lattice parameters.
We deduce $T_{c}$ by measuring the Meissner signal using a SQUID magnetometer 
in the field cooled mode with a constant applied magnetic field of 15 Oe
, the results of which are shown in Fig. \ref{squid}. We find that the onset 
temperature 
of the diamagnetic susceptibility, $T_{c}$, and the volume fraction agree 
with previous measurments in poly-crystalline
samples prepared in a similar fashion \cite{takagisusc,radaelli,takagidelta}. We note that the onset of the diamagnetic susceptibility is somewhat tailed in the case of $x=0.20$, consistent with \cite{takagisusc,radaelli,takagidelta}. In Ref. \cite{takagidelta} it is even argued that bulk superconductivity disappears in the overdoped region, which coincides with the disappearance of LRO (long range order) into the orthorhombic structural phase. As we present in section III, we also find a qualitative change in $^{63}1/T_{1}$ in the overdoped region. For these reasons we limit our determination of $^{63}x_{local}$ in section III to $x \leq 0.16$.

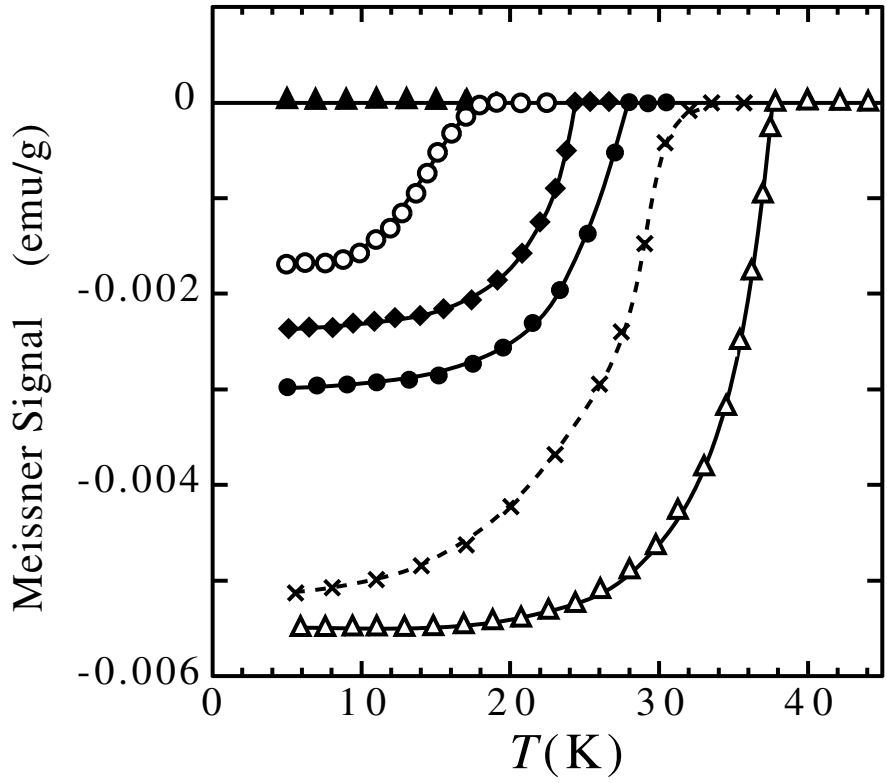
\begin{figure}
\caption{$x$ dependence of the lattice constants at 295 K deduced by 
X-ray diffraction along the $a_{o}$ ($\circ$), $b_{o}$ 
($\times$), and $c$ ($\bullet$) axes using orthorhombic  
notation. Lines are a guide for the eye.}
\label{latt}
\end{figure}

We deduce the lattice parameters with an X-ray 
diffractometer using the Cu 
$K\alpha^{1}$ line. The lattice 
parameters presented in Fig. \ref{latt} (the values of which we use 
in the analysis in section IV) show close agreement with 
previous results \cite{radaelli}. In this paper, we define $a_{o}$ and $b_{o}$ to be 
the in-plane lattice parameters of the orthorhombic cell and  
$[a_{o},b_{o},c]$ as the vector along the orthorhombic directions. 
Likewise, we define $a$ and $b(\equiv a$) as the lattice parameters in the
tetragonal cell and $[a,b,c]$ as the vector along the tetragonal 
direction. Our typical X-ray linewidths were comparable to the instrumental 
resolution with a {\it HWHM} of $\delta_{Instr} \sim 
0.025^{o}$ determined using a high quality Si standard.
Previous higher resolution X-ray diffraction experiments \cite{takagidelta}
(where $\delta_{Instr} \sim 0.01^{o}$) were able to use the observed X-ray linewidths and overlaps between 
different samples $x$ to deduce a lower 
bound to the compositional distribution with a {\it HWHM} of $\delta x_{Sr} \sim 0.01$. 
Only a lower bound to $\delta x_{Sr}$ could be inferred because X-ray diffraction 
takes a spatial average over length scales larger than tens of nm's 
\cite{takagidelta}. 
They attributed $\delta x_{Sr} \sim 0.01$ to 
imperfect mixing inherent to the solid state reaction. Similar 
conclusions were reached in Ref. \cite{radaelli} using the overlap of X-ray 
diffraction and neutron Bragg peaks between different $x$, where 
$\delta x_{Sr} \sim 0.015$ was deduced.

We confirmed the same order lower bound $\delta x_{Sr} \simeq 0.013$ in single crystal $x=0.15$
\cite{singerprl} using EMPA (electron micro-probe analysis) 
whose length scale is determined by the focus area of the beam $\sim 1
$ $\mu$m. This means that even high-quality single crystals whose 
spatial variation of $^{+2}$Sr content is as small as $\delta x_{Sr} \simeq 
0.013$ ($\ll ^{63\!\!}\Delta x_{local}$) over $\sim 1$ $\mu$m, exhibit a 
large variation in the local electronic states. Furthermore, we will 
demonstrate in section III that $^{63}\Delta x_{local}$ is temperature dependent 
and increases with decreasing temperature. This means that the 
variation in $^{+2}$Sr content from the solid state reaction alone cannot 
account for our findings. Instead, a thermodynamic process must be 
involved.

\subsection{NQR Background}

The essential condition for 
NQR is that the ground state spin $^{n}I$ of the nuclear isotope 
$n$ satisfy $^{n}I\geq 1$ \cite{abragam,slichter}.
In the case of Cu, there 
exist two stable isotopes $n = 63(65)$ with natural abundances of 
69 (31) \% respsectively, both with spin $^{(63,65)}I = 3/2$. This is both an 
advantage and a disadvantage. It is generally a disadvantage in the 
sense that their quadrupole moments $^{n}Q$ happen to be similar 
$^{63}Q/^{65}Q = 1.081$, and therefore 
their resonance spectrum tend to coincide. This is the case in La$_{2-x}$Sr$_{x}$CuO$_{4}$
for $x>0.05$ where the NQR spectrum from each isotope is broad enough 
that they merge in the temperature range of interest, and therefore 
$^{63}$Cu isotope enrichment is essential to separate 
multiple lines.

The observed NQR frequency $^{63}\nu_{Q}$ is 
given by the total EFG (electric field gradient) 
tensor $V_{tot}^{(\beta,\gamma)}$ (where $\beta$ and $\gamma$ are any 
two orthogonal spatial directions) surrounding the 
resonating $^{63}$Cu nucleus, and its quadrupole coupling constant 
$^{63}Q$. In terms of magnetic eigenstates $m_{I}$,
NQR for $I = 3/2$ corresponds to transitions between
$m_{I} = |\pm 1/2|$ and $m_{I}' = |\pm 
3/2|$ eigenstates \cite{abragam,slichter}. The EFG cannot split the $\pm$ 
degeneracy of the magnetic eigenstates since it is of charge origin.
The most general expression for the NQR tensor $^{63}\nu_{Q}^{(\beta,\gamma)}$ is given by 
\cite{abragam,slichter}:
\begin{equation}
    ^{63}\nu_{Q}^{(\beta,\gamma)} = \frac{e^{63\!}Q}{2h} \cdot V_{tot}^{(\beta,\gamma)},
    \label{gen}
\end{equation}
where $V_{tot}^{(\beta,\gamma)}$ is the traceless symmetric EFG tensor. $V_{tot}^{(\beta,\gamma)}$ 
can be rotated and diagonalised 
to point towards the principle axes $\alpha$ of the EFG. This leads
to the following relation
\begin{equation}
    ^{63}\nu_{Q}^{\alpha} = \frac{e^{63\!}Q}{2h} \cdot V_{tot}^{\alpha}.
    \label{mainvq}
\end{equation}
In the tetragonal phase, the principle axes $\alpha$ are along the crystal 
axes $[a,b,c]$, and satisfy the general traceless condition $V^{c}+V^{b}+V^{a}=0$. 
The direction of the largest component $|V^{c}|$ (where $|V^{c}|>|V^{b}|>|V^{a}|$) 
defines the {\it main principle axis} in the tetragonal phase. The asymmetry parameter $\eta$ ($0\leq \eta \leq 1$), defined as 
\begin{equation}
    \eta = \frac{V^{a}-V^{b}}{V^{c}},
    \label{eta}
\end{equation}
is typically small $\eta \leq 0.06$ for the planar Cu site in 
the cuprates \cite{penn} and for both A and B-sites \cite{song}. The observed NQR frequency $^{63}\nu_{Q}$ is then given by 
\begin{equation}
    ^{63}\nu_{Q} = \frac{e^{63\!}Q}{2h} \cdot V_{tot}^{c} \cdot \left(1+\eta^{2}/3 
    \right)^{1/2}.
    \label{mainvq21}
\end{equation}
In the present case, $\eta$ is small $(\eta \leq 0.06)$ and to a very good approximation
\begin{equation}
    ^{63}\nu_{Q} \simeq \frac{e^{63\!}Q}{2h} \cdot V_{tot}^{c}.
    \label{mainvq2}
\end{equation}

The value of $^{63}Q$ used varies between 
groups, but is generally taken 
to be either $^{63}Q = -0.211$ barns \cite{stern} which is based on 
theoretical calculation, or $^{63}Q = -0.16$ barns \cite{bleaney} 
which is based on ESR measurements in cuprate salts. Since we are 
dealing with experimental results, we choose to use $^{63}Q = -0.16$ barns 
\cite{bleaney}.  
As discussed in appendix A, however,
our final results of the length scale $^{63}R_{patch}$ is insensitive to the 
absolute value of $^{63}Q$.

In NQR, the direction along which the spin-lattice relaxation rate $^{63}1/T_{1}$ is 
measured is given by the main principle axis of the EFG (i.e. the $c$-axis), and cannot be changed externally.
The most general expression for $^{63}1/T_{1}$ in NQR with relaxation by a single magnetic process is given by \cite{moriya}
\begin{equation}
	        {}^{63} \frac{1}{T_{1}} = \frac{^{63\!}\gamma^{2}  k_{B} T}{\mu_{B}^{2}} 
	 \cdot \sum_{{\bf q}} 2 \left|^{63}A({\bf q})^{\perp} \right|^{2}
	\frac{\chi''_{\perp}({\bf q},\omega_{n})}{\omega_{n}},
\label{T1}
\end{equation}
where $\omega_{n}/2\pi = ^{63\!}\nu_{Q} (=  32 - 40$ MHz) is the NQR 
frequency, $^{63\!}\gamma/2\pi = 11.285$ MHz/Tesla is the nuclear gyro-magnetic 
ratio, and ${\bf q}$ is the reciprocal lattice vector (in tetragonal notation).

$^{63}A({\bf q})^{\perp}$ is the hyperfine form factor {\it perpendicular} to the main principle axis. The two perpendicular directions ($\perp$) in the present case both lie in the CuO$_{2}$ plane (i.e. in the $a-b$ plane). The ${\bf q}$ dependence of the in-plane value $^{63}A({\bf q})^{\perp}$ is given by
\begin{equation}
^{63}A({\bf q})^{\perp} = A_{\perp} + 2B \cdot \left( \cos(q_{x}a) + 
\cos(q_{y}a) \right),
\label{hyperfine}
\end{equation}
where $A_{\perp} \simeq 38 $ kOe/$\mu_{B}$ and $B \simeq 42 $ kOe/$\mu_{B}$ 
are found to be the same in La$_{2-x}$Sr$_{x}$CuO$_{4}$ 
\cite{mmplsco,imai} and 
YBa$_{2}$Cu$_{3}$O$_{y}$ \cite{takigawa1991} compounds.
$\chi''_{\perp}({\bf q},\omega_{n})$ (in emu/[mol f.u.]) is the in-plane imaginary part of the dynamic spin 
susceptibilty of the electrons. According to Eq. (\ref{hyperfine}), there 
is a large contribution to $^{63}1/T_{1}$ from the 
anit-ferromagnetic wavevector 
${\bf Q_{AF}} = (\pi/a,\pi/a)$ where $^{63}A({\bf Q_{AF}})^{\perp} = -139 $ 
kOe$/\mu_{B}$ \cite{imai}. Since  $\chi''_{\perp}({\bf q},\omega_{n})$ is peaked 
in the vicinity of ${\bf q=Q_{AF}}$, this implies
that $^{63}1/T_{1}$ is a local probe of the low-frequency 
(i.e. $\hbar \omega_{n} \ll k_{B} T$) in-plane anti-ferromagnetic spin fluctuations. 
$^{63\!}\gamma^{2}$ enters into Eq. (\ref{T1}) because the 
spin-lattice relaxation process is {\it magnetic} in origin, whereas the splitting of the 
energy levels $\omega_{n}$ is determined by the EFG which is of {\it charge} 
origin.

We measure $^{63}1/T_{1}$ by applying the following pulse sequence
\begin{equation}
\pi----t----[\pi/2-\tau-\pi ] -\tau - echo,
\label{pulse}
\end{equation}
where an r.f. pulse $\pi$ inverts the nuclear magnetization $M(t)$
at time $t$ prior to the spin-echo sequence in brackets. 
The time dependence of $M(t)$ 
is determined by taking the integral of the $echo$ in the time domain as a function of delay time $t$.
$^{63}1/T_{1}$ is then determined by fitting $M(t)$ to the
standard recovery form \cite{narath} 
\begin{equation}
M(t)  =  M(\infty) + \left( M(0) - M(\infty) \right) \cdot
  \left[ \exp \left(-\frac{3}{T_{1}}t \right) \right],
\label{mt}
\end{equation}
appropriate for NQR with $^{n}I=3/2$. 

The results of $^{63}1/T_{1}$ were independent 
of the r.f. pulse width $\pi$ in Eq. (\ref{pulse}). 
In the case of $x \lesssim 0.16$, $M(t)$ fit well to Eq. (\ref{mt}), implying that the spin-lattice relaxation rate is dominated by a 
single relaxation mechanism. In the case of $x \gtrsim 0.16$ below $\lesssim 100$ K, however, detailed measurements of $M(t)$ revealed a multi-exponential recovery, implying that $^{63}1/T_{1}$ is distributed at a fixed frequency.
In appendix B we use the multi-exponential recovery of $M(t)$ to deduce the intrinsic 
lattice broadening $\Delta \nu 
_{latt}^{T1} $ of the NQR line. In the 
case of $x = 0.16$, we  
find that $\Delta \nu _{latt}^{T1}= 0.62 (\pm 0.07)$ MHz, which is
consistent with our independent results of the EFG simulation $\Delta \nu 
_{latt} = 0.49 $ MHz presented in section IV. All $^{63}1/T_{1}$ data presented in this paper (except for appendix B) correspond to results of force-fits (in the case $x\gtrsim 0.16$) deduced using Eq. (\ref{mt}). The force-fit values from Eq. (\ref{mt}) correspond to the {\it average} value of the underlying distribution in spin-lattice relaxation rate.

As a result of the distribution in the spin-lattice relaxation rate, the value of $^{63}1/T_{1}$ determined using Eq. (\ref{mt}) 
showed a slight dependence on the pulse separation time $\tau$ in Eq. 
(\ref{pulse}). The most likely explanation is that using a long $\tau$ results in suppressed signal intensity from 
$^{63}$Cu nuclei with fast spin-spin relaxation rates ($^{63}1/T_{2}$). Since $^{63}$Cu nuclei with fast $^{63}1/T_{1}$ also have fast component in $^{63}1/T_{2}$ \cite{red}, one would expect a decrease in the average value of $^{63}1/T_{1}$ with increasing $\tau$.
In the case of $x=0.16$, increasing $\tau$ 
from 12 to 24 $\mu$s resulted in a uniform $\sim 10$ \% {\it decrease} 
for $^{63}1/T_{1}$ across the NQR line.
In order to retain the maximum possible signal 
intensity and measure the most representative value of 
$^{63}1/T_{1}$ at each frequency, we used the 
shortest possible value $\tau = 12 $ $\mu$s the experiment allows for {\it all}
measurements in this paper. Using the same experimental conditions throughout also allowed for a consistent determination of $^{63}x_{local}$.

Finally, we note that there are two relevant time-scales in the NQR experiment. The first is the "dynamic" time-scale associated with $^{63}1/T_{1}$. Since $^{63}1/T_{1}$ measures the spin fluctuations specifically at the NQR frequency $^{63}\nu_{Q}$ (Eq. (\ref{T1})), the appropriate "dynamic" time-scale is given by $1/^{63}\nu_{Q} \simeq 0.027$ $ \mu$s. 
The other relevant time-scale is the "static" one associated with the NQR spectrum. The "static" time-scale is given by the spin-echo refocusing time $2 \tau = 24 $ $\mu$s (Eq. (\ref{pulse})). $2 \tau$ defines the effective exposure time (or equivalently, an effective shutter speed of $1/(2\tau)$) for the NQR experiment. All dynamic effects with time-scales {\it shorter} than $2 \tau $ are unobservable through the NQR spectrum.

\section{Experimental Results}

\subsection{$^{63}1/T_{1}$ Results}

Before presenting the frequency dependence of $^{63}1/T_{1}$, we 
first present the temperature dependence of 
$^{63}1/T_{1,A}^{(0)}$ and $^{63}1/T_{1,B}^{(0)}$ taken at the CG of 
each spectrum. In Fig. \ref{T1A} and Fig. \ref{T1B} we present the 
temperature dependence of $^{63}1/T_{1,A}^{(0)}$ and $^{63}1/T_{1,B}^{(0)}$, respectively. While the solid curves through the 
$^{63}1/T_{1,A}^{(0)}$ data in Fig. \ref{T1A} are 
guides for the eye, the solid curves in Fig. \ref{T1B} represent 
$\left( \epsilon \cdot ^{63\!}1/T_{1,A}^{(0)} \right)$ data taken from Fig. \ref{T1A}. We plot Fig. \ref{T1B} 
as such to directly compare the temperature dependence of 
$\left( \epsilon  \cdot^{63\!}1/T_{1,A}^{(0)} \right)$
with $^{63}1/T_{1,B}^{(0)}$, where $\epsilon$ is a uniform scaling 
factor taken to be $\epsilon =[0.87,0.87,0.90,0.84]$ for $x = 
[0.20,0.16,0.115,0.07]$, respectively. 
As shown in Fig. \ref{T1B}, we find semi-quantitatively the same temperature 
dependence between the CG of the A and B-lines, with an overall 10 - 16 \% 
smaller value of $^{63}1/T_{1,B}$ compared to $^{63}1/T_{1,A}$, i.e. 
$^{63}T_{1,A}/^{63}T_{1,B} = 0.9-0.84$. 
Previous reports in $non$-isotope enriched samples \cite{itoh,itoh2} 
also found that $^{63}T_{1,A}/^{63}T_{1,B} < 1$.

At first glance of Fig. \ref{T1B}, our data for $x=0.16$ and $x=0.20$ below $\sim 150$ 
K may suggest that the ratio $^{63}T_{1,A}/ ^{63}T_{1,B}$ tends to 
decrease $\sim 10-15$ \% with 
decreasing temperature, similar to previous trends reported in 
\cite{itoh} for $x=0.20$.
We point out, however, that caution must be used when 
deducing any significance from temperature variations in 
$^{63}T_{1,A}/^{63}T_{1,B}$. 
The reason for this stems from the fact that spin-lattice relaxation rate is distributed at a fixed frequency, and that finite $\tau = 12$ $\mu$s conditions are used. As discussed in section II, the force-fit value $^{63}1/T_{1}$ will tend to underestimate the CG of the underlying distribution, the extent of which will depend on $\tau $ and the one-to-one correspondence between the spin-spin relaxation rate and spin-lattice relaxation rate for all $^{63}$Cu nuclear sites. However,
it is known that the spin-spin relaxation rate shows a complicated temperature dependence for both A and B-sites. In particular, the Gaussian-like component of the spin-spin 
relaxation rate ($^{63}1/T_{2G}$ \cite{T2G}) tends to get smaller with decreasing temperature while the Lorenztian-
like component ($^{63}1/T_{2L}$) becomes dominant. Below the onset of $^{63}$Cu wipeout
$T_{NQR}$, the spin-spin relaxation rate becomes totally Lorentzian-like 
\cite{hunt,curro,singer,hunt2} at both A and B-sites. Slight differences in the temperature dependence 
of $^{63}1/T_{2G}$ and $^{63}1/T_{2L}$ between the A and B-sites complicate the one-to-one correspondence with spin-lattice relaxation rate, and result in different underestimations of $^{63}1/T_{1}$ between the A and B-sites with decreasing temperature. Such factors must be considered when interpreting the temperature dependence in $^{63}T_{1,A}/ ^{63}T_{1,B}$ for $\lesssim 150 $ K and $x \gtrsim 0.16$. 

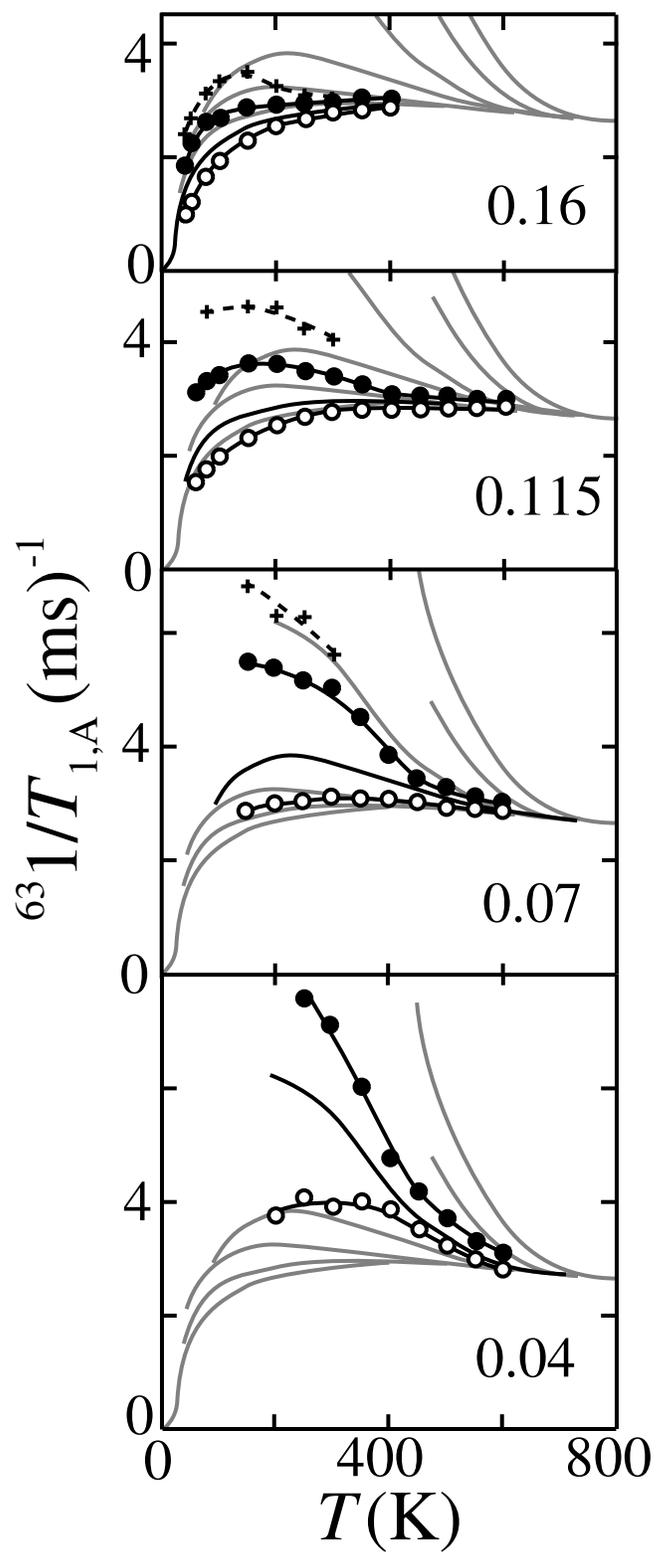
\begin{figure}
\caption{Temperature dependence of $^{63}1/T_{1,A}^{(-1/10)}$ ($+$), 
$^{63}1/T_{1,A}^{(-)}$ ($\bullet$), $^{63}1/T_{1,A}^{(0)}$ (black curve alone), and
$^{63}1/T_{1,A}^{(+)}$ ($\circ$) with nominal hole concentration
$x$ given in each panel. Grey curves are $^{63}1/T_{1,A}^{(0)}$ for $x=0.00$, 0,02, 0.04, 0.07 
,0.09, 0.115, and 0.16 where $^{63}1/T_{1,A}^{(0)}$ monotonically decreases 
with increasing $x$.}
\label{T1pm}
\end{figure}

In Fig. \ref{T1pm} we present the temperature dependence of $^{63}1/T_{1,A}$  
at various frequencies across the line, labeled according to Fig. 
\ref{300T1f2}. In the background of Fig. \ref{T1pm} we also show $^{63}1/T_{1,A}^{(0)}$ 
taken from Fig. \ref{T1A} which represent curves of 
constant $x$.
The most surprising discovery of the present work is that $^{63}1/T_{1}^{(+)}$, $^{63}1/T_{1}^{(-)}$ and 
$^{63}1/T_{1}^{(-1/10)}$ all show {\it qualitatively different} temperature 
dependence. For example, $^{63}1/T_{1}^{(+)}$ for $x=0.07$
exhibits semi-quantitatively the same behaviour as 
$^{63}1/T_{1,A}^{(0)}$ for $x=0.115$, while $^{63}1/T_{1}^{(-)}$ for 
$x=0.07$ exhibits semi-quantitatively the same behaviour as 
$^{63}1/T_{1,A}^{(0)}$ for $x=0.04$. 
This is consistent with the fact that the upper (lower) frequency side of the 
NQR spectrum for $x=0.07$ roughly coincides with the peak NQR frequency 
of $x=0.115$ ($x=0.04$) respectively, as shown in Fig. \ref{300T1f}. We also note that with 
decreasing temperature,
$^{63}1/T_{1}^{(+)}$, $^{63}1/T_{1}^{(-)}$ and 
$^{63}1/T_{1}^{(-1/10)}$ do not exactly follow $^{63}1/T_{1,A}^{(0)}$ for 
a given $x$, which indicates that $^{63}\Delta 
x_{local}$ is gradually growing with decreasing temperature.

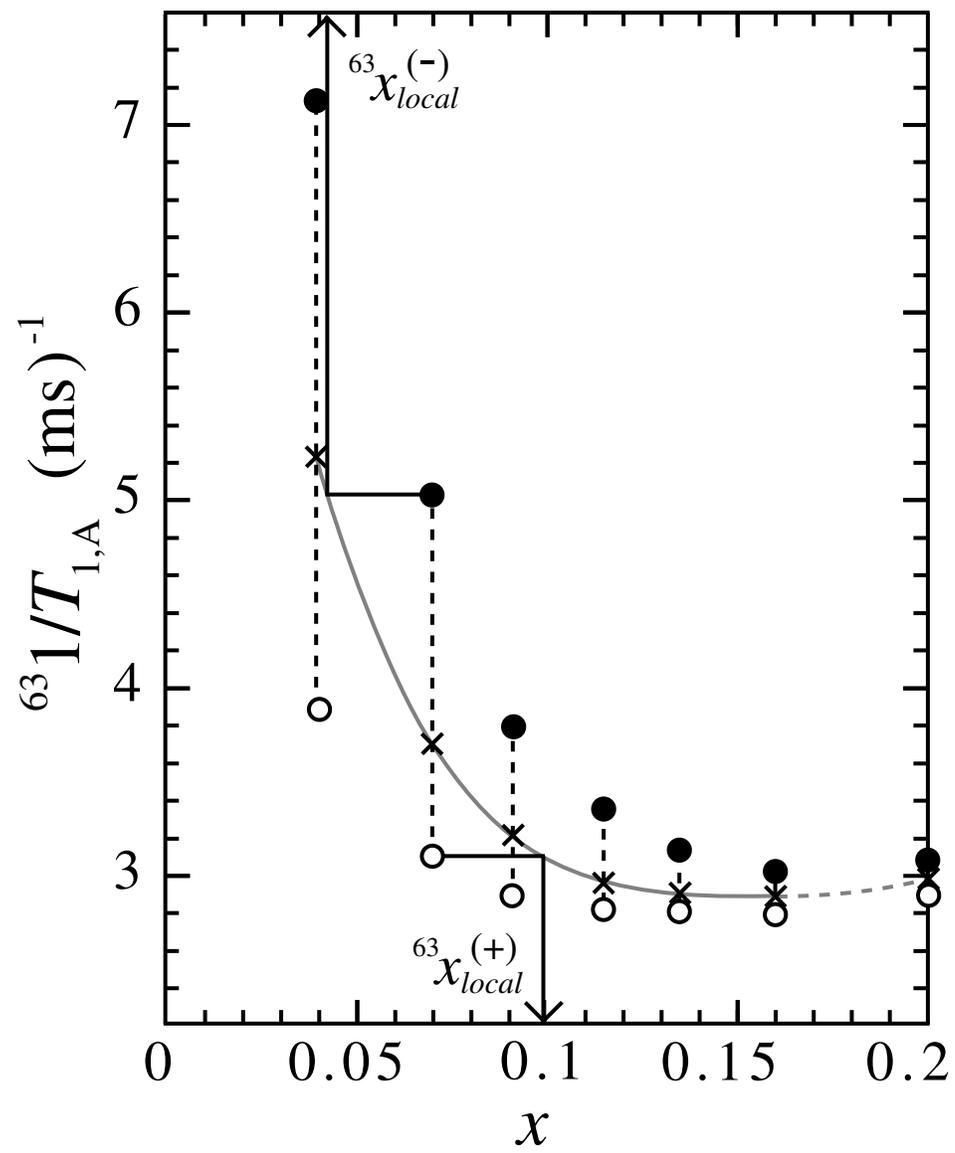
\begin{figure}
\caption{$x$ dependence of $^{63}1/T_{1,A}^{(-)}$ ($\bullet$), 
$^{63}1/T_{1,A}^{(0)}$ ($\times$),
and $^{63}1/T_{1,A}^{(+)}$ ($\circ$) at 295 K where the vertical 
dashed lines connect the data for each $x$.
Grey solid curve 
shows interpolation of $^{63}1/T_{1,A}^{(0)}$ for all $x$, while 
dashed grey line shows overdoped regime. The solid black horizontal 
and vertical lines 
illustrate how to extract $^{63}x_{local}$ in the case of $x=0.07$ 
according to $^{63}1/T_{1,A}^{(+)}$ at the upper frequency side of 
the $^{63}$Cu NQR spectrum defined as $^{63} x_{local}^{(+)}$, and according to $^{63}1/T_{1,A}^{(-)}$
at the lower frequency side defined as $^{63} x_{local}^{(-)}$.}
\label{300Kx}
\end{figure}

In Fig. \ref{300Kx} we illustrate the details of the process used to extract $^{63} 
x_{local}$ from the data in Fig. \ref{T1pm}. We first take the
data from Fig. \ref{T1pm} and plot $^{63}1/T_{1,A}^{(+)}$, 
$^{63}1/T_{1,A}^{(-)}$ and $^{63}1/T_{1,A}^{(0)}$ 
for each $x$ at a fixed temperature of 300 K. For clarity we connect 
the $^{63}1/T_{1,A}$ data for fixed $x$ by the dashed black lines.
We then create a smooth interpolation of $^{63}1/T_{1,A}^{(0)}$ 
for all $x$ shown by the solid grey curve. We show the procedure 
used to determine $^{63} x_{local}$ in the case of $x=0.07$ 
shown as the solid black vertical and horizontal lines. 
In the case of $^{63}1/T_{1}^{(+)}$, the horizontal black line 
determines what value of $^{63}1/T_{1,A}^{(0)}$ (solid grey curve) corresponds to $^{63}1/T_{1}^{(+)}$
, then the vertical line yields $^{63}x_{local}^{(+)}=0.10$ $(\pm 0.01)$ for the more
metallic regions. A similar proceedure for
$^{63}1/T_{1}^{(-)}$ yields $^{63}x_{local}^{(-)}=0.043$ $(\pm 0.008)$ for the more
insulating regions. The upper and lower 
frequency side therefore yield a consistent deviation $^{63}\Delta x_{local}=0.028 (\pm 
0.009)$ for $x=0.07$ at 300 K.
We then determine $^{63} x_{local}^{(+)}$ and $^{63}
x_{local}^{(-)}$ for all $x$ depending on available  
$^{63}1/T_{1,A}^{(0)}$ data and desired accuracy. We cannot 
determine $^{63}x_{local}^{(-)}$ for $x=0.04$ since we do not have 
$^{63}1/T_{1,A}^{(0)}$ data below $x=0.04$ at 300 K. The reason for this 
is that $^{63}1/T_{1}$ cannot be measured accurately for values 
$\gtrsim 7$ (ms)$^{-1}$. Furthermore, 300 K 
corresponds to the wipeout temperature $T_{NQR}$ for $x \leq 0.035$ 
\cite{hunt}, and we only determine $^{63}x_{local}^{(-)}$ 
from our data for $T \gtrsim T_{NQR}$ where full Cu signal intensity is observable. 
As for the limitations of $^{63}x_{local}^{(+)}$, we note that 
at 300 K, $^{63}1/T_{1,A}^{(0)}$ tends to merge to the same value of $\sim 3$ (ms)$^{-1}$ 
for $x \geq 0.09$ \cite{imai}. This implies that one needs to 
measure $^{63}1/T_{1}^{(+)}$ beyond the experimental uncertainties in order 
to get a reliable estimate of $^{63}x_{local}^{(+)}$. 

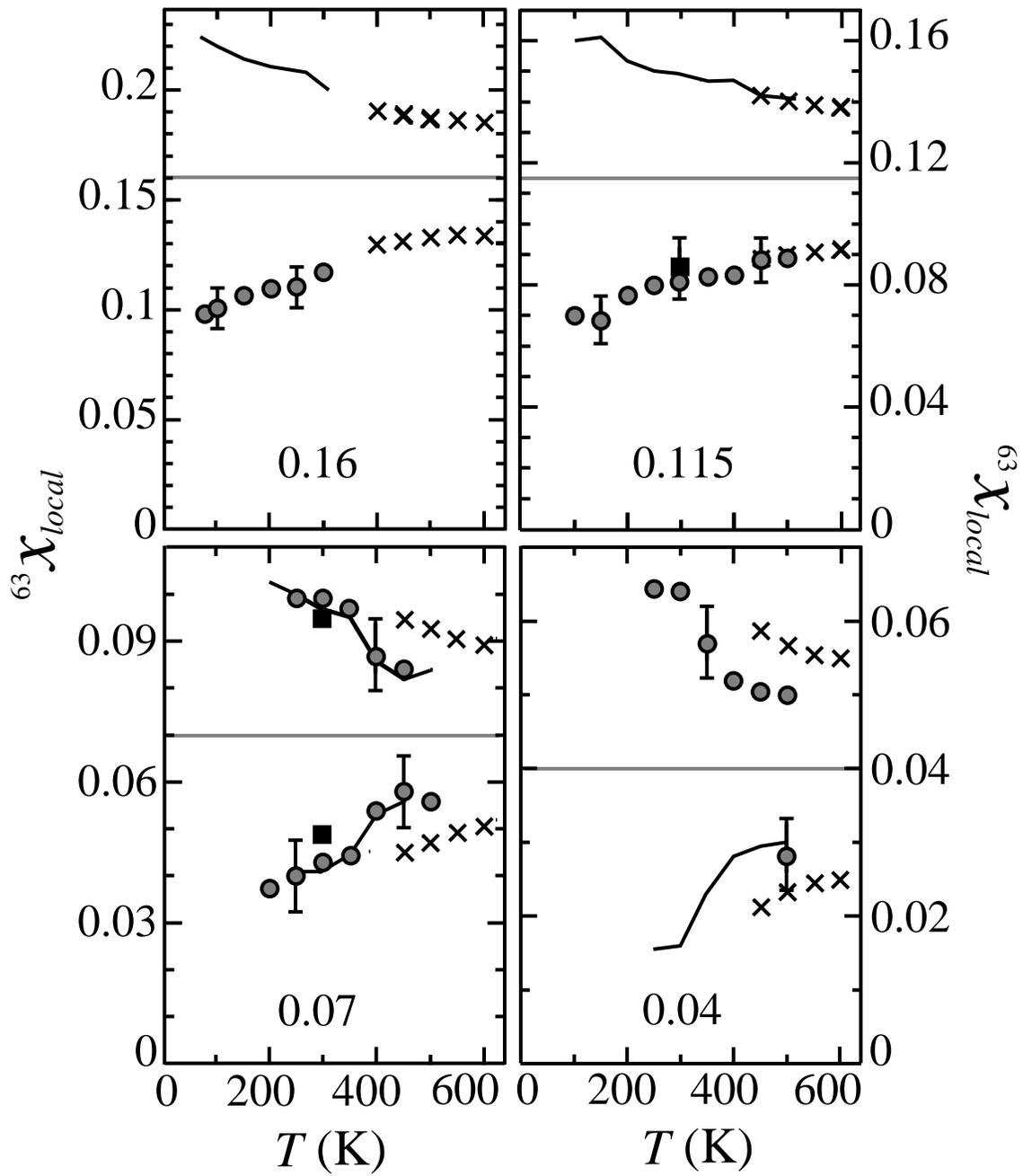
\begin{figure}
\caption{Temperature dependence of the local hole concentration $^{63}x_{local}$ 
(grey filled $\bullet$) deduced from the upper frequency side of 
the A-line using $^{63}1/T_{1,A}^{(+)}$ 
($^{63}x_{local}^{(+)} > x$), and also from the lower frequency side 
using $^{63}1/T_{1,A}^{(-)}$ ($^{63}x_{local}^{(-)}< x$), where $x$ is indicated 
in each section and shown as the grey horizontal lines.
Solid black lines show reflections of (grey filled 
$\bullet$) data through $x$ lines. Also shown is $^{63}x_{local}$ 
deduced using $^{63}1/T_{1,B}^{(+)}$ and $^{63}1/T_{1,B}^{(-)}$ at 
the B-line ($\blacksquare$), together with  
the upper boundary $^{63}x_{patch}$ ($\times$) deduced  
from calculated fit to the $^{63}$Cu NQR spectrum using a patch-by-patch model 
for the spatial variation. Representative error bars for $^{63}x_{local}$ are also shown.}
\label{deltax}
\end{figure}

The whole proceedure is repeated at different temperatures, the final 
results of which are summarized in Fig. \ref{deltax}. 
At different temperatures we 
find similar experimental limitations in determining $^{63}
x_{local}^{(+)}$ and $^{63} x_{local}^{(-)}$, however,
in cases such as $x=0.07$ where both can be determined, we consistently find that $^{63\!\!} \Delta
x_{local}^{(+)} \simeq ^{\;63\!\!}\Delta x_{local}^{(-)}$. 
In the overdoped region $x \geq 0.20$ we find a comparable frequency dependence 
of $^{63}1/T_{1}$ across the NQR line to that of $x=0.16$. However, as 
shown in Fig. \ref{300Kx}, we 
also find that $^{63}1/T_{1,A}^{(0)}$ starts to $increase$ with 
increasing $x$ for $x \geq 0.20$ at 300 K, and the same is true at all temperatures. 
As discussed in section II, we limit our determination of $^{63}x_{local}$ to the region $x \leq 0.16$
and for clarity we separate the data in Fig. \ref{300Kx} for $x > 0.16$ by using a dashed grey line.

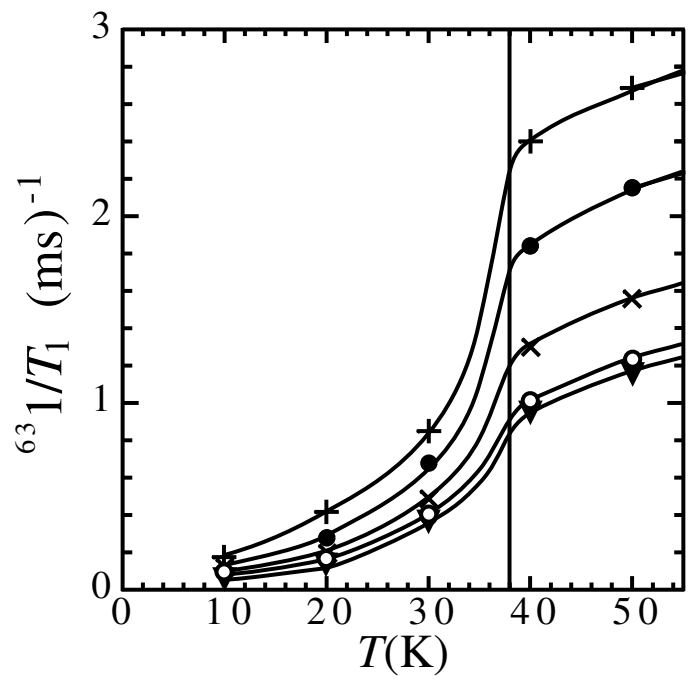
\begin{figure}
\caption{Temperature dependence of $^{63}1/T_{1}$ for $x=0.16$ across the
superconducting boundary at $T_{c}=38$ K (shown as grey verticle line) 
at various positions across the $^{63}$Cu NQR line including 
$^{63}1/T_{1,A}^{(-1/10)}$ ($+$), 
$^{63}1/T_{1,A}^{(-)}$ ($\bullet$), $^{63}1/T_{1,A}^{(0)}$ ($\times$),
$^{63}1/T_{1,A}^{(+)}$ ($\circ$), and $^{63}1/T_{1,B}^{(0)}$ 
($\blacktriangledown$). Black curves are a guide for the eye.}
\label{TcT1}
\end{figure}

In Fig. \ref{TcT1} we show the temperature dependence of $^{63}1/T_{1}$ for $x=0.16$ 
across the superconducting boundary $T_{c}$ = 38 K (Fig. \ref{squid})
at various positions across the NQR spectrum. Despite the fact that 
the spatial variation $^{63}x_{local}$ varies as much as  $0.10 \leq ^{63}
x_{local} \leq 0.22$ at $\sim 100$ K, all values of $^{63}1/T_{1}$ 
show a comparable fractional decrease below $T_{c}$ where the superconducting 
gap opens. In the case of $^{63}1/T_{1}^{(-)}$ where 
$^{63}x_{local}^{(-)}=0.10$, $T_{c}$ should be around 25 K, however, there is 
already a large drop in $^{63}1/T_{1}^{(-)}$ by 30 K. In the case of $^{63}1/T_{1}^{(-1/10)}$ where 
$^{63}x_{local}^{(-1/10)}=0.08$, $T_{c}$ should be as low as 20 K, however 
$^{63}1/T_{1}^{(-1/10)}$ shows a drop below 30 K.
This shows that the superconducting transition is a 
genuine bulk phenomenon which effects all patches, regardless of the local 
hole concentration $^{63}x_{local}$. 

\subsection{$^{63}\nu_{Q}$ Results}

In Fig. \ref{nuqall}(a) and (b) we show the temperature dependence of the 
resonance frequency at the CG of the NQR spectrum defined as
$\left<^{63}\nu_{Q}^{k} \right>$ where $k = (A,B)$ for the A and B-lines, 
respectively. We define the first and second moments of the NQR spectrum as such
\begin{eqnarray}
\left< ^{63\!}\nu_{Q}^{k} \right> &=& \frac{\sum_{j} ^{63\!}\nu_{j,Q}^{k} }{\sum_{j}} \nonumber \\
^{63\!}\Delta \nu_{Q}^{k} &= &\sqrt{\alpha_{0}} \cdot \sqrt{\frac{\sum_{j}
	\left(^{63\!}\nu_{j,Q}^{k}-\left<^{63\!}\nu_{Q}^{k} \right> \right)^{2}}{\sum_{j}}},
\label{meanwidth}
\end{eqnarray}
where $^{63}\nu_{j,Q}^{k}$ corresponds to the $j$th data point of the
observed NQR spectrum and $\sqrt{\alpha_{0}}  =  \sqrt{2Ln2} = 1.177$. For a Gaussian spectrum
(found to be the case for $x > 0.07$)
the use of the prefactor $\sqrt{\alpha_{0}}$ reduces $^{63}\Delta \nu_{Q}^{k}$
to the {\it HWHM} (half width at half maximum) exactly. 

The $\left<^{63}\nu_{Q}^{k} \right>$ data indicates two temperature regimes. Above $T_{Q}$ (black fit), we find a linear decrease of
$\left<^{63}\nu_{Q}^{A} \right>$ with decreasing temperature and a roughly constant 
value of $\left<^{63}\nu_{Q}^{B} \right>$. 
In section IV we attribute the temperature dependence above $T_{Q}$
to thermal contraction of the lattice constants within the tetragonal 
phase, and in appendix A we also use this temperature region to determine the anti-shielding factors needed for the EFG simulation. 
Below $T_{Q}$ (grey fit),
$\left<^{63}\nu_{Q}^{k} \right>$ increases with decreasing temperature for both A and B-lines. The increase in this region is 
attributed to the increasing local orthorhombic distortions \cite{imai} which in the 
case of $x \geq 0.04$ set in 
above the LRO structural phase transition temperature 
$T_{st} (\leq 515$ K) according to neutron diffraction \cite{yamada}.  
In the case of $x=0.20$, we find evidence 
for local orthorhombic distortion starting as high as 
$T_{Q} \simeq$  350 K where diffraction results show no sign of LRO. 
In the section V we fit the data below $T_{Q}$ and deduce the degree 
of tilting of the CuO$_{6}$ octahedra away from the $c$-axis (defined as 
$\theta_{local}$), and we find that there is
a sharp onset temperature for $\theta_{local}$ only the case of $x=0.0$. For $x> 
0.0$ we see a somewhat rounded transition into the local orthorhombic phase. 
We note that these local precursive effects to the LRO are 
consistent with pair distribution function analysis of neutron powder 
diffraction data \cite{bozin} and XAFS (X-ray Absorption Fine Structure) 
analysis \cite{haskel} 
which find evidence for local lattice distortions in the temperature region above 
$T_{st}$, and also into the overdoped regime $x \gtrsim 0.20$. 

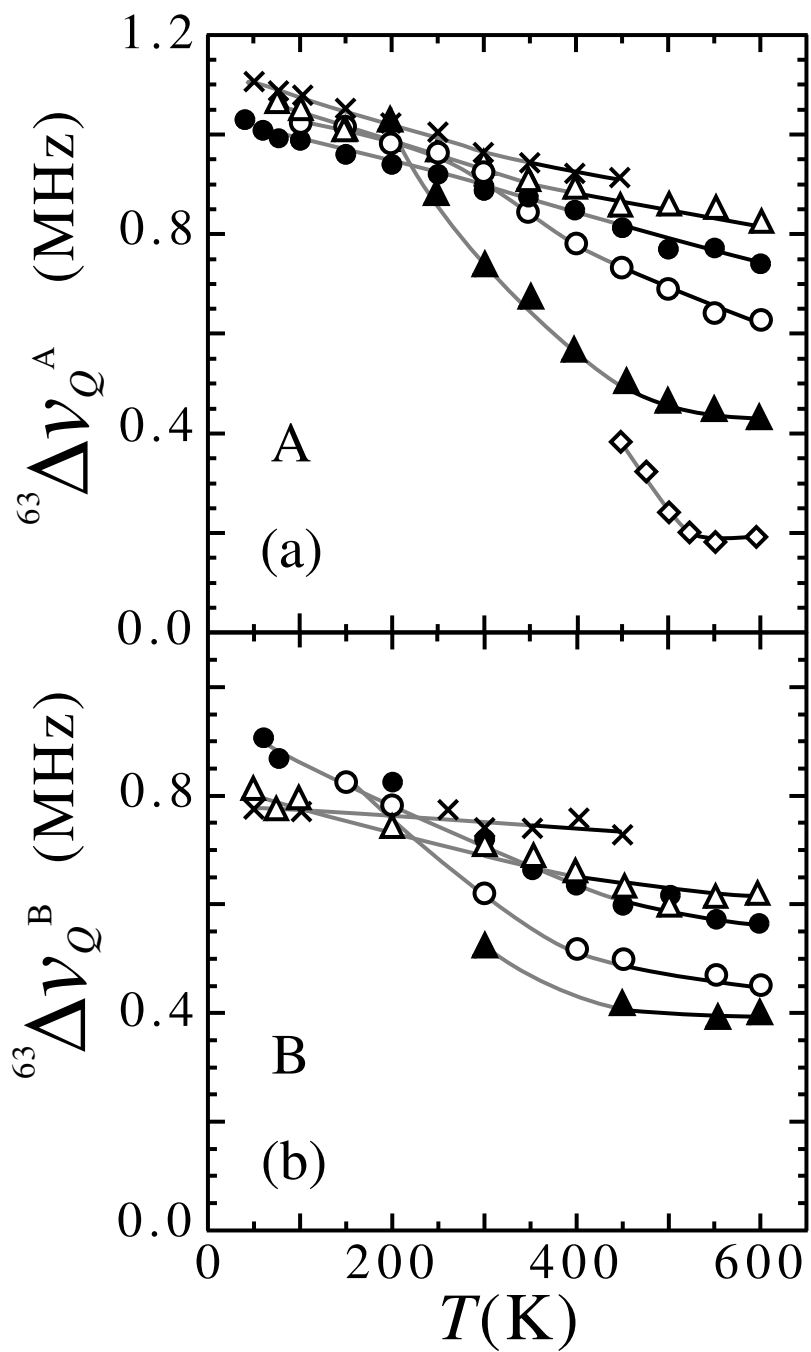
\begin{figure}
\caption{Temperature dependence of {\it HWHM}
of the $^{63}$Cu NQR spectrum 
(a) $^{63}\Delta\nu_{Q}^{A}$ at the A-line, and (b) 
$^{63}\Delta\nu_{Q}^{B}$ at the B-line for $x=0.20$ 
($\times$), $x=0.16$ ($\triangle$), $x=0.115$ ($\bullet$), $x=0.07$ ($\circ$),
$x=0.04$ ($\blacktriangle$), and $x=0.02$ ($\diamond$). 
Black and grey curves indicate temperature region above and below 
$T_{Q}$ respectively.}
\label{widthall}
\end{figure}

The temperature dependence of the observed {\it HWHM} $^{63}\Delta \nu_{Q}^{k}$ of the 
NQR spectra, defined using Eq. (\ref{meanwidth}), is
presented in Fig. \ref{widthall}. We also reproduce 
the temperature regions above and below $T_{Q}$, taken from Fig. 
\ref{nuqall}, as the black and grey curves
respectively. In section IV we use the $^{63}\Delta \nu_{Q}^{k}$ data 
for $T > T_{Q}$ to deduce a lower bound for the patch radius $^{63}R_{patch}$, 
and the corresponding an upper bound
$^{63}\Delta x_{patch} (\propto 1/^{63}R_{patch})$.

\section{analysis}

\subsection{EFG Background}

The goal of this section is to determine a 
static real-space model of the spatial variation in local hole 
concentration $^{63}x_{local}$. We shall use a point charge lattice summation 
incorporating randomness effects to 
calculate the inhomogeneous $distribution$ in the EFG
which constitutes the observed $^{63}$Cu NQR spectrum. To the best of 
our knowledge, such calculations incorporating randomness effects 
into the EFG calculation have not been reported. 
We shall deduce all parameters necessary for this task
in a self-consistent way by using our pool of NQR data, 
and we shall successfully account for the entire $^{63}$Cu NQR spectrum, 
including the B-line, with one adjustable parameter ($^{63}R_{patch}$) 
which defines the length scale of the spatial variation $^{63}x_{local}$.
Once $^{63}R_{patch}$ is optimized to fit the $^{63}$Cu NQR spectrum, we 
deduce an upper bound to the amplitude of the spatial variation across the 
plane $^{63}\Delta x_{patch} (\propto 1/ ^{63}R_{patch}$), which we show 
is consistent with $^{63}\Delta x_{local}$ determined in a model 
independent way from section III.

The distribution in the EFG is dominated by three mechanisms (a) the random substitution 
of donor ions, (b) the variation in local hole 
concentration $^{63}x_{local}$ over short length scales $^{63}R_{patch} \gtrsim 3.0$ 
nm, and (c) the distribution in local lattice distortions. Without {\it 
a priori} knowing the lattice distortions in (c), we attribute the whole NQR 
linebroadening to mechanisms (a) and (b). This implies that the deduced 
$^{63}R_{patch}$ we present is a lower bound and correspondingly $^{63} \Delta x_{patch} (\propto 1/^{63}R_{patch}) $
is an upper bound. However we shall demonstrate that $^{63} \Delta x_{patch} 
\sim ^{63}\Delta x_{local}$, which implies that using mechanisms (a) and (b) alone is 
justified in the temperature range $T > T_{Q}$. Below $T_{Q}$ where 
local orthorhombic distortions set in (Fig. \ref{nuqall}), we terminate our analysis of $^{63} 
\Delta x_{patch}$. The reason for this is that we also see an increase 
in the NQR linewidth (Fig. \ref{widthall}) below $T_{Q}$ in the case 
of $x \leq 0.07$, which suggests that ignoring mechanism (c) above is 
no longer justified for $T < T_{Q}$. In section V we make 
predictions about the magnitude of mechanism (c) by attributing the 
extra linebraodening for $T < T_{Q}$ to distributions in the lattice 
distortions. We note, however, that with the exception of section V, we ignore mechanism 
(c) in our analysis.

There are two approaches for computing the EFG.
The first is an $ab$ $initio$
approach \cite{martin,plib,husser} which involves quantum chemistry calculations,
and the second approach makes use of the experimental 
data \cite{penn,takigawa1989,shimizu,walstedtprb,singerprl} 
to deduce all the necessary parameters in an empirical way.
The $ab$ $initio$ approach has been extensively used to calculate
the EFG at the $^{63}$Cu site in La$_{2}$CuO$_{4}$ 
\cite{martin,plib,husser}, and 
gives a consistent value for the observed resonance frequency 
$\left<^{63}\nu_{Q}\right> \simeq 33 $ MHz \cite{imai} to within uncertainties in the 
quadrupole moment $^{63}Q$.
The basic idea behind the {\it ab initio} approach is that since
the EFG decreases rapidly $\sim 1/r^{3}$ away from the origin of the 
calculation, 
the most significant contributions should be from local EFG components. One can then 
justifiably separate a cluster of ions in the 
immediate vicinity of the Cu nucleus whose contributions are 
calculated using a full spin-polarized DF (density functional) 
or HF (Hatree-Fock) calculation, while the rest of the ions in the 
crystal are treated as point charges.
The smallest realistic cluster is typically CuO$_{6}$/Cu$_{4}$La$_{10}$,
where the central (CuO$_{6})^{-10}$ ionic cluster consists of 23 molecular orbitals 
for each spin projection made up of linear 
combinations of 5 $3d$ and 18 $2p$ 
atomic orbitals. Such calculations are found to depend 
on the cluster size used, where the larger clusters are more reliable 
yet involve increasingly complex calculations. They also depend on whether one uses DF 
or HF, however both successfully predict that the molecular 
orbital with the highest energy is the 
anti-bonding hybridization between the Cu $3d_{x^{2}-y^{2}}$ orbital 
and the O $2p_{x}$ and $2p_{y}$ orbitals. The DF method, however, 
predicts \cite{husser} more covalent-like bonding where the localized atomic 
spin density on the Cu is $\rho_{Cu} = 0.67$, while HF predicts 
more ionic-like bonding with $\rho_{Cu} = 0.90$.

Before discussing our new approach for calculating the inhomogeneous distribution of 
the EFG, let us clarify exactly what 
inhomogeneous broadening implies in the context of $^{63}$Cu NQR.
The linebroadening of the $^{63}$Cu NQR spectrum for $x >$ 0.02
is dominated by the $inhomogeneous$ 
distribution in the EFG. 
The inhomogeneous broadening may be pictured as such: say there is
a Cu nucleus $j$ lying in a particular EFG with value 
$V_{j}$ which resonates at frequency $\nu_{j}$, while a distant nucleus $k$ sits in a 
distinct EFG environment $V_{k}$ and resonates at $\nu_{k}$. As is predominantly the 
case across the sample, the separation  
between the two nuclei ${\bf r}_{(j,k)}$ 
is $larger$ than the range $\xi_{AF} 
\sim $3 $a$ of the indirect nuclear spin-spin coupling 
\cite{pennslicht} where $\xi_{AF}$ is the correlation length of the anti-ferromagnetic fluctuations.
In such cases, one may flip the $j$ nucleus using an r.f. pulse and observe 
its echo $without$ being effected by what the r.f. pulse does on
the $k$ nucleus. One then measures the $^{63}$Cu NQR spectrum by resonating at $\nu_{j}$ 
and effectively counting 
the number of nuclei $N_{j}$ in the sample with EFG values $V_{j}$, 
then changing the resonance frequency 
to $\nu_{k}$ and counting the number of nuclei $N_{k}$, thereby building up the 
$inhomogeneous$ NQR spectrum.

As one approaches the undoped limit $x < 0.02$, the random effects 
become less and less significant, and the 
linebroadening becomes predominantly $homogeneous$ in nature and dominated by 
the indirect nuclear spin-spin coupling \cite{slichter}. 
As we show in appendix B, in the case of pure homogeneous broadening 
for $x=0.0$, there is no frequency dependence in $^{63}1/T_{1}$ across the NQR 
spectrum, even in the orthorhombic phase $T<T_{st}$.
We note that the two $distinct$ length scales $\xi_{AF}$ and $^{63}R_{patch}$ 
which determine the 
linebroadening of the homogeneous and inhomogeneous NQR spectra, 
respectively, have qualitatively different temperature dependences. As 
we shall show, $^{63}R_{patch}$ mildly decreases with decreasing temperature, 
while $\xi_{AF}$ increases with decreasing temperature \cite{yamada}. 
Furthermore $^{63}R_{patch} \gg \xi_{AF}$ within the $x$ and temperature 
range of interest. 

\subsection{EFG Simulation}

In the experimental approach, the total EFG tensor 
$V_{tot}^{(\beta,\gamma),k}$ (see Eq. (\ref{mainvq2})) along the orthogonal 
spatial directions 
$(\beta,\gamma$) for the Cu site $k = (A,B)$ is segregated into an onsite $3d$ contribution 
$V_{3d}^{(\beta,\gamma)}$
and an offsite lattice contribution $V_{latt}^{(\beta,\gamma),k}$ as such
\begin{equation}
V_{tot}^{(\beta,\gamma),k} =  \zeta_{3d}^{k} \cdot V_{3d}^{(\beta,\gamma)}+ 
\zeta_{latt}^{k} \cdot V_{latt}^{(\beta,\gamma),k},
\label{sepvq}
\end{equation}
where we introduce the isotropic anti-shielding factor \cite{slichter}
$\zeta_{3d}^{k}$ originating from the local $3d$ electrons and the 
isotropic anti-shielding factor
$\zeta_{3d}^{k}$ orginating from the lattice.
We also leave the possibilty open that they depend on which Cu site $k = (A,B)$ 
is being calculated. In appendix A, we use a self-consistent approach to deduce that $\zeta_{latt}^{A} = 27.7$ and 
$\zeta_{3d}^{A} = 0.93$, while $\zeta_{latt}^{B} = 18.6$ 
and $\zeta_{3d}^{B} = 0.82$. We also show that the main principle value of the onsite EFG contribution 
$V_{3d}^{c}$ is negative while the lattice 
contribution $ V_{latt}^{c,k}$ is positive, resulting in an overall negative value of $V_{tot}^{c,k}$.

The anti-shielding
factors must be determined experimentally, while in the $ab$ $initio$ 
approach they are, in a sense, already accounted for. The 
anti-shielding factors are found $not$ to vary significantly between different classes of high $T_{c}$
cuprates \cite{shimizu}. Likewise, the anti-shielding factors 
deduced in a similar way for the planar oxygen site
\cite{takigawa1989,walstedtprb,tobe}
is also found to be consistent 
between different classes of cuprates, which supports the use 
of the experimental approach.

In accord with B. Bleaney and co-workers \cite{bleaney}, the main principle 
value of the onsite
contribution $V_{3d}^{c}$ arising
from the $3d_{x^{2}-y^{2}}$ hole is taken to be 
\begin{equation}
	 V_{3d}^{c} = -\frac{4}{7} e \left(1-4f_{\sigma}^{o}\right)
	 \left<r^{-3}_{3d}\right>,
	 \label{loccontr}
	 \end{equation}
where we have allowed for covalency in the form of 
$f_{\sigma}^{o}$ which represents the fractional spin denisty on the 
four neighbouring planar oxygen $2p_{\sigma}$ orbitals. 
In order to be 
consistent with our results of $^{17}$O hyperfine coupling 
analysis we use $f_{\sigma}^{o} = 0.076$ \cite{tobe} which favours more
covalent-like bonding.
Using ionic bonding with $f_{\sigma}^{o}=0$ results in an overall $\sim$ 14 \%
increase in $^{63}R_{patch}$ and corresponding $\sim$ 12 \% decrease 
in our theoretical estimate $^{63}\Delta x_{patch}$ over covalent-like bonding.
We note that $f_{\sigma}^{o} =0.076$ deduced from the oxygen hyperfine couplings
\cite{walstedtprb,tobe} is consistent
with Cu hyperfine coupling analysis using NMR \cite{takigawa1991} 
where $f_{Cu} = (1-4f_{\sigma}^{o})  = 0.7 $ 
was found. It is also generally accepted that the 
$ab$ $initio$ DF calculation which favors stronger covalency is a more 
realistic approach than the HF approach. 

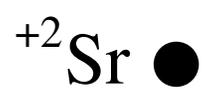
\begin{figure}
\caption{Section of the CuO$_{2}$ plane illustrating an example of the random placement of $^{+2}$Sr ions. The corner of each square 
represents a Cu site and the central Cu site ($\circ$) is taken as the origin of the EFG calculation. The (grey $\bullet$) symbols represent the planar O sites. 
The $^{+2}$Sr donor ions ($\bullet$) 
immdediatley above or below the Cu sites are shown, corresponding to 
the donors $d_{1}$ (see later). Also shown is a 
typical patch with a radius of
$^{63}R_{patch} = 3.0$ nm (=8 $a$) from the origin. For
this particular random configuration $\kappa$, the local hole 
concentration within the patch is $^{\kappa}x_{local} = 0.05$.}
\label{patchsimul}
\end{figure}

The lattice contribution $V_{latt}^{(\beta,\gamma)}$ to the 
total EFG is calculated using a point charge lattice summation
with fractional point charges. We use the standard 
expression for the summation
\begin{equation}
	V_{latt}^{(\beta,\gamma)} = 
	\sum_{j} q_{j}\left(3x_{\beta,j} x_{\gamma,j} -r_{j}^{2} 
	\delta_{\beta,\gamma}\right)/r_{j}^{5}, \\ 
\label{lattcontr}
\end{equation}
where the sum over $j$ refers to the sum over the surrounding point 
charges $q_{j}$ in the lattice a distance $r_{j}$ away. 
The charges $q_{j}$ are assigned as follows
\begin{eqnarray}
q_{Cu}& =& +\left(2-4f_{\sigma}^{o}\right)e  \nonumber \\
q_{O_{pl}} &= &-\left(2-2f_{\sigma}^{o} - x/2\right)e  \nonumber \\
q_{O_{ap}}& =& -2e \nonumber \\
q_{La} &=& +3e \nonumber \\ 
q_{Sr} &=& +2e,
\label{charges}
\end{eqnarray}
where $x$ is the nominal hole concentration, O$_{pl}$ stands for the planar oxygen site, O$_{ap}$ 
stands for the apical oxygen site and $e(>0)$ is the electronic charge 
(in emu). Note that we make use of the full difference in valency between the $^{+3}$La and 
$^{+2}$Sr ions. This implies that depending on the the geometrical 
distribution of the $^{+2}$Sr ions in the lattice, 
$V_{latt}^{(\beta,\gamma)}$ takes on different values. In Fig. 
\ref{patchsimul} we illustrate an example
of the random positioning of the $^{+2}$Sr ions surrounding the central 
Cu site from which the EFG is calculated.

The standard approach in all earlier works, except for 
\cite{singerprl}, has been to set 
$q_{La} = q_{Sr} = -(3-x/2)$ which effectively bypasses any randomness 
effects and results in delta function spectra of the EFG.
We start the EFG simulation by first taking 
$^{63}R_{patch} = \infty$ which uniformly places the donated holes $x$ 
from the $^{+2}$Sr ions with full mobility 
onto the planar O sites. The added holes reduce  
$|q_{O_{pl}}|$ in Eq. (\ref{charges}), bearing in mind that there are two planar 
oxygens per unit cell.
Placing the donated holes onto the planar oxygen sites as such is consistent 
with high energy spectroscopy studies \cite{fujimori}.

The positions of the
ions are allocated using an orthorhombic
cell (shown in Fig. \ref{illust}) which consists of $4 \times$(La$_{2-x}$Sr$_{x}$CuO$_{4}$)
formula units, each with the K$_{2}$NiF$_{4}$ structure 
\cite{ginsberg}. Using an orthorhombic cell will later permit us to incorporate the orthorhombic distortions in section V. 
The absolute values of the lattice constants $[a_{o},b_{o},c]$ 
were calibrated at 295 K using our data from Fig. \ref{latt}. 
The positions ${\bf r}_{j}(x,T)$ of the ions as a function of $x$ and 
$T$ were taken from a 
systematic X-ray powder diffraction study by P.G. Radaelli {\it et al.} 
\cite{radaelli}. The EFG calculation is 
naturally sensitive to inputs ${\bf r}_{j}(x,T)$, therefore we use
smooth interpolation of the values ${\bf r}_{j}(x,T)$ for different $x$
in order to avoid unnessecary scattering in the output 
$V_{latt}^{(\beta,\gamma)}$.
We also use the same thermal expansion coefficients 
\cite{radaelli,braden2}
$\alpha_{a} = +1.45\cdot 10^{-5}$ K$^{-1}$ and $\alpha_{c} = +1.42\cdot 
10^{-5}$  K$^{-1}$ for all $x$, consistent with corresponding 
thermal coefficients found in La$_{2-x}$Ba$_{x}$CuO$_{4}$ \cite{suzuki}.

The orthorhombic cell is then repeated in all space out to a radius of 50 \AA$^{ }$
from the origin $[0,0,0]$ of the EFG calculation.
The procedure for the random placement of the $^{+2}$Sr ions is carried out as 
follows: a different random number $\lambda$ with a flat probability distribution 
$0<\lambda<1$ is generated at each La site.
If $\lambda >x/2$ at a particular La site, a La ion with charge $+3e$ is assigned to 
that site, while if $\lambda< x/2$, a Sr ion with charge $+2e$ is 
assigned. In 
equation form this gives the following prescription
\begin{eqnarray}
	\lambda >x/2 &(\rightarrow)& +3e \nonumber \\
\lambda <x/2 &(\rightarrow)& +2e
\label{which}
\end{eqnarray}
at each La/Sr site. 
The factor 1/2 in the probability condition arises because there are 2 La/Sr sites per formula unit.
Once this is carried out at each La site, we have achieved the particular random configuration $\kappa$
of $^{+2}$Sr ions in the lattice, an illustration of which is given in 
Fig. \ref{patchsimul}.
In our formulation, we exclude the possibility of any $^{+2}$Sr - $^{+2}$Sr
clustering beyond probablilty theory, which is supported by 
the absence of any significant diffuse 
scattering in neutron diffraction experiments \cite{braden}. 

Once the random configuration $\kappa$ is determined, the summation in Eq. 
(\ref{lattcontr}) is carried out and 
$^{\kappa}V_{latt}^{(\beta,\gamma)}$ is diagonalised and the principle 
value $^{\kappa}V_{latt}^{c}$ is stored into a vector $P_{latt}$.
The calculation is computed in the tetragonal phase where the main principle axis of 
$^{\kappa}V_{latt}^{(\beta,\gamma)}$ is found to lie primarily along the 
$c$-axis (i.e.
$^{\kappa}V_{latt}^{c}$ is the main principle value). We note, however, that each configuration $\kappa$ causes a slight rotation of the 
the main principle axis away from the $c$-axis. Strictly speaking, for each configuration $\kappa$ one should then go one step further and take the projection of the main principle value \cite{abragam} along the $c$-axis (i.e. along the average main principle axis). In the case of $^{63}$Cu NQR, however, 
the projection only makes $\sim 2$ \% difference in $^{\kappa}V_{tot}^{c}$, and can be neglected. 

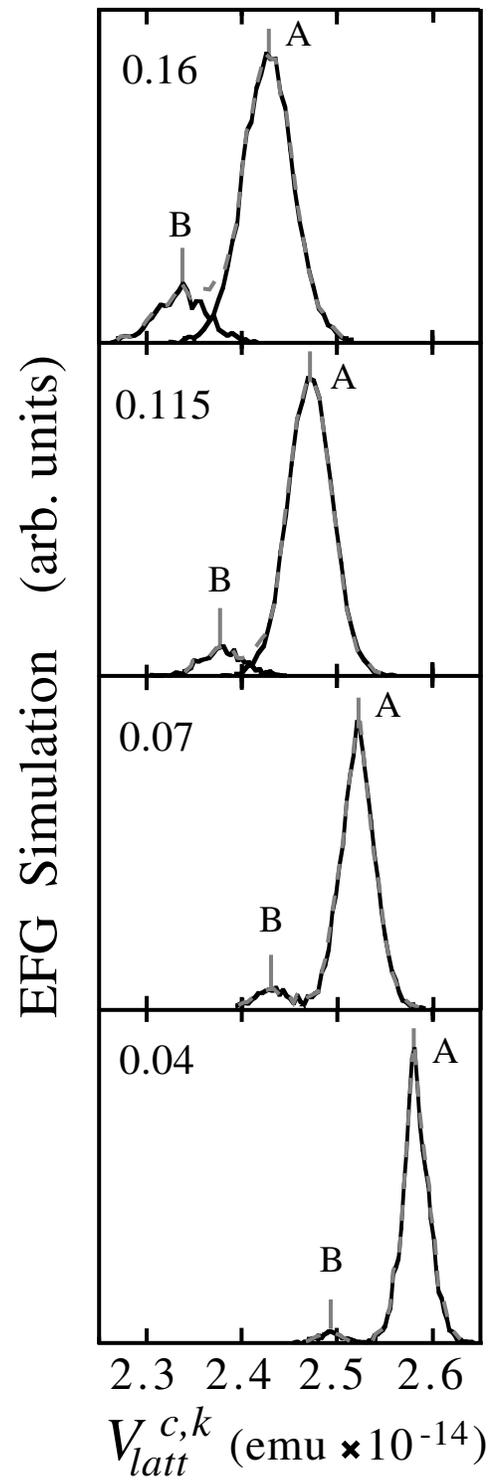
\begin{figure}
\caption{Results of the EFG simulation incorporating the random positioning 
of $^{+2}$Sr ions in the lattice. Grey dashed lines indicate spectrum of the principle 
value $V_{latt}^{c}$ of the EFG in [emu $\times 10^{-14}$] at 600 K for 
various $x$ shown in each part. Black lines show the same data but with the 
A and B-sites separated ($V_{latt}^{c,k}$, where $k=(A,B)$) within the calculation. 
Short grey vertical lines indicate 
CG position $\left<V_{latt}^{c,k}\right>$ used to calculate anti-shielding 
factors.}
\label{simul1}
\end{figure}

Next, the whole lattice 
is re-randomized and a new random
configuartion $\kappa'$ of $^{+2}$Sr ions is determined, the lattice summation 
in Eq. (\ref{loccontr}) is computed and the new value of 
$^{\kappa'}V_{latt}^{c}$ is stored into 
$P_{latt}$. This proceedure is re-iterated $\sim 10^{4}$ times until the dimension of
$P_{latt}$ is sufficiently large to create a histogram spectrum of 
$P_{latt}$ with $\sim 50$ bins across the lineshape, the results of which we show at 600 K in Fig. \ref{simul1} as the dashed grey lines.
Note that we always calculate the EFG for the Cu site at 
the origin using a different random configuration of $^{+2}$Sr ions 
for each run. An equivalent approach is to stick to one random configuration 
$\kappa$ for the whole sample
and calculate the EFG at each Cu site in the lattice thereby 
building up $P_{latt}$, however, this latter approach is computationally more intense. 
Since we are dealing with a purely random system, both methods are 
equivalent, therefore we use the former method.

We notice three immediate features about the results in Fig. \ref{simul1}. The first is the 
overall decrease in CG $\left<V_{latt}^{c}\right>$ with increasing 
$x$. This is a consequence of the change in lattice paramters and the 
decrease in $|q_{O_{pl}}|$ with increasing $x$. The second feature is the 
change in {\it HWHM} which is found to increase with increasing $x$
as a result of the 
increased amount of quenched disorder in the lattice.
The third feature in our lattice summation is the presence of the 
secondary peak known as the B-line.

\subsection{B-line}

The $^{63}$Cu NQR spectrum is known to show a secondary peak known as the 
B-line \cite{yoshimura}. Generally, the origin of the B-line has been 
attributed to structural effects from the dopant ions \cite{yoshimura,song},
but the limited range of $x$ investigated in earlier studies did not allow unambiguous 
identification of the origin of the B-line. In fact, there has been a 
persistent claim that the origin of the B-line is more exotic in 
nature \cite{hammel,martin,stat} and originates from an intrinsic response of the 
material to the presence of doped holes.
Such a conjecture that the B-line is an electronic rather than a 
structural effect was based on the observed similarity between the A-B line 
splitting $\sim 3$ MHz for both La$_{2-x}$Sr$_{x}$CuO$_{4}$ and 
La$_{2}$CuO$_{4+\delta}$ systems \cite{hammel,martin,stat}. It was 
thought that the same A-B line splitting between such distinct
systems was clear evidence that the second site was a result of the 
presence of the doped holes themselves, independent of the means of doping.
If this is the case, however, then it should equally well be 
observed in other 214 compounds such as 
La$_{2-x}$Ba$_{x}$CuO$_{4}$. However, a much larger A-B line splitting of  
$\sim 6$ MHz \cite{yoshimura,hunt2} is observed in 
La$_{2-x}$Ba$_{x}$CuO$_{4}$ where the $^{+2}$Ba ion is  
known to cause much larger ionic size effect disorder \cite{bozin} 
than the $^{+2}$Sr ion, and
which also stablizes an additional structural phase transition at low 
temperatures \cite{suzuki}.  
We also recall that the $^{63}$Cu NQR spectrum in 
La$_{2-x-y}$Eu$_{y}$Sr$_{x}$CuO$_{4}$ for $^{63}$Cu isotope enriched 
samples \cite{hunt2} shows a third structural peak (the C-line) whose 
fractional intensity is observed equal $y$. The C-line corresponds to a Cu 
nucleus directly above or below a Eu$^{3+}$ ion.

In Fig. \ref{fab} we present the experimental value of the relative intensity ($f_{B}^{exp}$)
of the B-line ($I_{B}$) to the total (${I_{A}+I_{B}}$) defined as such
\begin{equation}
    f_{B}^{exp} = \frac{I_{B}}{I_{A}+I_{B}},
    \label{bline}
    \end{equation}
where both $I_{A}$ and $I_{B}$ have been corrected for differences in spin-spin relaxation 
rates. Typically, $^{63}T_{2,A}/^{63}T_{2,B} \sim 0.85$ for both components of the spin-spin relaxation rate \cite{T2G}, similar to the ratio of $^{63}1/T_{1}$.
Our data in Fig. \ref{fab} shows that $f_{B}^{exp} \simeq x$ over a 
large range of $0.04 \leq x \leq 0.20$, which is consistent with the EFG simulation where $f_{B} = x$.
According to the EFG simulation, the Cu B-sites (illustrated in Fig. \ref{illust}) correspond to Cu nuclei located at ${\bf r} = [n_{a}a,n_{b}a,n_{c}c]$ (where 
$[n_{a},n_{b},n_{c}]$ are $\pm$ integers ranging from zero to $\infty$),
which have distinct EFG values due to the presence of $^{+2}$Sr ions 
located at positions ${\bf r} = [n_{a}a,n_{b}a,(n_{c} \pm 0.361)c]$. In this interpretation, the relative intensity of such Cu 
sites ($f_{B}$) is then given by the concentration of $^{+2}$Sr ions $x$, 
i.e. $f_{B} = x$. This site assignment for the B-site Cu nuclei can naturally account for the 
observed relative intensity $f_{B}^{exp} \simeq x$ (Fig. \ref{fab}).

Other evidence that the B-line is primarily structural in origin 
is the similarity in the frequency dependence of $^{63}1/T_{1,B}$ to $^{63}1/T_{1,A}$ (Fig. \ref{300T1f}).  
The similarity between in the frequency dependence of $^{63}1/T_{1,k}$
indicates that the length scale of the spatial
variation in $^{63}x_{local}$ is $larger$ 
than the average $^{+2}$Sr - $^{+2}$Sr distance $l_{Sr} = 
a/\sqrt{x}$, or equivalently the average B-site to B-site distance.
Later we show that the deduced length scale 
$^{63}R_{patch} \gtrsim 3$ nm is larger than $l_{Sr}$ in the region of interest $x>0.02$. Furthermore, as shown in Fig. \ref{deltax}, 
$^{63}x_{local}$ across the B-line is
consistent with $^{63}x_{local}$ across the A-line.

We also note that in Fig. \ref{T1B}, $^{63}T_{1,A}/ ^{63}T_{1,B}$ = 
0.9-0.84, i.e. that $^{63}1/T_{1,B}$ is uniformly suppressed 
compared with $^{63}1/T_{1,A}$. Our systmetic study up to $x=0.20$
showed that in the overdoped region $x \geq 0.20$, 
$^{63}1/T_{1}$ increases with increasing $x$. This implies that if 
there is hole localisation in the vicinity of a Cu B-site as claimed 
in Ref. \cite{hammel,martin,stat}, then $^{63}1/T_{1,B}$ at the 
B-site should be $larger$ than $^{63}1/T_{1,A}$ in the case of 
$x=0.16$. Fig. \ref{T1B} clearly shows that this is not the case, 
and further supports our interpretation that the B-line is primarily 
structural in origin. 
The most likely explanation for $^{63}T_{1,A}/^{63}T_{1,B} = 0.9-0.84$ (and $^{63}T_{2,A}/^{63}T_{2,B}$) is that
the hyperfine coupling constant $|^{63}A({\bf Q_{AF}})^{\perp}|$ is 5-10 \% lower 
at the B-site due to the increased lattice 
distortions in the vicinity of a $^{+2}$Sr ion.
It is known from XAFS 
\cite{haskelrapid} that the $^{+2}$Sr ion causes large structural 
distortions in its vicinity, which could well account for such changes in 
the hyperfine couplings. 

\subsection{EFG $\rightarrow ^{63}\nu_{Q}$}

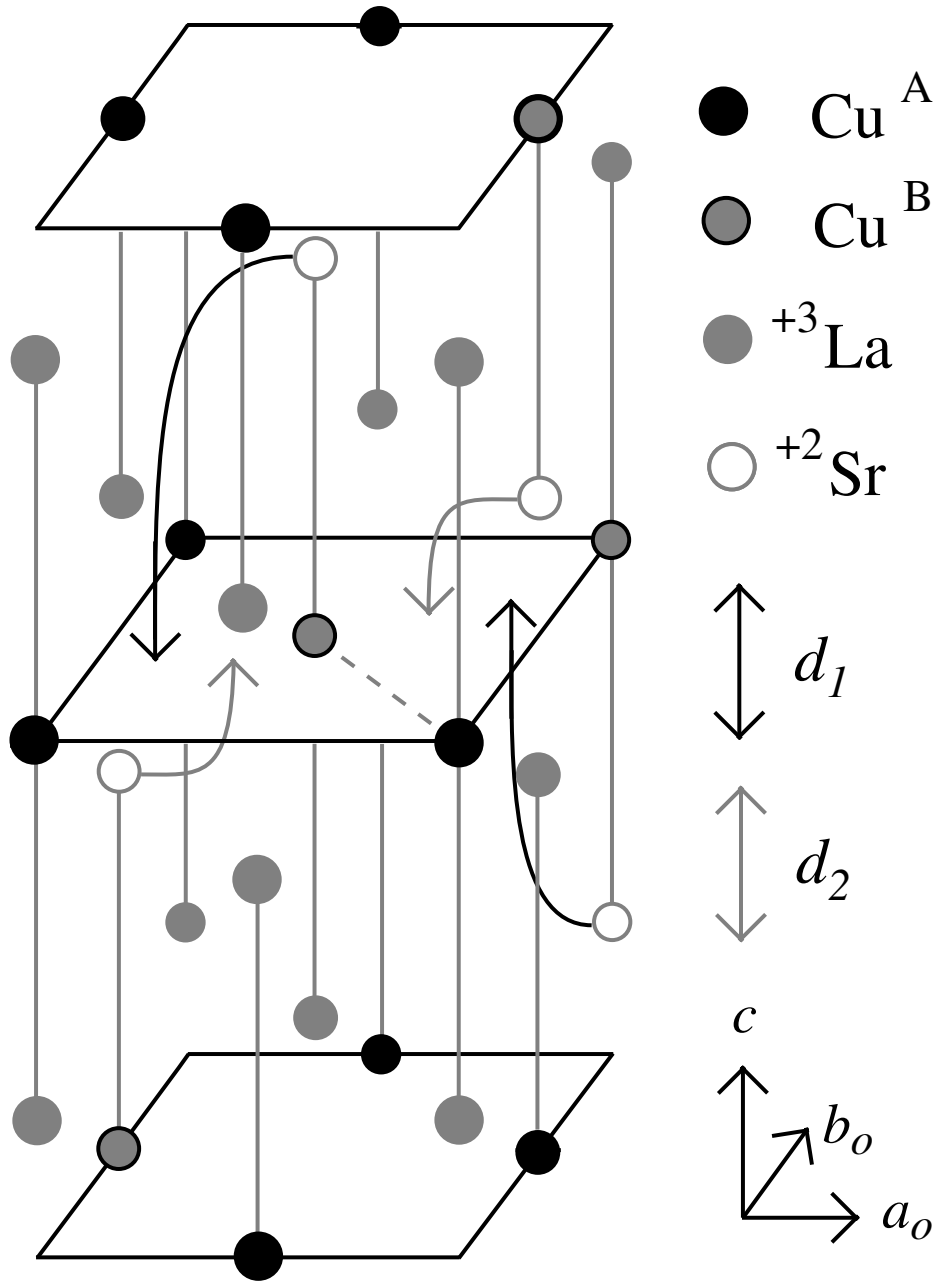
\begin{figure}
\caption{Illustration of the A-site Cu nucleus ($\bullet$) and
B-site Cu nucleus (grey filled $\circ$) determined according to adjacent positions of $^{+3}$La 
ion (grey $\bullet$) and $^{+2}$Sr ion (grey $\circ$), respectively.
Also shown are methods of doping into the central 
plane using $d_{1}$ $^{+2}$Sr sites (black curves) or
$d_{2}$ $^{+2}$Sr sites (grey curves). An 
orthorhombic cell $[a_{o},b_{o},c]$ is used, while dashed grey line illustrates Cu-Cu lattice spacing $a$ in tetragonal phase.}
\label{illust}
\end{figure}

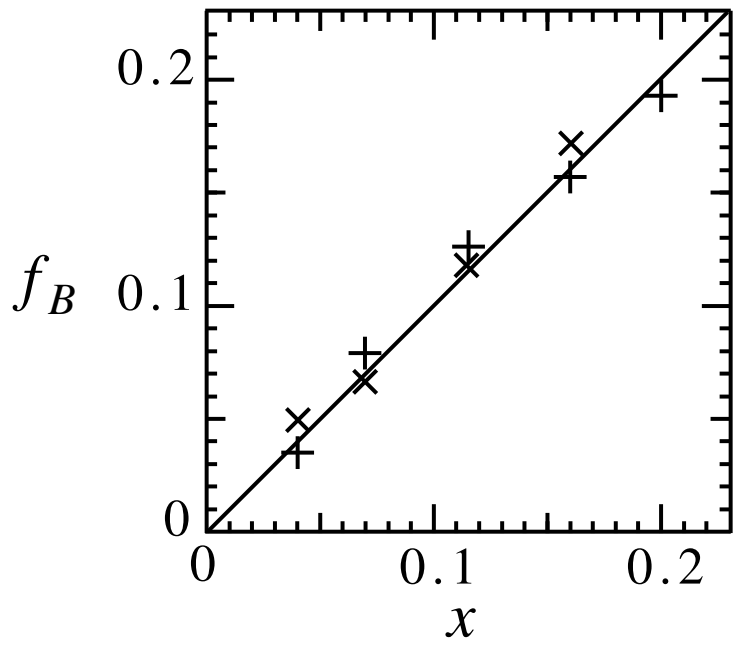
\begin{figure}
\caption{$x$ dependence of the fractional B-line intensity $f_{B}$
defined as $f_{B} = \frac{I_{B}}{I_{A}+I_{B}}$, where $I_{A}$ and $I_{B}$ 
have been corrected for differences between  
the spin-spin relaxation rates at the A and B-sites, respectively, at 300 K (+) and 600 K ($\times$). 
Solid line shows prediction from model $f_{B} = x$, and an illustration of  
the B-site is specifically shown in Fig. \ref{illust}.}
\label{fab}
\end{figure}

We now proceed with our simulation taking the B-site assignment shown in Fig. \ref{illust}.
We can separate the A and B-sites within the calculation by 
separating any randomly generated $^{\kappa}V_{latt}^{c}$ value with a $^{+2}$Sr ion 
in a position illustrated in Fig. \ref{illust}, and placing it into a new vector $P_{latt}^{B}$
, while placing all other generated $^{\kappa}V_{latt}^{c}$ values 
into another vector $P_{latt}^{A}$. This allows us to separate the A and B-lines as shown by the solid 
black lines in Fig. \ref{simul1}, and we attach the superscript $k = (A,B)$ to $^{\kappa}V_{latt}^{c,k}$. The first and second moments of the lattice contribution to the EFG are then computed as such
\begin{eqnarray}
\left<V_{latt}^{c,k} \right> &=&\frac{\sum_{j} P_{j,latt}^{k} }{\sum_{j}} \nonumber \\  
\Delta V_{latt}^{c} &= &\sqrt{\alpha_{0}} \cdot \sqrt{\frac{\sum_{j}
	\left(P_{j,latt}^{k}-\left<V_{latt}^{c,k}\right> \right)^{2}}{\sum_{j}}}.
\label{meanwidth2}
\end{eqnarray}
In order to determine the total EFG defined in Eq. (\ref{sepvq})
, we now determine the 
anti-shielding factors for both the A and B-lines from the CG data as such
\begin{equation}
\left<^{63}\nu_{Q}^{k} \right> = \frac{e^{63}Q}{2h} \cdot \left(
\zeta_{3d}^{k} \cdot V_{3d}^{c}+ 
\zeta_{latt}^{k}  \cdot \left<V_{latt}^{c,k}\right>\right),
\label{EFGconv}
\end{equation}
where we have used Eq. (\ref{sepvq}) and Eq. (\ref{meanwidth2}), along with the experimental data $\left<^{63}\nu_{Q}^{k} \right>$ (see appendix A for more details). 
Eq. (\ref{EFGconv}) and the anti-shielding factors convert the CG of the EFG simulation to the CG of the experimentally observed data. 

We are now in a position to convert each EFG value ($^{\kappa}V_{latt}^{c,k}$) stored in $P_{latt}^{k}$ into a frequency as such
\begin{equation}
^{\kappa}\nu^{k} = \frac{e^{63}Q}{2h} \cdot \left(
\zeta_{3d}^{k} \cdot V_{3d}^{c}+ 
\zeta_{latt}^{k} \cdot  {}^{\;\kappa}V_{latt}^{c,k}\right).
\label{EFGconv2}
\end{equation}
Using Eq. (\ref{EFGconv}) and Eq. (\ref{EFGconv2}), we can finally
predict the intrinsic lattice linebroadening ($\Delta \nu_{latt}^{k}$) as such
\begin{equation}
\Delta \nu_{latt}^{k} = \frac{e|^{63}Q|}{2h} \cdot \zeta_{latt}^{k} \cdot \Delta V_{latt}^{c}.
\label{broad}
\end{equation}
First, note that in Eq. (\ref{broad}) and Eq. (\ref{meanwidth2}) we have $not$ attach a superscript $k=(A,B)$ to 
$\Delta V_{latt}^{c}$. This is because the EFG simulation predicts the $same$ width $\Delta V_{latt}^{c}$
for both the A and B-lines. 
Second, note that we have assumed that the total width from the calculation $\Delta 
\nu_{latt} ^{k}$ is proportional to the intrinsic lattice width $\Delta V_{latt}^{c}$
with NO corresponding width from the onsite contribution, i.e. $\Delta 
V_{3d}^{c} =0$. We justify this by the fact that the 
ratio of widths between the A and B-lines is experimentally
found to be $^{63}\Delta\nu_{Q}^{B}/ ^{63\!\!}\Delta \nu_{Q}^{A} \simeq 0.63$ in the temperature region 
$T>T_{Q}$. According to appendix A, this ratio is closer to the
ratio of the anti-shielding factors for the lattice contribution 
$\zeta_{latt}^{B}/\zeta_{latt}^{A} = 0.67$ than for the onsite contribution
$\zeta_{3d}^{B}/ \zeta_{3d}^{A} = 0.89$. Note that the anti-shielding factors 
are deduced without any information about the widths, therefore taking $\Delta 
V_{3d}^{c} =0$ is justified.

Up to this point in the EFG simulation, we have 
taken care of the linebroadening due to the random placement of $^{+2}$Sr ions in the lattice (mechanism (a)), however, we
are still assuming that the holes are 
uniformily distributed across the plane, i.e. $^{63}R_{patch} = \infty$. 
Clearly this is not the whole picture since our calculation of $\Delta 
\nu_{latt}^{k} $ underestimates the observed linewidth $^{63}\Delta \nu_{Q}^{k}$ 
by a factor $\sim 2$, as can be seen in Fig. \ref{300T1f2}.

\subsection{$^{63}R_{patch}$ Simulation}


The introduction of the spatial variation in local hole concentration 
$^{63}x_{local}$ into the EFG calculation is an extension of what we have 
described in arriving at the intrinsic width $\Delta 
\nu_{latt}^{k} $. We start by randomly positioning the $^{+2}$Sr donor 
ions into the $^{+3}$La sites as described earlier.
Next, we define a patch (i.e. a circle) of radius $^{63}R_{patch}$
around the Cu site at the origin, as shown in Fig. \ref{patchsimul}.
We then determine the total number of Cu sites $N^{R_{patch}}$ within
$^{63}R_{patch}$ by multipying the area ($\pi ^{63\!}R_{patch}^{2}$) the patch covers 
by the areal Cu density ($1/a^{2}$) as such
\begin{equation}
N^{R_{patch}}  =  \frac{\pi ^{63\!}R_{patch}^{2}}{a^{2}},
\label{NCu}
\end{equation}
where we round $N^{R_{patch}}$ to the nearest integer. These 
$N^{R_{patch}}$ Cu sites within the radius $^{63}R_{patch}$ 
will uniformly share the donated holes from neighbouring $^{+2}$Sr ions.
Next, we determine how many donated holes $N_{Sr}^{\kappa}$ are to be 
shared within $^{63}R_{patch}$, where $N_{Sr}^{\kappa}$ itself is the random 
number to be calculated.
As shown in Fig. \ref{illust}, there are 
two distinct Sr/La sites (labeled $d_{1}$ and $d_{2}$), from which we can donate 
holes. This comes about because there are two distinct Sr/La sites 
within the unit cell.
For donation of holes from $d_{1}$, we accept a donor hole into $^{63}R_{patch}$ 
provided there is a $^{+2}$Sr ion which satisfies the position (in tetragonal notation)
\begin{eqnarray}
{\bf r}^{d_{1}} &=& [n_{a} a, n_{b} a, \pm 0.361 c]  \nonumber \\
\sqrt{n_{a}^{2}+n_{b}^{2}} &<& ^{63}R_{patch}/a 
\label{d1}
\end{eqnarray}
relative to the origin.
For donation of holes from $d_{2}$, we accept a donor hole into $^{63}R_{patch}$ 
provided there is a $^{+2}$Sr ion which satisfies the position
\begin{eqnarray}
{\bf r}^{d_{2}} &= &[(n_{a}+1/2) a, (n_{b}+1/2) a, \pm 0.139 c] \nonumber  \\
\sqrt{n_{a}^{2}+n_{b}^{2}} &<& ^{63}R_{patch}/a,
\label{d2}
\end{eqnarray}
and in both cases, $n_{a}$ and $n_{b}$ are $\pm$ integers ranging 
from zero to $\infty$. 
We do not a priori know which site $d_{1}$ or $d_{2}$ are the donor 
sites, or if the donors come from a combination of $d_{1}$ and 
$d_{2}$. We therefore treat both or combinations of both as equal 
possibilities. Technically, 
the n.n. (nearest neighbour) $d_{2}$ is the closest hole donor to the 
central Cu nucleus when occupied by a $^{+2}$Sr ion.
The n.n. $d_{1}$ site on the other hand has more bonding with 
the central Cu via the apical oxygens. The n.n $d_{1}$ 
site has the largest effect on $^{63} \nu_{Q}$ when occupied by a $^{+2}$Sr ion
and in particular gives rise to the B-site.

Once we have counted all the $N_{Sr}^{\kappa}$ satisfying either Eq. 
(\ref{d1}) or Eq. (\ref{d2}) for a particular random configuration of 
$^{+2}$Sr ions $\kappa$, we deduce the local 
hole concentration $^{63}x_{local}^{\kappa}$ as such
\begin{equation}
^{63}x_{local}^{\kappa} = N_{Sr}^{\kappa} / N^{R_{patch}}.
\label{xlocal}
\end{equation}
We now carryout the EFG summation in Eq. (\ref{loccontr}) for the 
particular configuration of $^{+2}$Sr ions $\kappa$, where the planar oxygen charge $q_{O_{pl}}$ in Eq. (\ref{charges}) is replaced by
\begin{equation}
q_{O_{pl}}^{\kappa} = -\left(2-2f_{\sigma}^{o} - 
^{63\!}x_{local}^{\kappa}/2 \right)e 
\label{qnew}
\end{equation}
for all planar oxygens within the patch radius $^{63}R_{patch}$. All other
planar oxygen sites outside the radius $^{63}R_{patch}$, including those on 
other CuO$_{2}$ planes, are assigned the uniform hole concentration $x$. 
Using this scheme effectively ignores 3-d correlations of $^{63}x_{local}^{\kappa}$ 
from neighbouring CuO$_{2}$ planes, however, such effects are small 
and of higher order.

As before, we then convert $^{\kappa}V_{latt}^{c,k}$ into a frequency using Eq. (\ref{EFGconv2})
and we store $^{\kappa}\nu^{k}$ into a vector $P_{R}^{k}$.
The position of the $^{+2}$Sr ions are then re-randomized to obtain a new 
random configuration $\kappa'$, 
then a new value of $^{\kappa'} \nu^{k}$ 
is deduced and stored into $P_{R}^{k}$. 
This procedure is repeated 
$\sim 10^{4}$ times until a sufficient number of data points exist in 
$P_{R}^{k}$ so as to create a histogram spectrum of the
lineshape over $\sim$50 bins.
The CG of the spectra $\left<P_{R}^{k}\right>$ coincide with 
the experimental data $\left<^{63}\nu_{Q}^{k}\right> $ (Eq. (\ref{EFGconv})), while a new {\it HWHM} ($\Delta \nu_{R_{patch}}^{k}$) is determined as such
\begin{equation}
\Delta \nu_{R_{patch}}^{k} = \sqrt{\alpha_{0}} \cdot \sqrt{\frac{\sum_{j}
	\left(P_{j,R}^{k}-\left<^{63}\nu_{Q}^{k}\right> \right)^{2}}{\sum_{j}}}.
\label{meanwidth3}
	\end{equation}

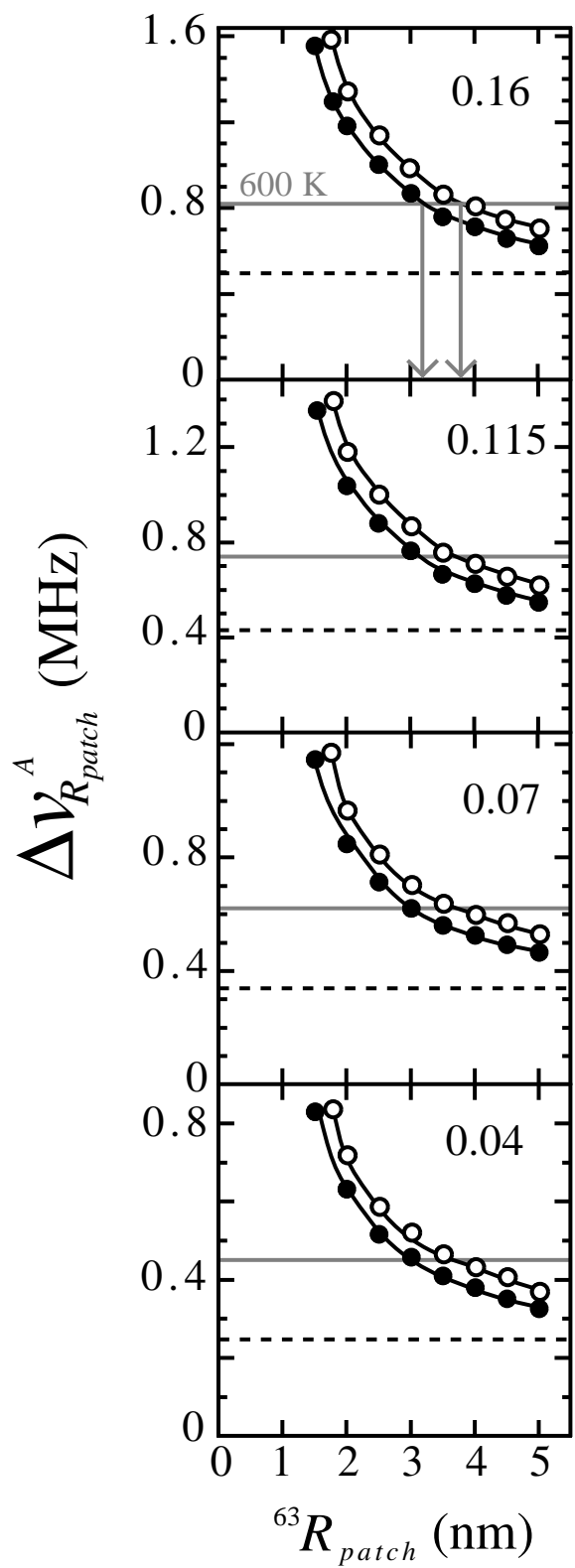
\begin{figure}
\caption{Calculated value of {\it HWHM} $\Delta \nu_{R_{patch}}^{A}$ (MHz) 
for the A-line as a function patch radius 
$^{63}R_{patch}$ (nm) for different $x$ shown in each part using the 
$^{+2}$Sr dopant ions $d_{1}$
$\Delta \nu_{R_{patch},d_{1}}^{A}$
($\circ$), and $d_{2}$ dopants $\Delta \nu_{R_{patch},d_{2}}^{A}$
($\bullet$). Solid black curves through data are guides for 
the eye. Grey solid horizontal line shows experimental data $^{63}\Delta \nu_{Q}^{A}$ at 
600 K while dashed black line shows intrinsic lattice width $\Delta 
\nu_{latt}^{A}$ reached in the limit $^{63}R_{patch} = \infty$.}
\label{corr}
\end{figure}

We then repeat this whole procedure and calculate $\Delta \nu_{R_{patch}}^{k}$
for various values of $^{63}R_{patch}$, the results of which
are shown in Fig. \ref{corr} for the A-line. The 
results for the B-line are the same but with a uniform decrease
$\Delta \nu_{R_{patch}}^{B}/\Delta \nu_{R_{patch}} ^{A} = \zeta_{latt}^{B} / 
\zeta_{latt}^{A} (= 0.67$) for all $^{63}R_{patch}$. Also shown in Fig. \ref{corr} is how to deduce 
the best fit value of $^{63}R_{patch}$ to the lineshape at 600 K. We match 
experimental values of $\Delta \nu_{Q}^{k}$ (shown by the grey 
horizontal line) to the calculation shown by the 
solid black curves, and extrapolate $^{63}R_{patch}$ shown by the grey 
vertical arrows. 

Note that according to the arrows, $^{63}R_{patch}^{d_{2}}$ deduced using
dopants from $^{+2}$Sr ions at $d_{2}$ is 15 \% smaller than $^{63}R_{patch}^{d_{1}}$
deduced using dopants from $^{+2}$Sr ions at $d_{1}$. Put in another 
way, when we do the calculation for fixed $^{63}R_{patch}$, 
we deduce a $\sim$ 15 \% smaller value of $\Delta \nu_{R_{patch}^{d_{2}}}^{k}$ 
compared to $\Delta \nu_{R_{patch}^{d_{1}}}^{k}$.
The reason for $\Delta \nu_{R_{patch}^{d_{2}}}^{k} < \Delta \nu_{R_{patch}^{d_{1}}}^{k}$
can be explained by a correlation between the 
local hole concentration $^{63}x_{local}^{\kappa}$ on the planar 
oxygens and the $^{+2}$Sr donor ions in n.n. positions to the origin.
The result of adding one hole from a n.n. 
$^{+2}$Sr ion amounts to two separate 
shifts to the resonance frequency. The first shift is from the added hole itself 
which decreases of magnitude of the surrounding planar oxygen charge $|q_{O_{pl}}|$ 
and results in a large
$increase$ in the resonance frequency, regardless of the hole's origin. We shall 
define this shift as $\delta \nu^{pl} (>0)$.
The second effect of adding one hole from a n.n. 
$^{+2}$Sr donor ion is from the n.n. $^{+2}$Sr ion itself which results in a 
shift $\delta \nu^{Sr}$, however, the sign of shift $\delta \nu^{Sr}$ depends on
whether it is at $d_{2}$ or $d_{1}$.

Let us imagine scenario (1) where the added 
hole specifically comes from a $^{+2}$Sr ion in the n.n.n. (next nearest neighbour) $d_{1}$ site. 
We do not use the n.n. $d_{1}$ site in this case since this is the 
B-line which, according to the simulation, shows the $same$ difference
$\Delta \nu_{R_{patch}^{d_{2}}}^{B} / \Delta \nu_{R_{patch}^{d_{1}}}^{B} (\simeq 0.85$) as the A-line.
It can be shown that a $^{+2}$Sr ion in the n.n.n. $d_{1}$ site
causes a small positive shift $\delta \nu^{d_{1}} >0$, therefore the total 
shift $\delta \nu ^{(1)}$ in scenario (1) is given by
$\left(\delta \nu^{pl} + \delta \nu^{d_{1}}\right)  = \delta \nu ^{(1)} (>0)$ where 
both components are positive and enhance the total effect.
Next let us imagine scenario (2) where the $^{+2}$Sr ion from which the added hole  
came from is specifically located at a n.n. $d_{2}$ position to the origin. It can be shown that placing a 
$^{+2}$Sr ion at the n.n. $d_{2}$ position causes a small $negative$ 
shift $\delta \nu^{d_{2}} <0$
to the resonance frequency. The total shift $\delta \nu^{(2)}$ in scenario 
(2) is therefore given by
$\left(\delta \nu ^{pl} + \delta \nu ^{d_{2}}\right)  = \delta \nu ^{(2)} (>0)$ 
where the individual components counteract each other.
Therefore, in scenario (1) the overall positive shift to the resonance frequency 
is enhanced while in scenario (2) it is reduced. After many random runs which
include the independent random effects from distant $^{+2}$Sr ions, 
scenario (1) will lead to a larger overall spread, therefore
$\Delta \nu_{R_{patch}^{d_{2}}}^{k} < \Delta \nu_{R_{patch}^{d_{1}}}^{k}$.

\subsection{$^{63}R_{patch}$ Results}

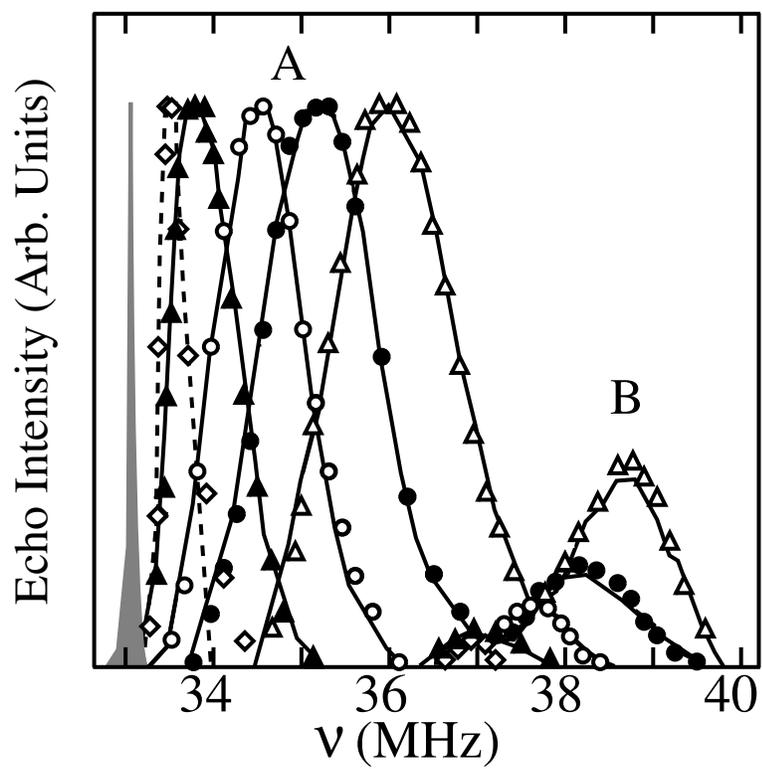
\begin{figure}
\caption{$^{63}$Cu NQR lineshape at 600 K for $x=0.16$ ($\triangle$), $x=0.115$ ($\bullet$), $x=0.07$ ($\circ$),
$x=0.04$ ($\blacktriangle$), $x=0.02$ ($\diamond$) and $x=0.0$ (grey 
region). Solid lines are best fits to the spectra using the patch-by-patch model and $d_{1}$ 
dopants. The lower bound to the patch radius $^{63}R_{patch} = 3.0-4.0$ nm used for the solid line fits
are shown in Fig. \ref{600Kx}.}
\label{600Kf}
\end{figure}

In Fig. \ref{600Kf} we show the results for the best fit to 
the lineshape data at 600 K. Note that the observed asymmetry of the A and B-lines for $x \geq 0.04$ are well reproduced within the model without any additional parameters.
The fits shown in Fig. \ref{600Kf}
are deduced using $d_{1}$ dopants with $^{63}R_{patch}^{d_{1}}$, however 
the {\it same} fits can be deduced using $d_{2}$ dopants with 
$^{63}R_{patch}^{d_{2}}$.
The best fit values of $^{63}R_{patch}^{d_{1}}$ and $^{63}R_{patch}^{d_{2}}$ 
are shown in Fig. \ref{600Kx}(a). In Fig. \ref{600Kf} we also show the data and fit for $x=0.02$.
As can be seen, however, the asymmetry in the experimentally observed
line-profile for $x=0.02$ is somewhat larger that our patch 
model can predict, hence we show the fit as a dashed line. 
It is interesting to note that in the range $x \leq 0.02$, 
$^{63}R_{patch}^{d_{1}}$ and $^{63}R_{patch}^{d_{2}}$ 
tend towards the average donor separation distance $l_{Sr} 
=a/\sqrt{x}$, as shown in Fig. \ref{600Kx}. If $^{63}R_{patch} \simeq 
l_{Sr} $, our model implies that each patch only covers one 
$^{+2}$Sr ion on average. In such a senario we no longer expect our model 
to account for the data. In fact, the critical region $x = 0.02$ which
separates the spin-glass phase and the 
anti-ferromagnetic phase for $x \leq 0.02$ \cite{cho} is known to 
show complex behaviour \cite{rigamonti}. 

We also note in Fig. \ref{600Kf} that the intensity of the B-line 
is slightly underestimated by the calculation compared with the 
experimental data. 
This is simply due to the fact 
that the spin-spin relaxation rate is experimentally $\sim$ 15 \% 
smaller at the B-site together with the fact that
we are measuring the data points using $\tau = 
12$ $\mu$s. After correcting for differences in the spin-spin relaxation rate, the 
observed relative intensity of the B-line decreases slightly and is 
found to go as $f_{B}^{exp} \simeq 
x$ (Fig. \ref{fab}), which is then consistent with the simulation.
In order to fit the NQR spectrum in Fig. \ref{600Kf}, we have 
attributed all of the extra line broadening to $^{63}R_{patch}$. 
In Fig. \ref{600Kx}(b) we compare the experimentally observed width 
$^{63}\Delta \nu_{Q}^{A}$ with the calculated intrinsic width 
$\Delta \nu_{latt}^{A}$.

In Fig. \ref{300KX} we show the same plot as Fig. \ref{600Kx} except
at 300 K, where the $^{63}R_{patch}$ are deduced in a similar way to 
600 K shown in Fig. \ref{corr}. At 300 K, we see an effect 
not present at 600 K, namely that the observed linewidth $^{63}\Delta \nu_{Q}^{A}$
for $x \leq 0.115$ in Fig. \ref{300KX}(b) no longer 
decreases with decreasing $x$ as it does at 600 K. Also shown in Fig. \ref{300KX} is 
the onset of the LRO orthorhombic phase
\cite{yamada} shown as the dashed 
lines through the data points. The result of the observed enhancement 
of the linewidths $^{63}\Delta \nu_{Q}^{A}$
for $x \leq 0.115$ in Fig. \ref{300KX}(b) results in a large drop in the 
calculated $^{63}R_{patch}$ shown in Fig. \ref{300KX}(a). Comparing 
the $x$ dependence in $^{63}R_{patch}$ within the tetragonal phase at 
600 K and at 300 K for $x \geq 0.115$ seems to suggest 
that $^{63}R_{patch}$ is more or less $x$ independent $provided$ that 
the local lattice distortions do not dramatically enhance the observed NQR 
linebroadening.

The effects of the orthorhombic distortions can also be seen in Fig. 
\ref{corrT} where we plot the temperature dependence of the average 
$^{63}R_{patch}$ defined as such
\begin{equation}
    ^{63}R_{patch} = \left( ^{63}R_{patch}^{d_{1}}+ 
    ^{63}R_{patch}^{d_{2}}\right)/2 .
    \label{ave}
    \end{equation}
We chose to plot the average value $^{63}R_{patch}$ alone for 
clarity. This corresponds to a simulation where half of the donors come from $d_{1}$ 
sites and half come from $d_{2}$ sites.
In Fig. \ref{corrT} the 
black and grey lines correspond to regions above and below $T_{Q}$ 
taken from Fig. \ref{nuqall}, while the dashed black line shows the onset of LRO 
$T_{st}$ \cite{yamada}. In the case of $x=0.04$ we clearly see a dip in $^{63}R_{patch}$
below $T_{Q}\simeq T_{st}$ which is most likely due to the fact 
that we have neglected distributions in the local orthorhombic distortions. 
In section V we make the reasonable assumption that 
$^{63}R_{patch}$ for $x=0.04$ continues to decrease lineaarly (defined as $^{63}R_{patch}'$) with decreasing 
temperature below $T_{Q}$, similar to $x \geq 0.115$ samples. Assuming that $^{63}R_{patch}'$ is the underlying length scale will allow us to estimate the distribution in local orthorhombic distortions.

We also remark that similar patch-by-patch models of disordered doping 
in the CuO$_{2}$ plane have been independently used 
to account for the smeared density of 
states found in photoemission peaks
as a result of a distribution in the superconducting gap \cite{bala}.
In such models the only free parameter used to account for the smoothness of 
the density of states is the length  
scale of the model which is determined by the in plane superconducting coherence length 
$\xi_{ab} \sim 2.0-3.0$ nm. It is interesting to note that in the case of $x=0.16$, the lower bound
$^{63}R_{patch} \gtrsim 2.5 \pm 0.5$ nm in the region of $T_{c}$ is comparable 
to $\xi_{ab} \simeq 3.3$ nm in thin films \cite{suzukicoh}.

\subsection{$^{63}\Delta x_{patch}$ Results}

In the temperature region above $T_{Q}$, we can also deduce an upper 
boundary $^{63} x_{patch}$ for the spatial variation 
$^{63}x_{local}$ without any further computation. We proceed 
as described above, but for
each random lattice configuration $\kappa$, we also store
each value $^{63}x_{local}^{\kappa}$ (used to compute $^{\kappa} 
\nu_{Q}^{k}$) into a new vector $X_{local}^{k}$.
We then build up a vector $X_{local}^{k}$ of randomly distributed 
local hole concentrations for each lattice configuration $\kappa$.
The mean of the set $X_{local}^{k}$ is just the nominal value 
$\left<X_{local}^{k}\right> = x$, while the effective {\it HWHM} $^{63}\Delta x_{patch}$ (defined in an 
analogous manner to Eq. (\ref{meanwidth2})) is found to be independent of which 
site $k=(A,B)$ is computed.
By deducing $^{63}\Delta x_{patch}$ over a large range of the free parameters  
$\left[x,^{63}R_{patch}\right]$, we find our calculated values of $^{63}\Delta 
x_{patch}$ fit well to the Binomial theorem 
\begin{eqnarray}
^{63}\Delta x_{patch} &=& \sqrt{\alpha_{0}} \cdot
\sqrt{\frac{x(1-x)}{N^{R_{patch}}} } \nonumber \\  
 &=& \frac{a}{^{63}R_{patch}} \cdot  \sqrt{x(1-x)} \cdot \sqrt{\frac{\alpha_{0}}{\pi}},
\label{halfbin}
\end{eqnarray}
where we have used the relation in Eq. (\ref{NCu}) for the second line.
In Fig. \ref{deltax} we show the temperature dependence of $^{63}
x_{patch}$ using Eq. (\ref{equan2}) and Eq. (\ref{halfbin}) for
$^{63}R_{patch}$ defined in Eq. (\ref{ave}). Using 
$^{63}R_{patch}^{d_{1}}$($^{63}R_{patch}^{d_{2}}$) instead of Eq. 
(\ref{ave}) results in a small $-(+)7$ \% variation in $^{63}\Delta 
x_{patch}$, respectively. The fact that $^{63}x_{patch} \sim 
^{63}x_{local} $ suggests that (a) our patch-by-patch model is 
appropriate in describing the spatial variation in $^{63}x_{local}$, (b) our estimate 
of the intrinsic lattice broadening $^{63}\Delta \nu _{latt}$ is 
correct in the temperature region above $T_{Q}$, and (c) the static value $^{63}x_{patch} $ deduced 
from the lineshape analysis is consistent with the low frequency dynamic value $^{63}
x_{local}$ deduced using $^{63}1/T_{1}$.

\section{Orthorhombic Lattice Distortions}

\subsection{Local Order Parameter}

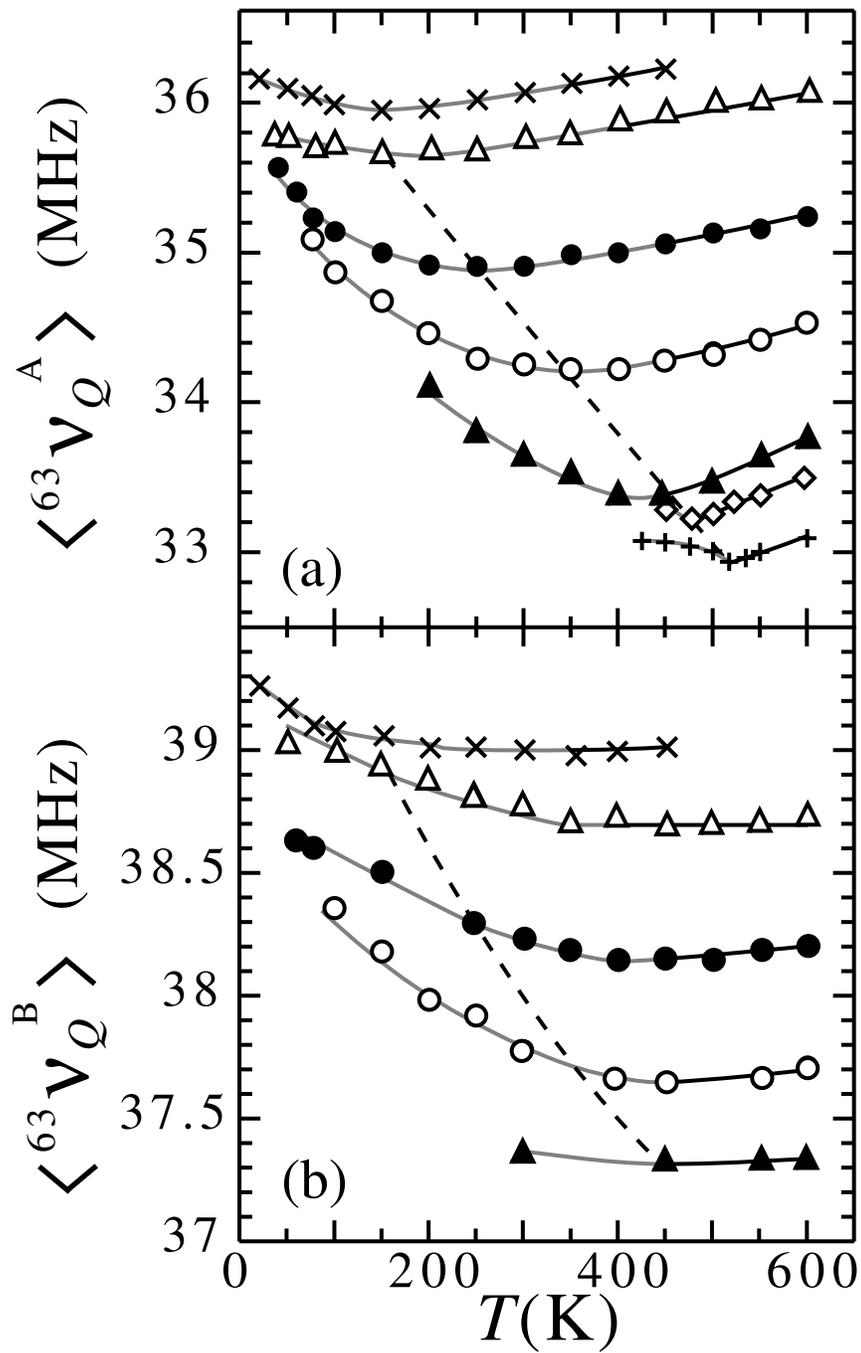
\begin{figure}
\caption{Temperature dependence of CG of the $^{63}$Cu NQR spectrum 
(a) $\left<^{63}\nu_{Q}^{A}\right>$ at the A-line, and (b) 
$\left<^{63}\nu_{Q}^{B}\right>$ at the B-line for $x=0.20$ 
($\times$), $x=0.16$ ($\triangle$), $x=0.115$ ($\bullet$), $x=0.07$ ($\circ$),
$x=0.04$ ($\blacktriangle$), $x=0.02$ ($\diamond$), and $x=0.0$ (+). 
Black curves show fit above $T_{Q}$ where 
$^{63}x_{patch}$ is deduced, while grey curves show fit below $T_{Q}$ 
where local orthorhombic distortions are deduced. Dashed black line 
shows onset of orthorhombic phase according to LRO \cite{yamada}.}
\label{nuqall}
\end{figure}

In the orthorhombic phase,
the CuO$_{6}$ octahedra are tilted by an angle $\theta_{c} \leq 
5^{o}$ \cite{radaelli}
away from the $c$-axis towards the $b_{o}$-axis, and the primitive lattice constants in the 
plane are split ($b_{o} > a_{o}$). We can calculate the effect of 
these distortions to the EFG by placing the ions in an orthorhombic cell (see Fig. 
\ref{illust}) with appropriate orthorhombic modulations
\cite{ginsberg,radaelli} and compute the changes to the $^{63}\nu_{Q}$.
However, we already know from Fig. \ref{nuqall} that $^{63}\nu_{Q}$ starts to increase above 
the structural phase transition temperature $T_{st}$ deduced from
neutron and X-ray diffraction results \cite{radaelli}. These bulk 
probes detect LRO, whereas NQR is a local probe, therefore 
it is more appropriate to deduce the local orthorhombic 
distortions from our $^{63}\nu_{Q}$ data, and then compare them with LRO results.

The orthorhombic distortions cannot {\it a priori} be deduced from the NQR 
data alone. We must first take certain know relations from LRO 
results and apply them to NQR. In particular, we shall use the neutron diffraction 
results reported by P.G. Radaelli {\it et al} \cite{radaelli} which 
demonstrate that certain measured 
orthorhombic parameters such as the orthorhombic splitting $(b_{o} - 
a_{o})$, the octahedron tilting angle $\theta_{c} (\leq 5^{o}$) which measures the departure angle 
of the O$_{ap}$-Cu-O$_{ap}$ bond
from the $c$-axis, and the ``scissors'' angle $\theta_{scis} (\leq 
90^{o}$) which measures the O$_{pl}$-Cu-O$_{pl}$ angle, are all related 
to powers of the underlying orthorhombic {\it order parameter} we define as 
$\theta_{LRO}(x,T)$. 

At low temperatures ($\lesssim 70$ K) where lattice fluctuations are small,  
mean-field theory gives a good account of the observed lattice distortions \cite{radaelli}. In light of this, we fit the data in Ref. \cite{radaelli} to a mean-field like form and deduced
\begin{eqnarray}
\theta_{c} &  \simeq & \theta_{LRO}(x,T) \nonumber \\
\theta_{scis} &\simeq & 90 - 4 \cdot  10^{-2} \cdot  
\left[\theta_{LRO}(x,T)\right]^{2} \nonumber \\
(b_{o} - a_{o}) & \simeq & 6 \cdot  10^{-4} \cdot \frac{(b_{o} +  a_{o})}{2} \cdot 
\left[\theta_{LRO}(x,T)\right]^{2} 
\label{constraint}
\end{eqnarray}
where all angles are in degrees. These set of equations provide a
constraint for the orthorhombic distortions which determine
the positions ${\bf r}_{j}(x,T)$ of the $j$th ion in the lattice. We now 
{\it assume} that the LRO constraints in Eq. (\ref{constraint}) also apply to the 
$local$ constraints which determine the local position ${\bf r}_{j}(x,T)$ 
of the $j$th ion in the lattice (i.e. 
$\theta_{LRO}(x,T) \rightarrow \theta_{local}(x,T)$ in Eq. (\ref{constraint})).

The EFG lattice summation in Eq. (\ref{lattcontr}) is now carried out in a similar way as 
described in section IV but with the orthorhombic 
distortions incorporated into the positions of the ions. The 
positions ${\bf r}_{j}(x,T)$ of the ions are deduced using the absolute values of the lattice 
constants $a_{o}$ and $c$ at 295 K from Fig. \ref{latt} along with their 
corresponding thermal coefficients $\alpha_{a}$ and $\alpha_{c}$ as 
before, however, we allow $b_{o}$ to vary according to the
constraints in Eq. (\ref{constraint}) and the unknown local structural parameter 
$\theta_{local}(x,T)$. We position the La/Sr ions and the apical 
oxygen ions according to the tilting angle $\theta_{c} = 
\theta_{local}(x,T)$ towards the $b_{o}$ direction, while the planar oxygens 
are positioned according to $\theta_{sc}$ in Eq. (\ref{constraint}) and $\theta_{local}(x,T)$. We now have all the ionic positions ${\bf 
r}_{j}(x,T)$ as a function of $\theta_{local}(x,T)$. 
Just as in section IV, we then compute the EFG lattice summation and resonance frequency $^{\kappa}\nu^{k}$ (Eq. (\ref{EFGconv2})) for the random configuration $\kappa$
and patch radius $^{63}R_{patch}$, however, the EFG lattice summation is taken using the tilted ionic positions specified by $\theta_{local}(x,T)$. We store $^{\kappa}\nu^{k}$ into a vector $P_{\theta}^{k}$ and re-randomize the lattice until $P_{\theta}^{k}$ is sufficiently large to build a histogram spectrum. We define the new CG $\left<P_{\theta}^{k}\right>$ incorporating finite $\theta_{local}$ as $ \left<\nu_{\theta}^{k}\right>$.

The results of the simulation suggest that the increase in the CG
due to $\theta_{local}$ $alone$ goes as 
\begin{equation}
\left<\nu_{\theta}^{k}\right> - \left<\nu_{\theta=0}^{k}\right> \simeq  c^{k} \cdot \left< \theta_{local}^{2}  \right>, 
\label{deltatheta}
\end{equation}
where $c^{A} \simeq $ + 0.14 MHz/[deg.$^{2}$] and $c^{B} (\sim \frac{\zeta_{latt}^{B}}{\zeta_{latt}^{A}} \cdot c^{A}) \simeq $ + 0.08 MHz/[deg.$^{2}$] are $x$ and $T$
independent within $\sim \pm 15$ \%. Note that NQR does not
distinguish between $\pm \theta_{local}$, but only the average square magnitude 
$\left< \theta_{local}^{2}  \right>$ across the whole sample.
Another conclusion we deduce from the calculation is that
after diagonalizing the EFG tensor, the main principle axis is found to be
tilted an angle $|\theta_{NQR}|$ away from the $c$-axis (where 
$\theta_{NQR} \simeq 0.9 \cdot \theta_{local}$), while the EFG asymmetry parameter $\eta$ remains 
small $\eta < 0.06$. 

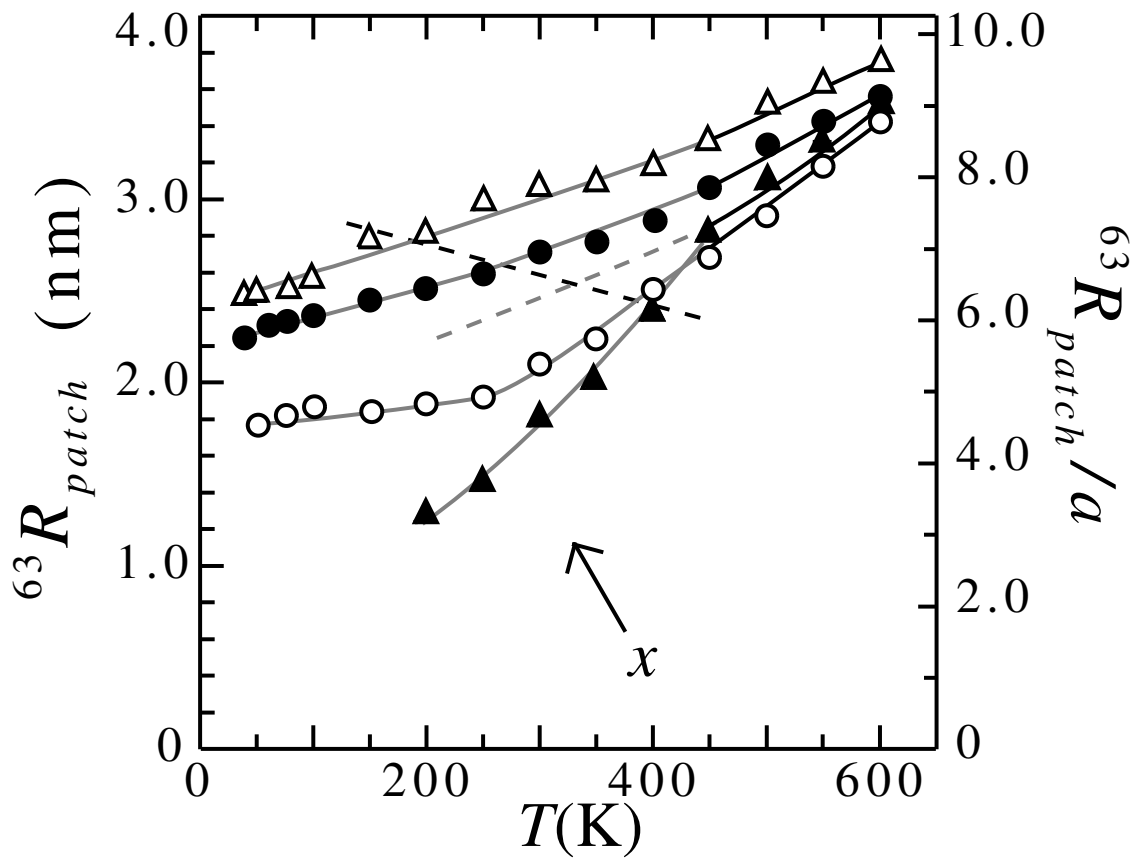
\begin{figure}
\caption{Temperature dependence for the square of the $local$ structural parameter  
(or equivalently, the square of the
CuO$_{6}$ octahedron tilting angle) 
$\left<\theta_{local}^{2}\right>$ (in deg.$^{2}$) for $x = 0.20$ ($\times$), $x=0.16$ ($\triangle$),
$x=0.115$ ($\bullet$), $x=0.07$ ($\circ$),
$x=0.04$ ($\blacktriangle$) and $x=0.0$ (+) deduced from the data in 
Fig. \ref{nuqall}.}
\label{theta}
\end{figure}

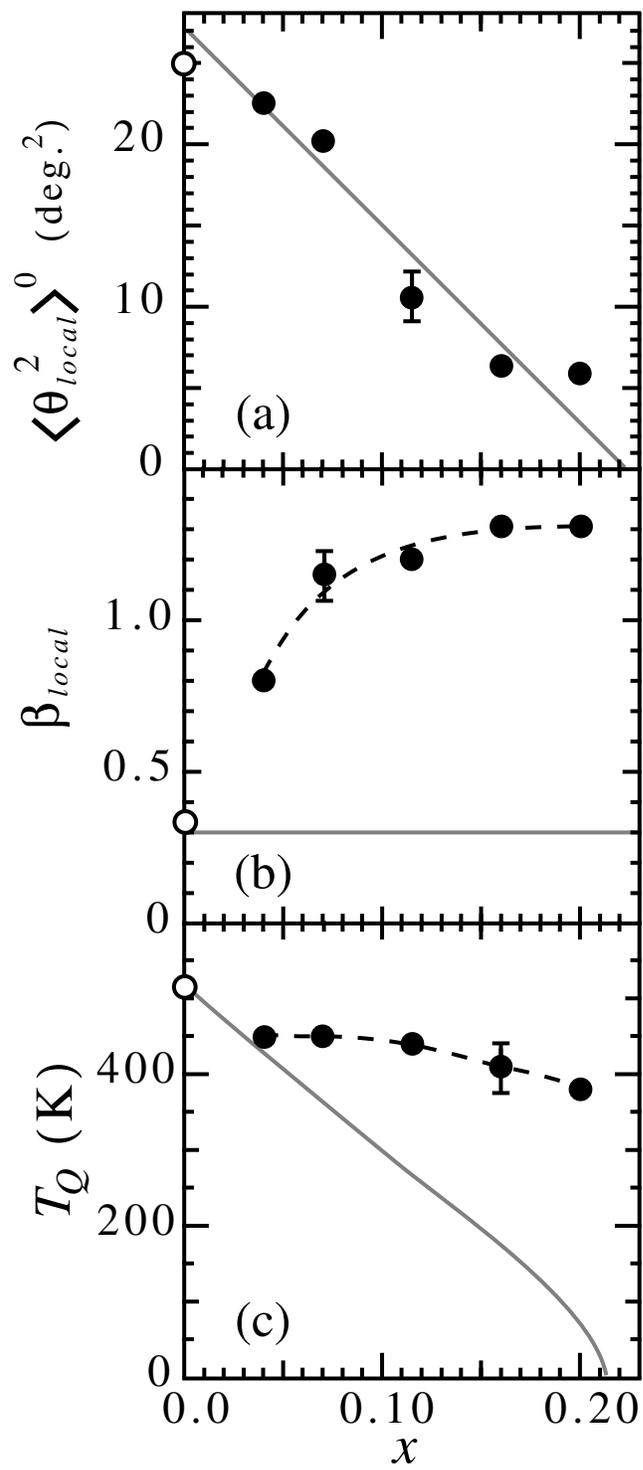
\begin{figure}
\caption{$x$ dependence for the coefficients of the local structural parameter 
$\left<\theta_{local}^{2}\right>$ including
(a) maximum square value $\left<\theta_{local}^{2}\right>^{0}$ (deg.$^{2}$), (b) reduced temperature 
coefficient $\beta_{local}$, and (c) onset temperature $T_{Q}$ (K) for $x=0.0$ ($\circ$) and 
$x>0.0$ ($\bullet$).
Dashed lines are guides for the eye. Solid grey
lines are LRO patameters (a) $\left<\theta_{LRO}^{0}\right>^{2}$ (deg.$^{2}$) \cite{radaelli}, 
(b) $\beta$ \cite{ginsberg,beta} and (c) $T_{st}$ \cite{yamada}.}
\label{thetasumm}
\end{figure}

We are now in a position to deduce the average $\left< 
\theta_{local}^{2}  \right>$ from our 
experimental data $\left<^{63}\nu_{Q}^{k}\right>$ (Fig. \ref{nuqall}).
In the spirit of the LRO results, we deduced the temperature 
dependence of $\left< \theta_{local}^{2}  \right> $ (presented in Fig. \ref{theta}) from the $\left<^{63}\nu_{Q}^{k}\right>$ data assuming the paramteric form 
\begin{equation}
   \left< \theta_{local}^{2}  \right> = \left< \theta_{local}^{2}  
   \right>^{0} \cdot 
   \left(\frac{T_{Q}-T}{T_{Q}}\right)^{2 \beta_{local}},
\label{order}
\end{equation}
where $ \left< \theta_{local}^{2}\right>^{0} $ 
is the square averaged tilting angle at $T=0$ K,
$\beta_{local}$ is the reduced temperature 
coefficient, and $T_{Q}$ is 
local onset temperature for the orhorhombic distortions. All three free parameters in Eq. (\ref{order}) depend on $x$ and are optimized to fit $\left<^{63}\nu_{Q}^{k}\right>$ (in Fig. \ref{nuqall}). The best fit to the $\left<^{63}\nu_{Q}^{k}\right>$ are shown as the black and grey curves in Fig. \ref{nuqall}, and the corresponding temperature dependence of $\left< 
\theta_{local}^{2}  \right>$ is shown in Fig. \ref{theta}.
In Fig. \ref{thetasumm} we 
summarize the $x$ dependence of best fit values
$ \left< \theta_{local}^{2} \right>^{0}$, $\beta_{local}$ and $T_{Q}$.
We compare these parameters with equivalent mean field parameters 
$\left<\theta_{LRO}^{0}\right>^{2}$, $\beta$ and $T_{st}$ according to LRO 
\cite{radaelli,yamada}, shown as the solid grey lines in Fig. \ref{thetasumm}.

In the case of $x=0.0$, all three parameters deduced by NQR are consistent 
with equivalent LRO parameters deduced by neutron and X-ray diffraction. 
We find that $\beta_{local} \simeq 0.34$ according to NQR for $x=0.0$, which is consistent with the LRO critical exponent $\beta \simeq 0.30$ according to neutron and X-ray scattering for all $x$ \cite{beta,ginsberg}. A similar critical exponent is also found in La$_{2-x}$Ba$_{x}$CuO$_{4}$ where $\beta = 0.33$ \cite{suzuki}, which are all consistent with the theoretical predictions from a 3d $X-Y$ model.

For $x\geq 0.04$ we find consistent
values between $ \left< \theta_{local}^{2} \right>^{0}$ and 
$\left<\theta_{LRO}^{0}\right>^{2}$.
In the case of $\beta_{local}$ and $T_{Q}$ for $x > 0.04$, however, we find 
qualitative differences with the LRO parameters $\beta$ and $T_{st}$, respectively.
The fact that $\beta_{local} \sim 1$
for $x > 0.04$ indicates that locally, the transition into the 
orthorhombic phase is smeared out in temperature. The fact 
that $T_{Q} \gtrsim 400$ K indicates that the local onset temperature remains 
large and roughly $x$ independent. 

In order to account for the differences between local and LRO structural results one must first recall that 
our NQR results measure the local square deviation $\left< \theta_{local}^{2} \right>$, while the diffraction results measure the coherent average $\left< \theta_{LRO} \right>$ across the sample. 
One must also leave open the possibility that the orthorhombic structural transition for $x>0.04$ is partly  order-disorder and partly displacive in nature.
In an order-disorder dominated structural transition (as proposed by other
local probes \cite{bozin,haskel} in the region $x>0.15$),
$\theta_{local}$ is argued to be finite and temperature {\it independent} in the LRO tetragonal 
phase ($T>T_{st}$), while $\left< \theta_{LRO} 
\right> = 0$ when averaged over large ($\gtrsim 10$ nm) length scales. In such a scenario, one can picture 
certain short length scale regions ($\ll 10$ nm) which align along the positive $b_{o}$ direction where $\left< \theta_{local} 
\right> > 0$, while an equal number of other regions align along the negative $b_{o}$ direction with $\left< \theta_{local} 
\right> < 0$, therefore $\left< \theta_{LRO} 
\right> = 0$. The temperature dependence in an order-disorder type transition does not involve a $T$ dependence in the magnitude of $| \theta_{local}| $ itself. Instead, the $T$ dependence lies in the coherent orientation of the local regions towards a particular direction, for arguments sake say along positive $b_{o}$ direction. The coherent average of the first moment $\left< \theta_{LRO} \right> $ will then be finite and positive ($\left< \theta_{LRO} 
\right> > 0$), and will {\it appear} to increase with decreasing temperature as more and more local regions in the sample choose to align along the positive $b_{o}$ direction. 

In a displacive dominated structural transition (as argued for $x \leq 0.04$), the magnitude of $| \theta_{local}| $ itself increases with decreasing temperature below $T_{st}$ and $all$ the local regions in the sample align along the same $b_{o}$ direction. Our new NQR results combined with LRO results support a displacive structural transition for $x\lesssim 0.04$, while a mixture of displacive and order-disorder type
transition occurs in the region $x> 0.04$. For $x>0.04$ below $T<T_{Q}$, NQR indicates that locally the octahedra start to tilt in a displacive manner
(i.e. the tilting angle of each octahedra increases 
with decreasing temperature below $T<T_{Q}$ and $ \left< 
\theta_{local}^{2} \right> >0$). In the region $T_{st} < T <T_{Q}$, $\theta_{local}$ continues to increase 
yet there continue to be an equal number of regions with positive and negative tilting with respect to the $b_{o}$ direction, and therefore $\left< \theta_{LRO} \right> = 0 $. In the region $T< T_{st}$, 
$\theta_{local}$ continues to increase with 
decreasing temperature, however, the local regions start to 
coherently align along one direction (say along the positive $b_{o}$ direction), and $\left< \theta_{LRO} \right> > 0 $.
This interpretation is entirely consistent with other local probes \cite{bozin,haskel}, but
in addition we have 
deduced the onset temperature $T_{Q}$ for the local orthorhombic distortions. 
In the case of $x=0.20$ we find that $T_{Q} \sim 350$ K even though the LRO structural transition disappears \cite{yamada}.

\subsection{Distribution in Local Parameter}

In all of our analysis of the $^{63}$Cu NQR linebroadening thus far, we have ignored potential linebroadening from local lattice distortions. In the previous part of this section, we deduced the temperature dependence of the square $average$ distortions $ \left< 
\theta_{local}^{2} \right> $ across the sample using the CG data $\left<^{63}\nu_{Q}^{A} \right>$. We shall now estimate the corresponding $distribution$ of $\theta_{local}$ using the {\it HWHM} of the NQR spectrum, $^{63}\Delta \nu_{Q}^{A}$.
As discussed earlier, the three linebroadening 
mechanisms for the $^{63}$Cu NQR lineshape are (a) the random placement of $^{+2}$Sr ions in the lattice
(b) the patch-by-patch variation in hole concentration 
$^{63}x_{local}$, 
and (c) the local lattice distortions. In section IV we used mechanisms (a) and (b) to account for the 
experimental data, and deduced the lower bound $^{63}R_{patch} $ for the spatial variation $^{63}x_{local}$. We shall now use all three mechanisms (a),(b) and (c) to account for the observed {\it HWHM} data $^{63}\Delta \nu_{Q}^{A}$ in the case of $x=0.04$. 
Note that we cannot {\it a priori} separate (a),(b) and (c), therefore in order to obtain the contribution from (c), we must first make reasonable assumptions for (a) and (b).

In Fig. \ref{4distra} we plot the temperature dependence of the observed data 
$^{63}\Delta \nu_{Q}^{A}$ for $x=0.04$ above and below $T_{Q} \simeq T_{st} \simeq 450 $ 
K taken from Fig. \ref{widthall}. For $T>T_{Q}$, we 
attribute all of the observed linebroadening to mechanism (a) and (b) given by $\Delta \nu_{R_{patch}}^{A}$. For $T<T_{Q}$, however, we make  
the ansatz that the $\Delta \nu_{R'_{patch}}^{A}$ continues to increase only
linearly, as shown in Fig. \ref{4distra}. 
$\Delta \nu_{R'_{patch}}^{A}$ in Fig. \ref{4distra}
results from using a different estimate $^{63}R_{patch}'$ for the patch radius, shown as the grey dashed line in Fig. \ref{corrT} for $x=0.04$. The motivation for using $^{63}R_{patch}'$ for $x=0.04$ is that the temperature dependence follows that of $x \gtrsim 0.115$ where the overall amplitude of the orthorhombic distortions ($\left<\theta^{2}_{local} \right>^{0}$ in Eq. (\ref{order})) are smaller. 
 
In order to simulated the NQR spectrum, we now proceed through the simulation as in section IV, however, we calculate the local
hole concentration $^{63}x_{local}^{\kappa}$ for a random 
configuration $\kappa$ using the patch radius $^{63}R_{patch}'$ (Fig. \ref{corrT}), {\it instead} of $^{63}R_{patch}$. Along with each simulation run $\kappa$,
we also {\it independently} 
choose a random value of $|\theta_{local}^{\kappa}|$ taken from a trial PDF (probability distribution 
function), examples of which are shown in Fig. \ref{4distrb}. The positions of 
all the ions in the random lattice configuration $\kappa$ are calculated 
given the value
$|\theta_{local}^{\kappa}|$ and Eq. (\ref{constraint}) (with $\theta_{LRO} \rightarrow \theta_{local}^{\kappa}$).
We then take the lattice summation in Eq. (\ref{lattcontr}) using the
the random positions of the ions with $|\theta_{local}^{\kappa}|$, 
together with the random local hole concentration $^{63}x_{local}^{\kappa}$.
The main principle value of the diagonalized resonance 
frequency $^{\kappa} \nu^{k}$ is then computed and stored into a vector 
$P_{\theta}$, then a new random lattice $\kappa'$ is re-generated along 
with a new value of $|\theta_{local}^{\kappa'}|$
according to the trial PDF. The corresponding resonance frequency $^{\kappa'}\nu^{k}$ is calculated and stored into $P_{\theta}$. As in section IV, taking account of the random fluctuations of the main principle axis around the average NQR axis $\theta_{NQR}$ ($\sim 0.9 \cdot \theta_{local}$) makes only minor $\sim 2$ \% corrections to $^{\kappa} \nu^{k}$. As before, we repeat this 
proceedure $\sim 10^{4}$ times 
until $P_{\theta}$ is sufficiently large to produce a histogram NQR spectrum with $\sim $ 50 bins. Since $^{63}R_{patch}'$ (see Fig. \ref{corrT}) is already determined, the remaining free parameter in the simulation is now the PDF for $|\theta_{local}|$.

The simulation is repeated using various trial PDF's of $|\theta_{local}|$ until the 
the $same$ simulation of the NQR spectrum using $^{63}R_{patch}$ from section IV is obtained. A good starting point for the PDF turns out to be a Gaussian distribution in the squared variable $\theta_{local}^{2}$ centered at the values $ \left< 
\theta_{local}^{2} \right> $ taken from Fig. \ref{theta} for $x=0.04$. The width of the Gaussian distribution can then be optimized to best reproduce the observed NQR spectrum. After various trials, the final results of the optimized PDF's for $\left|\theta_{local}\right| (\equiv \sqrt{\theta_{local}^{2}})$ are shown in Fig. \ref{4distrb} at various temperatures $T \lesssim T_{Q}$.

\section{Conclusions}

We have presented a detailed systematic study of the spatial variation in 
electronic states using $^{63}$Cu NQR in 
$^{63}$Cu isotope enriched poly-crystalline samples of
La$_{2-x}$Sr$_{x}$CuO$_{4}$ for $0.04\leq x \leq 0.16$.
By analysing the extent of the frequency dependence of $^{63}1/T_{1}$ 
across the inhomogeneous linebroadening of the $^{63}$Cu NQR spectrum, we 
determined the spatial variation in hole concentration 
$^{63}x_{local}$ given by
$^{63}x_{local} = x \pm ^{63}\Delta x_{local}$, 
where $^{63}\Delta x_{local}$ was defined as the characteristic 
amplitude or extent of the spatial variaiton
away from nominal hole concentration $x$. We showed that $^{63}
x_{local}(\neq x)$ is a thermodynamic effect whose extent
increases with decreasing temperature below 500-600 K and reaches 
values as large as $^{63}\Delta x_{local}/x \simeq 0.5$ in the 
temperature region $\gtrsim 
150$ K. Furthermore, we showed that the spatial variation 
$^{63}x_{local}$ is an $intrinsic$ effect in
La$_{2-x}$Sr$_{x}$CuO$_{4}$ by confirming the $same$ extent 
of the inhomogeneity in our poly-crystalline samples to that in high-quality 
single crystals \cite{tobe}. To the best of our knowledge these 
reports, together with those in 
Ref.'s \cite{singerprl,tobe},
are the first of its kind to show the temperature dependence 
of the intrinsic inhomogeneity in La$_{2-x}$Sr$_{x}$CuO$_{4}$, or any 
other system with quenched disorder. 
In Ref. \cite{tobe} the results of the inhomogeneous electronic 
state $^{17}x_{local}$ determined using $^{17}$O NMR in high-quality 
La$_{2-x}$Sr$_{x}$CuO$_{4}$ crystals 
are shown to be consistent with the $^{63}$Cu NQR 
result $^{63}x_{local}$. We believe that this is a powerful 
demonstration of the validity of our analysis in this work, as well 
as in \cite{singerprl,tobe}. Moreover, consistency between $^{17}$O NMR
and $^{63}$Cu NQR results indicate that the spatial variation in 
the spin and charge channels are correlated across most of the Brillouin zone. 

We have argued that there are two essential ingredients for the spatial 
variation $^{63}x_{local}(\neq x)$. They are as follows: (a) that the $^{+2}$Sr 
donor ions are positioned with random probability throughout the lattice and 
(b) that there is a short length scale for the spatial variation $^{63}x_{local}$ 
of the order $^{63}R_{patch} \gtrsim 3.0-4.0$ nm. Using 
(a) and (b) above we succesfully fit the 
inhomogeneous $^{63}$Cu NQR spectrum (including the B-line)
using a patch-by-patch model for the spatial variation $^{63}x_{local}$
with the patch radius $^{63}R_{patch}$ as the only adjustable 
parameter. We assumed that no extra $^{63}$Cu NQR 
linebroadening from the lattice distortions existed, therefore the 
calculation for $^{63}R_{patch}$ resulted in a lower bound.
As shown in Ref. \cite{tobe}, however, our lower bound estimate 
$^{63}R_{patch}$ is shown to be consistent with $^{17}R_{patch}$ 
deduced from the inhomogeneous $^{17}$O NMR spectrum, which justifies 
our assumptions used to determine $^{63}R_{patch}$ in the temperature 
region $T>T_{Q}$.
Within the EFG simulation, we also deduced an upper 
boundary to the spatial variation in electronic states 
$^{63}x_{patch}$ given by $^{63}x_{patch} = x \pm ^{63}\Delta 
x_{patch}$, where the extent of the spatial variation 
$^{63}\Delta x_{patch}$ was shown to depend on $^{63}R_{patch}$ as 
such $^{63}\Delta x_{patch}\propto 1/^{63}R_{patch}$ (i.e. 
$^{63}R_{patch}= \infty$ implied the homogenous doping limit $^{63}x_{patch} = x$). 
Within the temperature region $T>T_{Q}$ we found the consistent result 
$^{63}x_{patch} \gtrsim ^{63}x_{local}$, which justified
our EFG simulation and short length scale model of the inhomogeneous electronic state. 

$^{63}R_{patch}$ was deduced 
assuming no clustering of the $^{+2}$Sr ions in the lattice beyond probability 
theory, however, we note that even in the case of 
clustering over length scales 3-4 nm, the presence of a short length scale 
is still needed to account for the observed spatial variation in electronic states. 
We also argued that the short length scale implied that any experimental probe which 
averages over spatial extents larger than $^{63}R_{patch} \gtrsim 3.0-4.0$ 
nm results in an $underestimate$ of the extent of the 
inhomogeneous electronic state. 
NQR, however, is a local probe, which 
makes it $ideal$ for detecting short $\sim$nm variations in the 
electronic state. Our findings have a strong impact on the general, 
but now incorrect, view that doped holes are homogeneously distributed in the 
CuO$_{2}$ planes of La$_{2-x}$Sr$_{x}$CuO$_{4}$. 

We have also compared our local orthorhombic
structural distortions results to LRO structural results taken 
from neutron and X-ray diffraction \cite{radaelli,yamada} techniques
which probe coherence phenomena over length scales 
larger than ten's of nm \cite{takagidelta}. In the case of $x=0.0$ 
we found good agreement between NQR and LRO results, however, for $x \gtrsim 0.04$ 
we presented evidence that the local orthorhombic distortions started at elevated 
temperatures $T_{Q}(\gtrsim 400$ K) $above$ the onset of LRO $T_{st}$. 
We also found that the local distortions were somewhat smeared below 
$T_{Q}$ compared with $x=0.0$. Our local structural results were found to be 
consistent with a mixture 
of order-disorder and displacive transitions in the 
region $x \gtrsim 0.04$, consistent with other local probes \cite{bozin,haskel}.
Our NQR studies at elevated temperatures 
have now quantified the onset of the local orthorhombic distortions 
at $T_{Q} (\gtrsim 
400$ K) in the region $0.04 \leq x \leq 0.20$. 

It is known that in the $x<1/8$ region of the phase diagram, 
the temperature dependence of the wipeout fraction $^{63}F(T)$ (defined as the fracton of unobservable 
Cu nuclei that are wiped-out)
is somewhat tailed \cite{hunt,hunt2,singerprl} below $T_{NQR}$
in both La$_{2-x}$Sr$_{x}$CuO$_{4}$ and its Eu and Nd co-doped 
materials. In the case of 
La$_{2-x}$Sr$_{x}$CuO$_{4}$ for $x=0.04$, the onset temperature for Cu
wipeout is rather high at $T_{NQR} \simeq 300$ K, yet 
half the Cu nuclei are still observable (i.e. $^{63}F(T) \sim 
1/2$) at $\sim T_{NQR}/2$, 
implying that $T_{NQR}$ does not signify the onset of a $collective$ 
phenomenon for $x < 1/8$. This may indicate the formation of a small number of 
local moments in the CuO$_{2}$ plane which cause the wipeout of the 
NQR signal for neighbouring Cu sites, as in a classic spin-glass.

Rather than invoking a spin-glass type model, however, we can now naturally account for the tailed wipeout for $x<1/8$
using the temperature dependence of the growing spatial variation in electronic 
states $^{63}x_{local}$. In the temperature region 
$T_{NQR} \sim 300$ K for $x= 0.04$, the characteristic local hole concentration 
$^{63}x_{local}$ for insulating patches
is inferred to be as low as $^{63}x_{local} \sim 0.015$ (Fig. \ref{deltax}). 
Recall that $^{63}1/T_{1}$ for $x=0.02$ at $\sim 300$ K is already too 
large for the Cu signal to be observable, therefore the more insulating patches for $x= 0.04$ at 
300 K begin to wipe-out. As one cools the $x=0.04$ sample below 300 K, there are 
a growing number of patches with lower and lower hole concentration which are also 
wiped-out, leading to an gradual increase in the wipeout fraction 
$^{63}F(T)$ with decreasing temperature. The fact that 
$^{63}x_{local}$ only gradually fans out with decreasing temperature naturally
explains the tailed nature of the wipeout 
fraction $^{63}F(T)$ below $T_{NQR}$ in the underdoped region $x < 1/8$. 

In Eu and Nd co-doped 
La$_{2-x}$Sr$_{x}$CuO$_{4}$
for $x<1/8$, the same tailed feature in $^{63}F(T)$
is found together with an inflection point $T_{NQR}^{inflec}  (< T_{NQR})$ below which 
$^{63}F(T)$ increases rapidly with decreasing temperature. In the case of Eu and Nd co-doping with $x<1/8$, $T_{NQR}^{inflec}$
coincides with the onset temperature $T_{LTT}$ of an additional structural 
phase transition with tetragonal symmetry known as the 
LTT (low temperature tetragonal) phase $and$ the onset of charge 
ordering $T_{charge}$. In the case of La$_{1.6-x}$Nd$_{0.4}$Sr$_{x}$CuO$_{4}$, $T_{charge} \simeq T_{LTT} \simeq 70$ K \cite{ichi}.
This coincidence suggests that the glassy slowing of the 
stripe modulation below $T_{NQR}^{inflec} (\simeq T_{charge}) $ is accelerated by the tetragonal symmetry 
compatible with the charge stripe symmetry \cite{tranquada}. 
$^{63}$Cu isotope enriched samples of Eu and Nd co-doped 
La$_{2-x}$Sr$_{x}$CuO$_{4}$ are found to have comparable frequency dependence 
of $^{63}1/T_{1}$ across its inhomogeneous $^{63}$Cu NQR spectrum. In 
the Eu and Nd co-doped cases, therefore, we can naturally
account for the tailed wipeout in the temperature region 
$T_{NQR}^{inflec}<T<T_{NQR}$ by the growing inhomogeneity in electronic states 
with decreasing temperature, while 
for  $T < T_{NQR}^{inflec} (\simeq T_{charge}$), the wipeout is dramatically accelerated due to the 
pinning of charge stripes by the lattice symmetry.
Although there is no LRO into a LTT
phase in La$_{2-x}$Sr$_{x}$CuO$_{4}$ which can stablize the
charge stripes, there is evidence for an incipient structural transition 
into the LTT phase \cite{kimura}, which may suggest a similar phenomenon 
is occuring in La$_{2-x}$Sr$_{x}$CuO$_{4}$.

\appendix

\section{Anti-shielding factors}

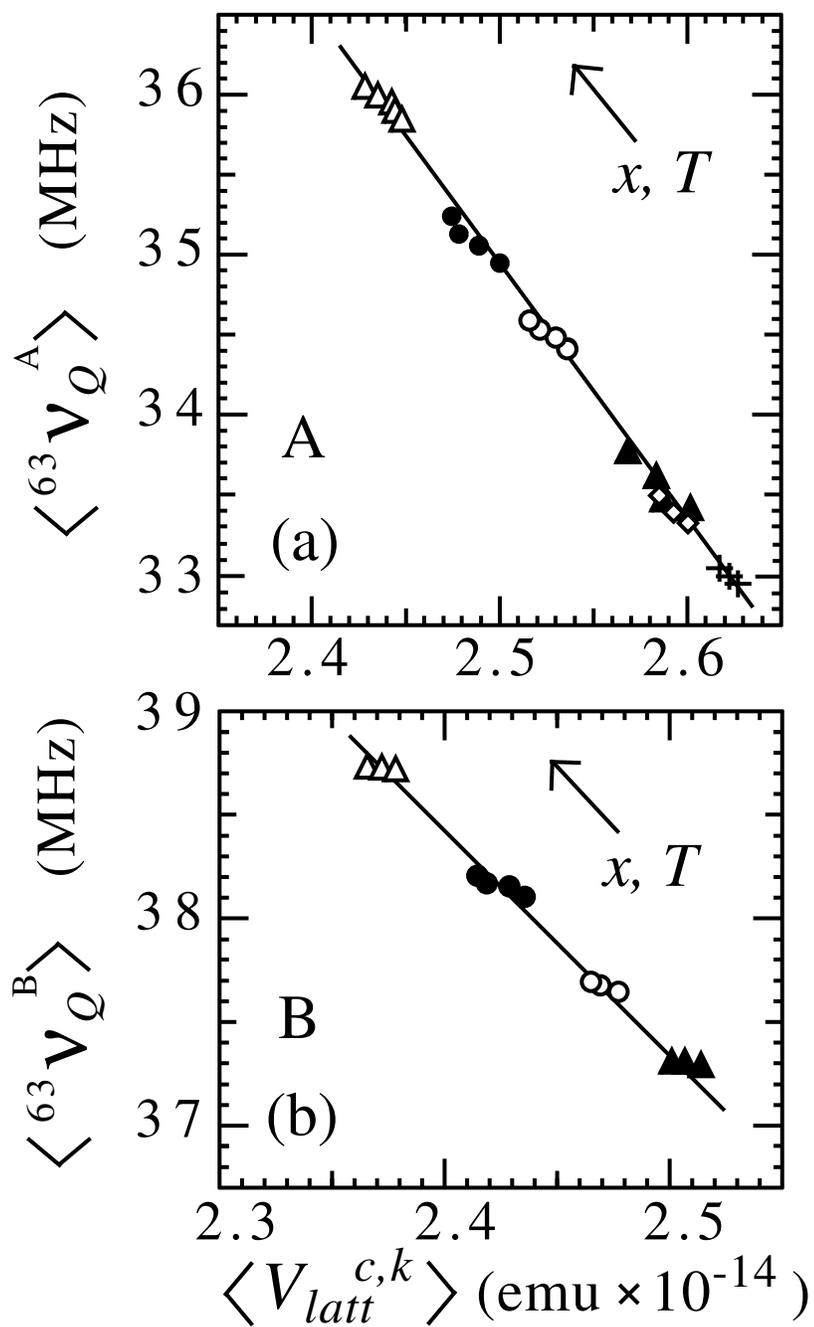
\begin{figure}
\caption{(a) Cross-plot of 
experimental data $\left< ^{63}\nu_{Q}^{A} \right>$ taken from Fig. \ref{nuqall}(a) against 
the calculated EFG at CG $\left< V_{latt}^{c,A}\right> $ taken from Fig. 
\ref{simul1} with $x$ and $T$ as implicit parameters, both for the A line. (b) Same 
cross-plot as in (a) but for the B-line.}
\label{gammas}
\end{figure}

In this appendix we deduce the anti-shielding factors \cite{slichter} $\zeta_{latt}^{k}$ and 
$\zeta_{3d}^{k}$, where $k = (A,B)$.
We first assume that both anti-shielding factors are isotropic and independent of $x$ and $T$,
but are allowed to vary between the A and B-lines.
We produce a cross-plot (shown in Fig. \ref{gammas}(a)) of the
resonance frequency at the CG of the
A-line $\left<^{63}\nu_{Q}^{A}\right>(x,T)$ (Fig. \ref{nuqall}(a)) against 
the CG of the lattice contribution to the EFG $\left<V_{latt}^{c,A}\right>(x,T)$ (Fig. 
\ref{simul1}). Likewise, we produce the same cross-plot for the B-line shown 
Fig. \ref{gammas}(b). Note that we only use data in temperature region 
above $T_{Q}$ (Fig. \ref{nuqall}) 
where local orthorhombic distortions are absent.

Then we deduce the linear fit using the standard empirical form 
\cite{penn,shimizu}:
\begin{equation}
    \left<^{63}\nu_{Q}^{k}\right> =\Lambda_{3d}^{k} - \Lambda_{latt}^{k} \cdot
    \left<V_{latt}^{c,k}\right>
\label{dedgammas}
\end{equation}
where $k = (A,B)$ and all quantities are overall positive.
The best fit to the data shown by the solid 
black lines in Fig. \ref{gammas} indicates that $\Lambda_{3d}^{A}$ = 
75.3 MHz and $\Lambda_{latt}^{A} = 16.0$ MHz/[emu$\times 10^{-14}$] for the A-line,
and that $\Lambda_{3d}^{B}$ = 
64.6 MHz and $\Lambda_{latt}^{B} = 10.7$ MHz/[emu$\times 10^{-14}$] for the B-line. 

Using Eq. (\ref{EFGconv}) we express the 
$\Lambda$'s in terms of the unknown anti-shielding factors as such
\begin{eqnarray}
\Lambda_{3d}^{k} & =& \frac{4e^{2}|^{63}Q|}{14 h} 
\left(1-4f_{\sigma}^{o}\right) 
\left<r^{-3}_{3d}\right> \cdot \zeta_{3d}^{k} \nonumber \\
\Lambda_{latt}^{k} & = & \frac{e|^{63}Q|}{2h} \cdot \zeta_{latt}^{k}
\label{Lambdazeta}
\end{eqnarray}
Using $f_{\sigma}^{o} = 0.076$ \cite{tobe} 
, $^{63}Q = -0.16 $ barns \cite{bleaney}, and the bare 
value $\left<r^{-3}_{3d}\right> =7.5$ a.u. \cite{bleaney} for free cupric 
ions, we find that for the A-line $\zeta_{latt}^{A} = 27.7$ and 
$\zeta_{3d}^{A} = 0.93$, while for the B-line $\zeta_{latt}^{B} = 18.6$ 
and $\zeta_{3d}^{B} = 0.82$. These set of results are consistent with 
previous estimates \cite{penn,shimizu}.

Converting the above analysis using $^{63}Q = -0.211$ barns 
\cite{stern} instead of $^{63}Q = -0.16 $ barns \cite{bleaney} results 
in a uniform 24.2 \% decrease in our estimates of the $\zeta$'s. 
However, note that what really matters in converting from calculation 
to experiment are the $\Lambda$'s in Eq. (\ref{dedgammas}). Likewise, when 
accounting for the linebroadening in section IV to deduce $^{63} \Delta x_{patch}$,
it is also the $\Lambda$'s which ultimately 
matter, not the $\zeta$'s, therefore our end result for $^{63} \Delta 
x_{patch}$ and $^{63}R_{patch}$ are independent of what value of $^{63}Q$ we use.

\section{Magnetization Recovery}

In this appendix we quantify the observed 
multi-exponential recovery of the magnetization $M(t)$ in the case of $x=0.16$. The fact that
the recovery is multi-exponential implies that $^{63}1/T_{1}$ is {\it distributed} at a
{\it fixed} frequency on the NQR line. We can naturally account for the distribution 
in $^{63}1/T_{1}$ at fixed frequency by recalling that the $^{63}$Cu NQR line has an underlying intrinsic lattice broadening ($\Delta \nu _{latt}^{k}$) which will tend to smear out the frequency dependence of $^{63}1/T_{1}$ across the NQR line. Smearing out the frequency dependence of  $^{63}1/T_{1}$ results in a distribution of $^{63}1/T_{1}$ values at a fixed frequency. Using the the multi-exponential recovery data, we estimate the extent of the underlying lattice broadening $\Delta \nu _{latt}^{T1} = 0.62 (\pm 0.07) $ MHz in the case of $x=0.16$ at the A-line. We find that $\Delta \nu _{latt}^{T1}$ is consistent with the lattice broadening $\Delta \nu_{latt}^{A}=
0.49 $ MHz found independently using our point charge calculation (section IV), which also adds weight to our EFG simulation of the random lattice. 

The experimentally 
observed nuclear recovery $M(t)$ can be fit to Eq. (\ref{mt}), where 
$M(0)$, $M(\infty)$ and $^{63}1/T_{1}$ are free parameters of the 
fit. The bare recovery is then defined as $I(t)$ where
\begin{equation}
    I(t) =  \frac{M(t)  -  M(\infty)} {  M(0) - M(\infty) }.
\label{norm}
\end{equation}
In the case of single exponential recovery, $I(t)$ is then found to be a straight line on a semi-log plot, an example of which is shown in Fig. \ref{recova} where we present $I(t)$ in the case $x=0.0$ at 475 K for various positions across its NQR spectrum. 
As discussed earlier, the $x=0.0$ spectrum is homogeneously broadened, and therefore shows no frequency dependence of $^{63}1/T_{1}$ across its 
line (i.e. $I(t)$ has the same slope at different frequencies), nor does it show any sign of having a distribution at a fixed frequency.
We also note that the data in Fig. \ref{recova} is taken at 
475 K which is the orthorhombic phase, therefore we can also rule out 
the possibility that the orthorhombic distortions are the $direct$ cause of 
observed frequency dependence in $^{63}1/T_{1}$.

In Fig. \ref{recovb} we show the bare recovery $I(t)$ at similar 
positions across the A-line for $x=0.16$ at 100 K. First note that the slopes in the recovery get steeper
towards lower frequency. This corresponds to the frequency dependence in $^{63}1/T_{1}$ where $^{63}1/T_{1,A}^{(-)} > 
^{63}1/T_{1,A}^{(0)} > ^{63}1/T_{1,A}^{(+)}$. Second, note that $I(t)$ shows finite curvature on a semi-log plot. This implies that there is a distribution in recoveries $I(t)$ which can be fit to the general form
\begin{equation}
    I(t) =  \frac{1}{\sum_{j} a_{j}} \cdot  \sum_{j} a_{j} \exp \left(-\frac{3}{T_{1,j}}t 
    \right).
\label{bare}
\end{equation} 
Fitting the recoveries to a single exponential 
form (i.e. taking $a_{j}$ =1 and $1/T_{1,j} = 1/T_{1}$ in Eq. 
(\ref{bare})) results in the force fit values $^{63}1/T_{1,A}^{(-)}$, $
^{63}1/T_{1,A}^{(0)}$ and $^{63}1/T_{1,A}^{(+)}$. These force fit values represent the average of the underlying distribution in $^{63}1/T_{1,A}$.

The dashed lines in Fig. \ref{recovb} show the best fit
result from a sum of exponentials $1/T_{1,j}$ 
whose distribution coefficients $a_{j}$ are presented as the dashed lines in Fig. \ref{100K}. 
The distribution coefficients $a_{j}$ in Fig. \ref{100K} are 
determined as such
\begin{eqnarray}
    a_{j}^{(-)}& =& \exp \left( -\frac{(P_{T1}- 
    1/T_{1,j})^{2}}{(\Delta_{P}^{(-)})^{2}} \right)  \label{aso}  \\
        a_{j}^{(+)}& =& \exp \left(-\frac{(P_{T1}- 
    1/T_{1,j})^{2}}{(\Delta_{P}^{(+)})^{2}} \right),
    \label{as}
    \end{eqnarray}
where Eq. (\ref{aso}) is used for $1/T_{1,j} < P_{T1}$ and Eq. 
(\ref{as}) is used for $1/T_{1,j} > P_{T1}$. The three free parameters used to fit $I(t)$ are therefore 
the peak of the distribution $P_{T1}$ and the two widths $\Delta_{P}^{(-)}$ 
and $\Delta_{P}^{(+)}$.
When optimizing the three parameters in Eq. (\ref{as}), we also use the constraint that the CG of the distribution (shown as the solid vertical lines 
in Fig. \ref{100K})
coincide with the force-fit value $^{63}1/T_{1,A}$ (presented in Fig. \ref{1563}) deduced with a single 
exponential fit.

We then deduce the second moment of the distributions in Fig. \ref{100K} 
in an analogous manner to Eq. (\ref{meanwidth}) and Eq. (\ref{meanwidth2}),
and we find that the {\it HWHM} of the 
distributions are $\Delta 1/T_{1,A}^{(-)} = 0.33$ (ms)$^{-1}$, $\Delta 
1/T_{1,A}^{(0)} = 0.27$ (ms)$^{-1}$ and $\Delta 1/T_{1,A}^{(+}) = 0.20$ (ms)$^{-1}$. 
Note that the {\it HWHM} of the distributions are $\sim $ 15 \% of 
average value
$^{63}1/T_{1,A}$, which implies that accurate recovery data $I(t)$ is 
needed down to $I(t) \sim 0.001$. 

We can now get an estimate of the underlying intrinsic lattice 
broadening $\Delta \nu_{latt}^{T1}$ by 
converting the {\it HWHM} $\Delta 1/T_{1,A}$ deduced above into equivalent
frequency widths. In order to do so 
we take a local derivative $\beta_{\nu}$ of the underlying frequency dependence of 
$^{63}1/T_{1,A}$ at a frequency $\nu$ defined as such
\begin{equation}
\beta^{\nu} = \left |\frac{ \delta 1/T_{1,A}}{\delta \nu} \right|_{\nu},
\label{fudge}
\end{equation}
where $1/T_{1,A}$ is the interpolated value of $^{63}1/T_{1,A}$ shown in Fig. 
\ref{1563} as the solid line. Using $^{63}1/T_{1,A}$ in units of (ms)$^{-1}$ and 
$\nu$ in units of MHz, we find that $\beta^{(-)} = 0.62$, $\beta^{(0)} = 
0.39$, and $\beta^{(+)} = 0.29$, from which we deduce the {\it HWHM} (in MHz)
\begin{eqnarray}
\Delta \nu _{latt}^{(-)} & = & \Delta 1/T_{1,A}^{(-)}/\beta^{(-)} = 0.53 \nonumber \\
\Delta \nu _{latt}^{(0)} & = & \Delta 1/T_{1,A}^{(0)}/\beta^{(0)} = 0.69 \nonumber \\
\Delta \nu _{latt}^{(+)} & = & \Delta 1/T_{1,A}^{(+)}/\beta^{(+)} = 0.68
\label{whatnow}
\end{eqnarray}
for the lower half intensity, center and upper half 
intensity of the spectrum, respectively. Finally, we deduce the
average value $\Delta \nu _{latt}^{T1}$ of the three estimates in 
Eq. (\ref{whatnow}) to be $\Delta \nu _{latt}^{T1} = 0.62 (\pm 0.07)$ MHz. This is 
consistent with $\Delta \nu _{latt}^{A} = 0.49$ MHz deduced independently 
using the point charge calculation, as shown in Fig. \ref{1563}.

We note that our whole analysis in this appendix is derived using fixed
$\tau =12$ $\mu$s conditions. As discussed in section II and III, however, there exists a small $\tau$ dependence on the underlying distribution in the spin-lattice relaxation rate. Such effects could account for the 25 \% difference between the lattice width $\Delta \nu _{latt}^{T1}$ derived from Eq. (\ref{whatnow}) and the point charge calculation $\Delta \nu _{latt}^{k}$ in section IV.
\begin{acknowledgements}
The work at M.I.T. was supported by NSF DMR 98-08941 and 99-71264. 
\end{acknowledgements}

\begin{figure}
\caption{Temperature dependence of field-cooled Meissner signal
in 15 Oe for poly-crystalline La$_{2-x}$Sr$_{x}$CuO$_{4}$ with $x=0.20$ 
($\times$), $x=0.16$ ($\triangle$),
$x=0.115$ ($\bullet$), $x=0.09$ ($\blacklozenge$), $x=0.07$ ($\circ$), and 
$x=0.04$ ($\blacktriangle$). Lines are a guide for the eye.}
\label{squid}
\end{figure}

\begin{figure}
\caption{Temperature dependence of $^{63}1/T_{1,A}^{(0)}$ at the 
CG of the A-line for $x=0.20$ (solid grey line), $x=0.16$ ($\triangle$),
$x=0.115$ ($\bullet$), $x=0.09$ ($\blacklozenge$), $x=0.07$ ($\circ$), 
$x=0.04$ ($\blacktriangle$), $x=0.02$ ($\diamond$), and $x=0.0$ (+). Solid black curves are a 
guide for the eye. All data (in this Fig. and in Fig. \ref{T1B})
are taken above the $^{63}$Cu wipeout 
temperature $T_{NQR}$ using a fixed pulse separation time of $\tau =$ 12 
$\mu$s.}
\label{T1A}
\end{figure}

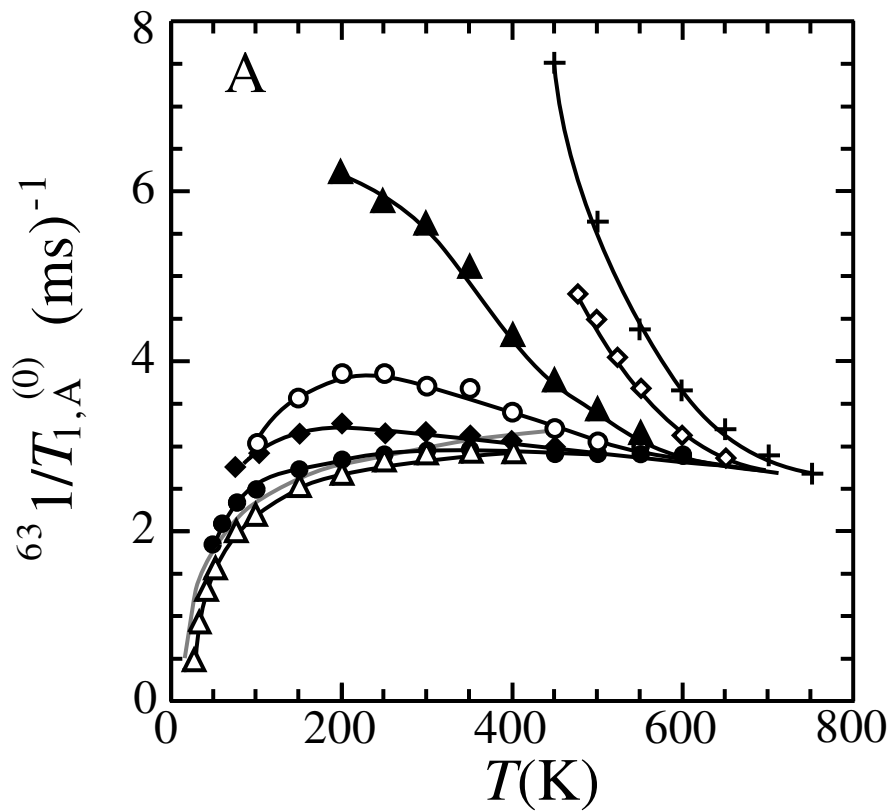
\begin{figure}
\caption{Temperature dependence of $^{63}1/T_{1,B}^{(0)}$ 
at the CG of the B-line
where the same data symbols from Fig. \ref{T1A} are used, together with the 
black solid lines which represent $\left( \epsilon \cdot ^{63}1/T_{1,A}^{(0)} \right)$ data
taken directly from Fig. \ref{T1A} with a uniform scaling factor 
$\epsilon$, where $\epsilon 
=[0.87,0.87,0.90,0.84]$ for $x = [0.20,0.16,0.115,0.07]$, respectively. Grey 
dashed line represents $^{63}1/T_{1,B}^{(0)}$ for $x=0.20$.}
\label{T1B}
\end{figure}

\clearpage

\begin{figure}
\caption{(a) Lower bound estimate of the patch radius $^{63}R_{patch}$ (in units of nm on left 
axis and in multiples of lattice spacing $a$ on right axis) used 
to fit spectra in Fig. \ref{600Kf} at 600 K using the $^{+2}$Sr donor ions $d_{1}$ ($\circ$) 
or $d_{2}$ ($\bullet$) as donors. ($\diamond$) corresponds to 
$^{+2}$Sr-$^{+2}$Sr separation $l_{Sr} = a/\sqrt{x}$. (b) 
Experimentally deduced {\it HWHM} $^{63}\Delta \nu_{Q}^{A}$ ($\blacksquare$) of spectra shown 
in Fig. \ref{600Kf} along with calculated intrinsic lattice 
broadening $\Delta \nu_{latt}^{A}$ for the A-line. All lines are a 
guide for the eye. Also shown is $\Delta \nu _{latt}^{T1}$ ($\times$) 
for $x=0.16$ deduced in appendix B.}
\label{600Kx}
\end{figure}

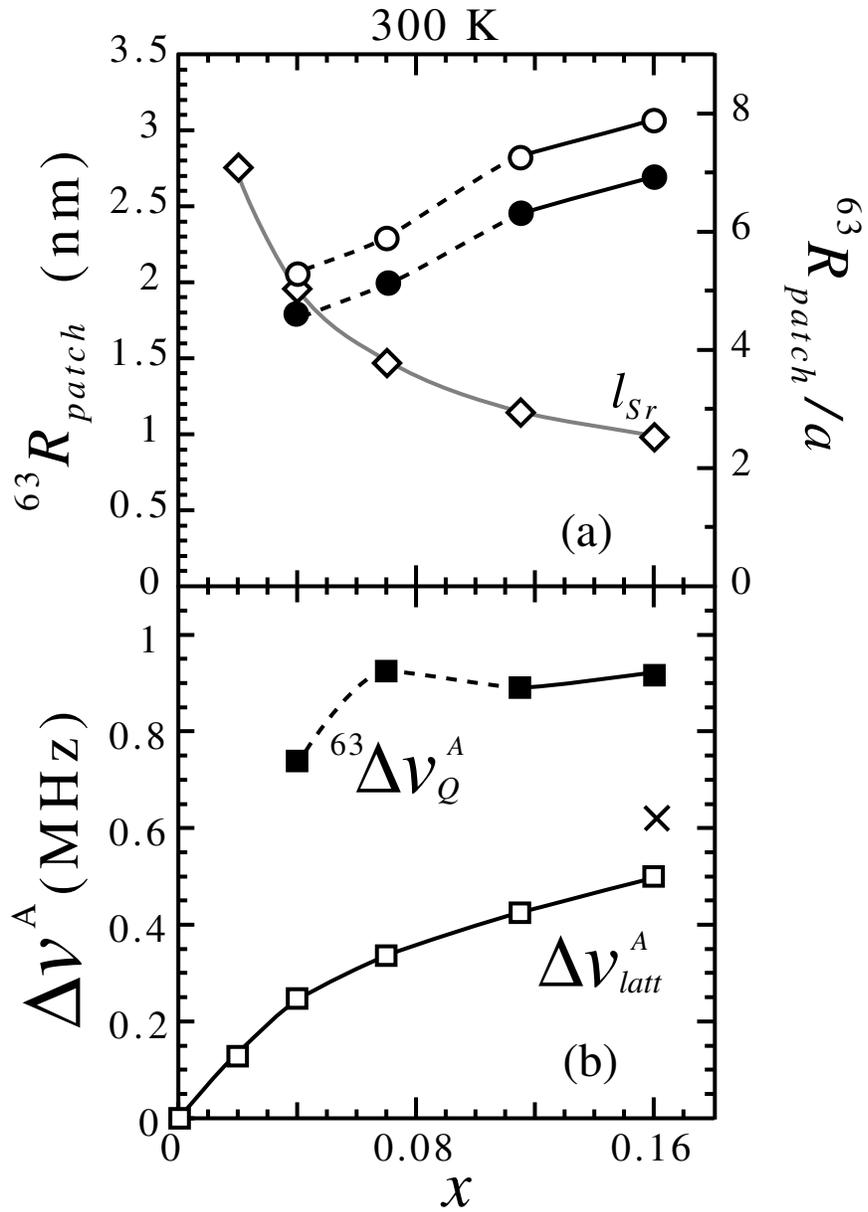
\begin{figure}
\caption{Same plot and symbols as Fig. \ref{600Kx} but at 300 K.
Dashed black lines corresponds to orthorhombic structural phase for 
$x\leq 0.115$ according to LRO \cite{yamada}. All other lines are a guide for the eye.}
\label{300KX}
\end{figure}

\begin{figure}
\caption{Temperature dependence of lower bound estimate to the patch 
radius $^{63}R_{patch}$ (defined according to average in Eq. 
(\ref{ave})) deduced from the width of the $^{63}$Cu NQR spectrum shown in Fig. 
\ref{widthall} for $x=0.16$ ($\triangle$), $x=0.115$ ($\bullet$), $x=0.07$ ($\circ$), and
$x=0.04$ ($\blacktriangle$). Lines are a guide for the eye. Grey lines indicate the temperature region 
$T < T_{Q}$ where local orthorhombic distortions are present (Fig. \ref{nuqall}). Dashed black line corresponds to LRO 
structural temperature $T_{st}$ \cite{yamada}. Dashed grey line for $x=0.04$ corresponds to $^{63}R'_{patch}$ (see text) for $T < T_{Q}$.}
\label{corrT}
\end{figure}

\begin{figure}
\caption{Temperature dependence of experimental {\it HWHM} $^{63}\Delta \nu_{Q}^{A}$ for 
$=0.04$ ($\blacktriangle$) taken from Fig. \ref{widthall}, where black and 
grey lines show temperature region above and below $T_{Q}=T_{st} 
=450 $ K. Also shown is
$\Delta \nu_{R_{patch}'}^{A}$ ($\circ$) for $T < T_{Q}$ determined using $^{63}R_{patch}'$ taken from Fig. \ref{corrT}.
The remaining linebroadening due to orthorhombic distortions alone, $\Delta \nu_{\theta}^{A}$ ($\times$), are also estimated (see Ref. \cite{expl}).}
\label{4distra}
\end{figure}

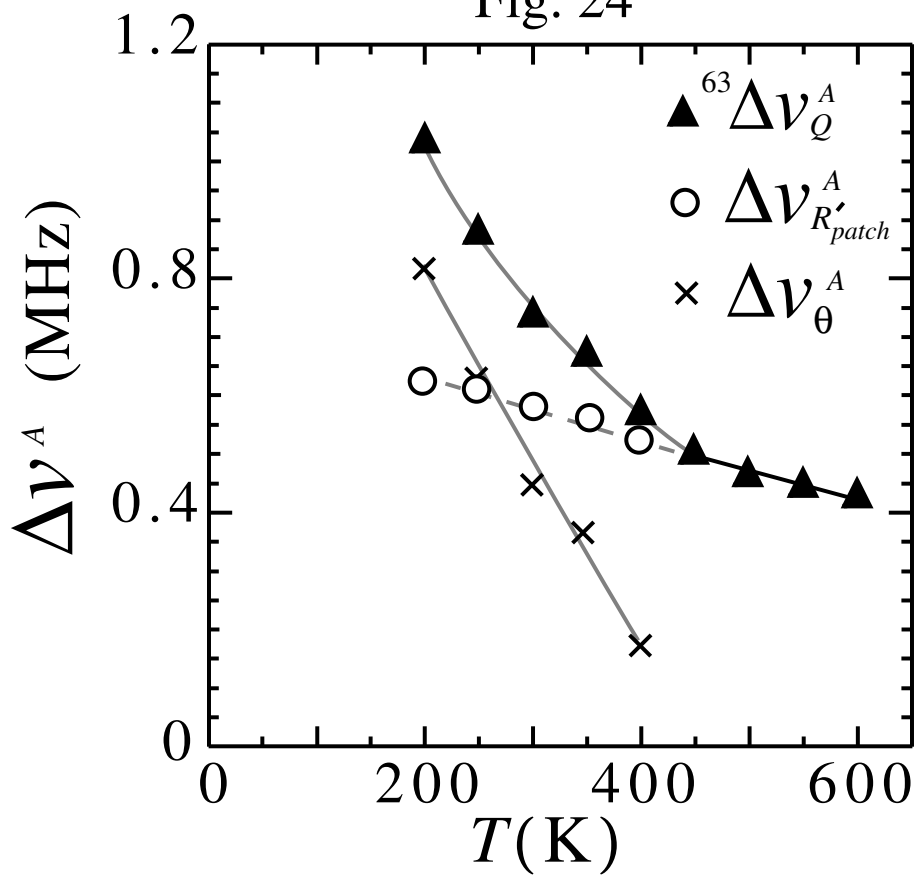
\begin{figure}
\caption{Probability Distribution Function for the magnitude of local tilting angle 
$|\theta_{local}|$ for $x=0.04$ at various temperatures $T < T_{Q}$. At each temperature, the PDF's shown result in the best fit to the NQR lineshape assuming $^{63}R'_{patch}$ from Fig. \ref{corrT}.}
\label{4distrb}
\end{figure}

\begin{figure}
\caption{Time dependence of bare recovery $I(t)$ deduced from 
magnetisation $M(t)$ (see Eq. (\ref{bare}) for $x=0.0$ at 475 K ($< T_{st}$),
taken at various positions across NQR spectrum
including $^{63}1/T_{1,A}^{(0)}$ (grey $\times$), $^{63}1/T_{1,A}^{(+)}$ ($\circ$) and 
$^{63}1/T_{1,A}^{(-)}$ ($\bullet$). Solid line is best fit to a single exponential recovery.}
\label{recova}
\end{figure}

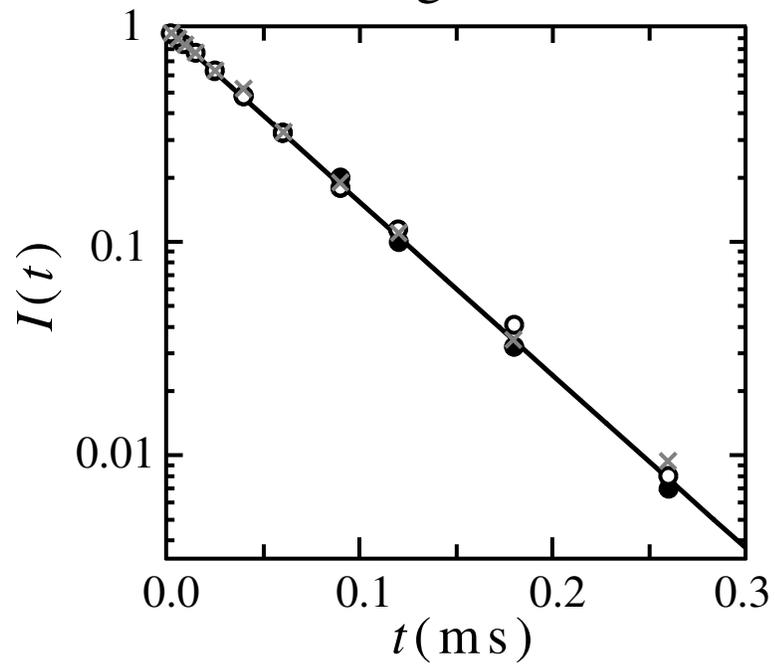
\begin{figure}
\caption{Same plot and symbols as Fig. \ref{recova} but for $x=0.16$ at 100 K. 
Solid lines are force fits using a 
single exponential recovery,
while dashed lines are best fits using multiple exponential recoveries whose distribution is
shown in Fig. \ref{100K}.}
\label{recovb}
\end{figure}

\begin{figure}
\caption{Distribution coefficients $a_{j}$ of
$1/T_{1,j}$ (dahsed lines) deduced from best 
fit to recoveries $I(t)$ shown 
in Fig. \ref{recovb} for $x=0.16$ at 100 K.
Solid vertical lines show both force fit 
values $^{63}1/T_{1,A}^{(-)}$
$^{63}1/T_{1,A}^{(0)}$ (grey line) and $^{63}1/T_{1,A}^{(+)}$ in order of 
increasing frequency shown by arrow which coincide with CG of distributions.}
\label{100K}
\end{figure}

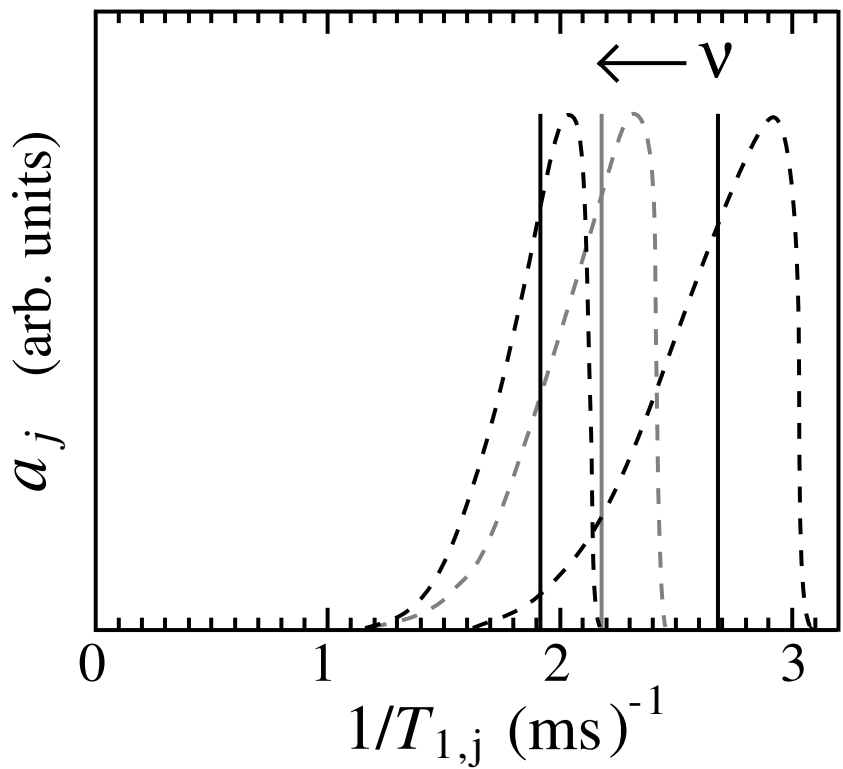
\begin{figure}
\caption{$^{63}$Cu NQR spectrum of $x=0.16$ at 100 K ($\bullet$) 
along with force fit values $^{63}1/T_{1}$ ($\circ$) at various positions across the 
line. Grey lineshape shows fit using
patch-by-patch model with a lower bound for the patch radius of $^{63}R_{patch} = 2.7 \pm 0.2$ nm,
while grey dashed lines shows intrinsic linebroadening (with {\it HWHM} of
$\Delta \nu_{latt}^{k}$) according to EFG simulation (section IV).
Black lineshape shows intrinsic linebroadening (with {\it HWHM} of $\Delta \nu_{latt}^{T1}$) deduced 
from recoveries in Fig. \ref{recovb} at the A line.
Solid curves through ($\circ$) are interpolations between data points.}
\label{1563}
\end{figure}


\begin{thebibliography}{999}
\bibitem{red}The Redfield contribution ($^{63}1/T_{2R}$) to the spin-spin relaxation rate ($^{63}1/T_{2}$) \cite{slichter} is given by $^{63}1/T_{2R} = \left(\beta+R \right) \cdot ^{63}1/T_{1}$, where $R \sim 4.5$ is the $T_{1}$ anisotropy and $\beta =2$ in the case of $^{63}$Cu NQR.
\bibitem{T2G}The spin-spin relaxation rate is determined by applying the spin-echo sequence (shown in brackets in Eq. (\ref{pulse})) and measuring the magnetisation ($M(\tau)$) as a function of $\tau$. $M(\tau)$ is then fit to the standard form $M(\tau) = M(0) \cdot \exp\left(-\frac{2\tau}{T_{2L}} - \frac{(2\tau)^{2}}{2(T_{2G})^{2}}\right)$, where $^{63}1/T_{2L}$ is the Lorentzian-like component (which includes the Redfield contribution \cite{red}) and $^{63}1/T_{2G}$ is the Gaussian-like \cite{slichter} component due to the indirect nuclear spin-spin coupling.
\end{thebibliography}
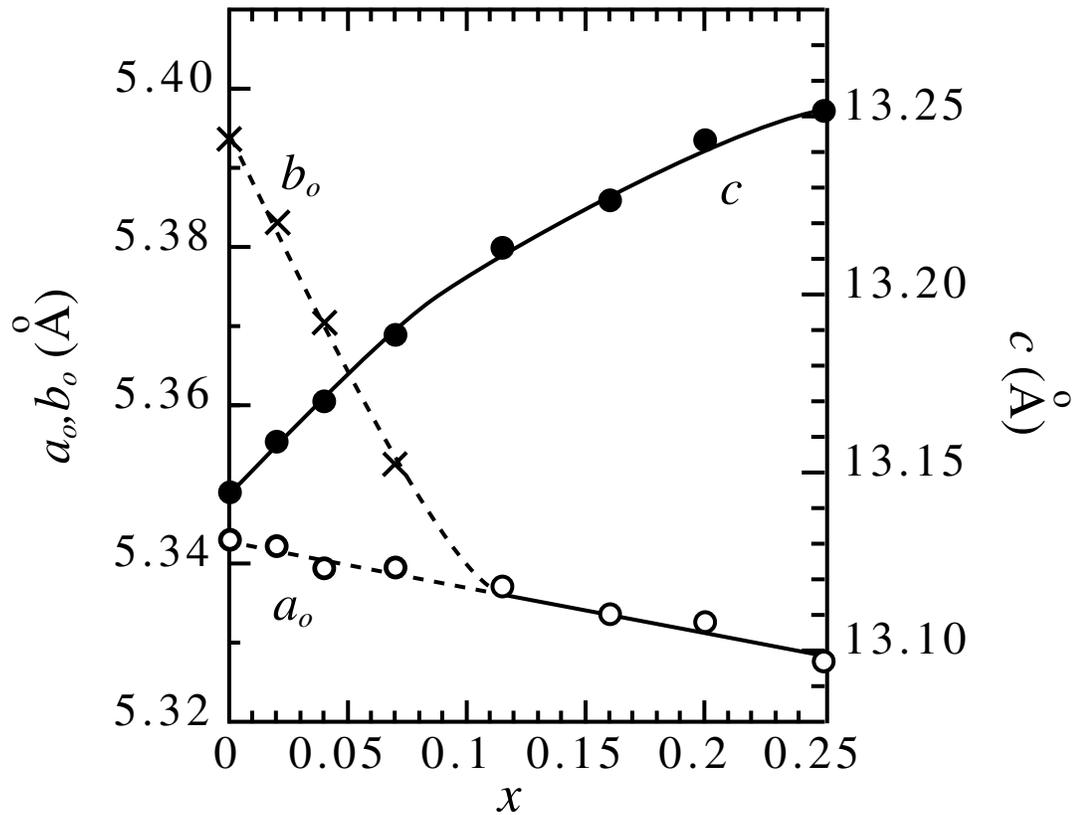
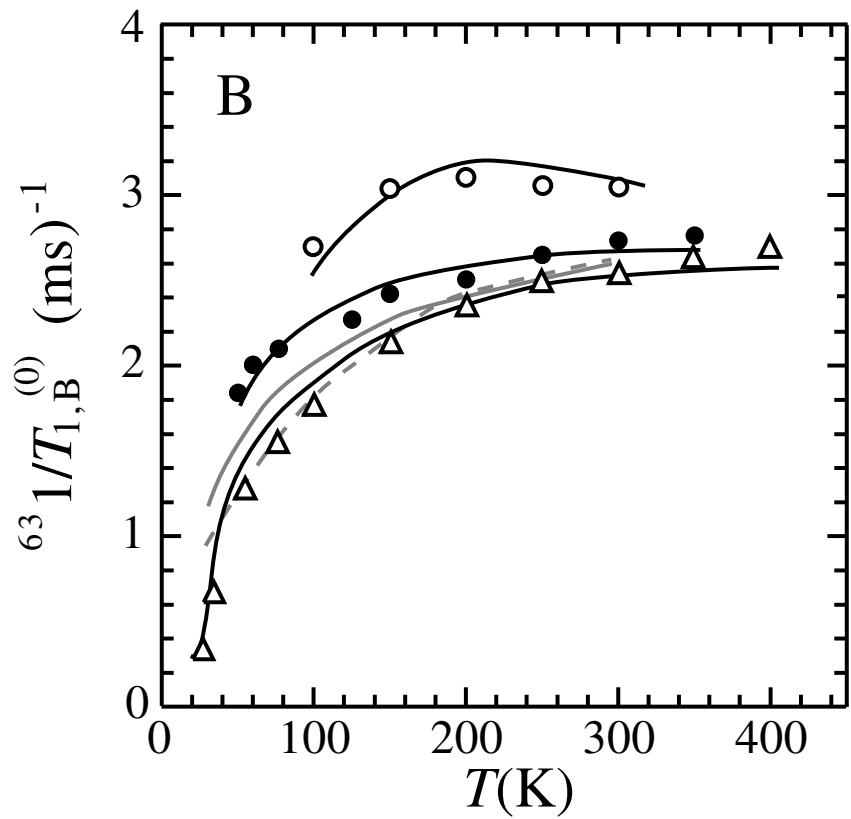
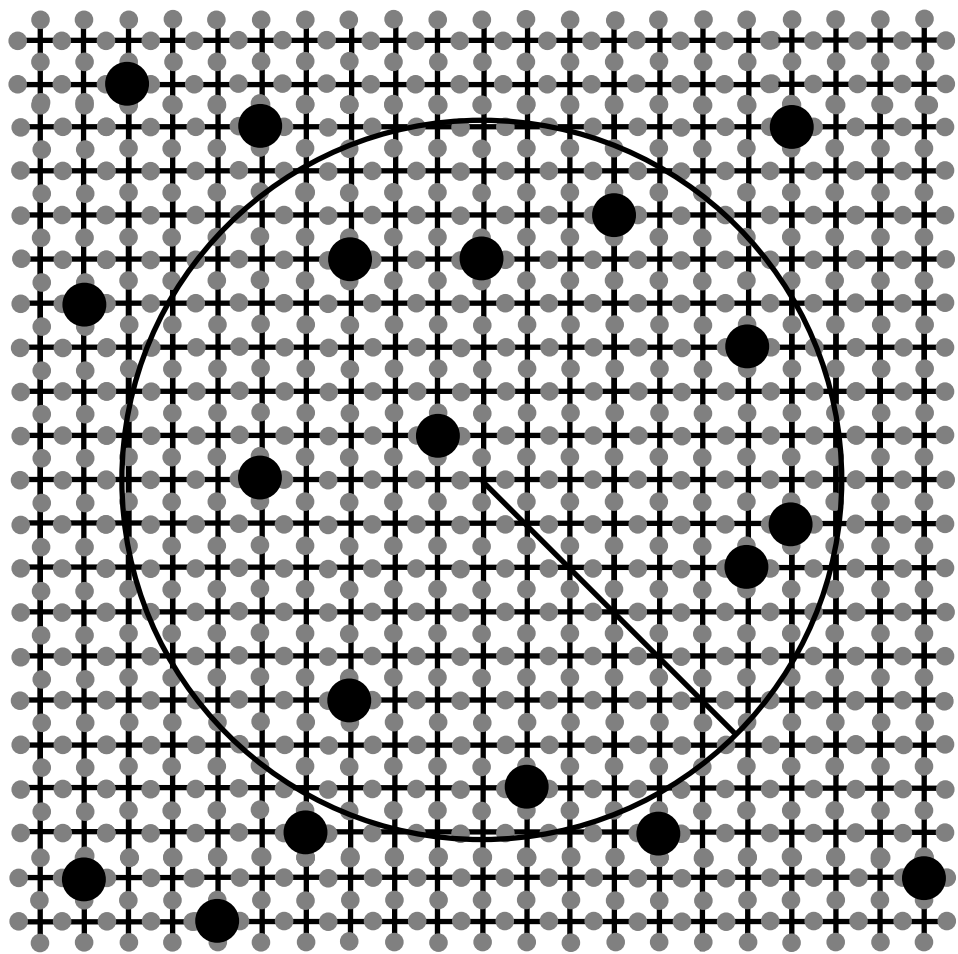
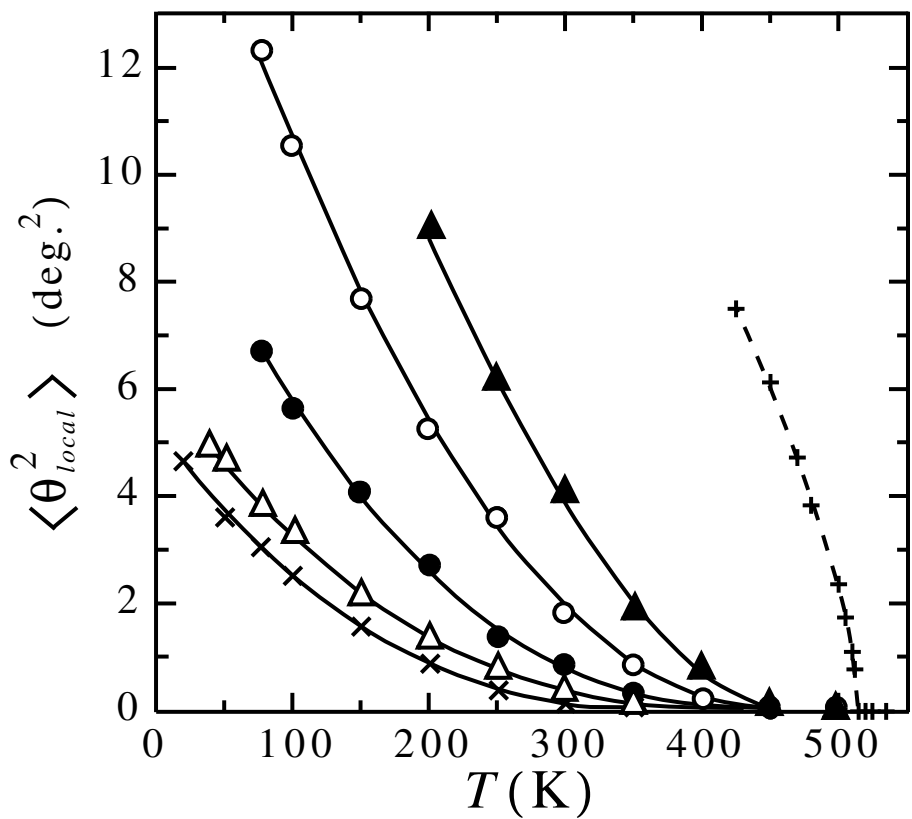
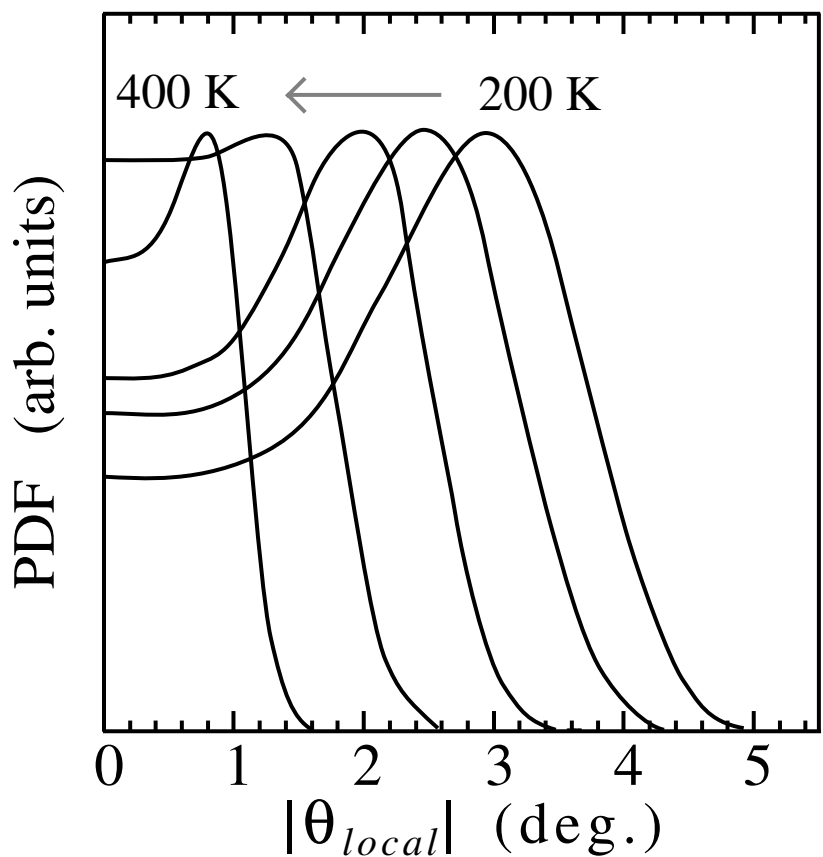
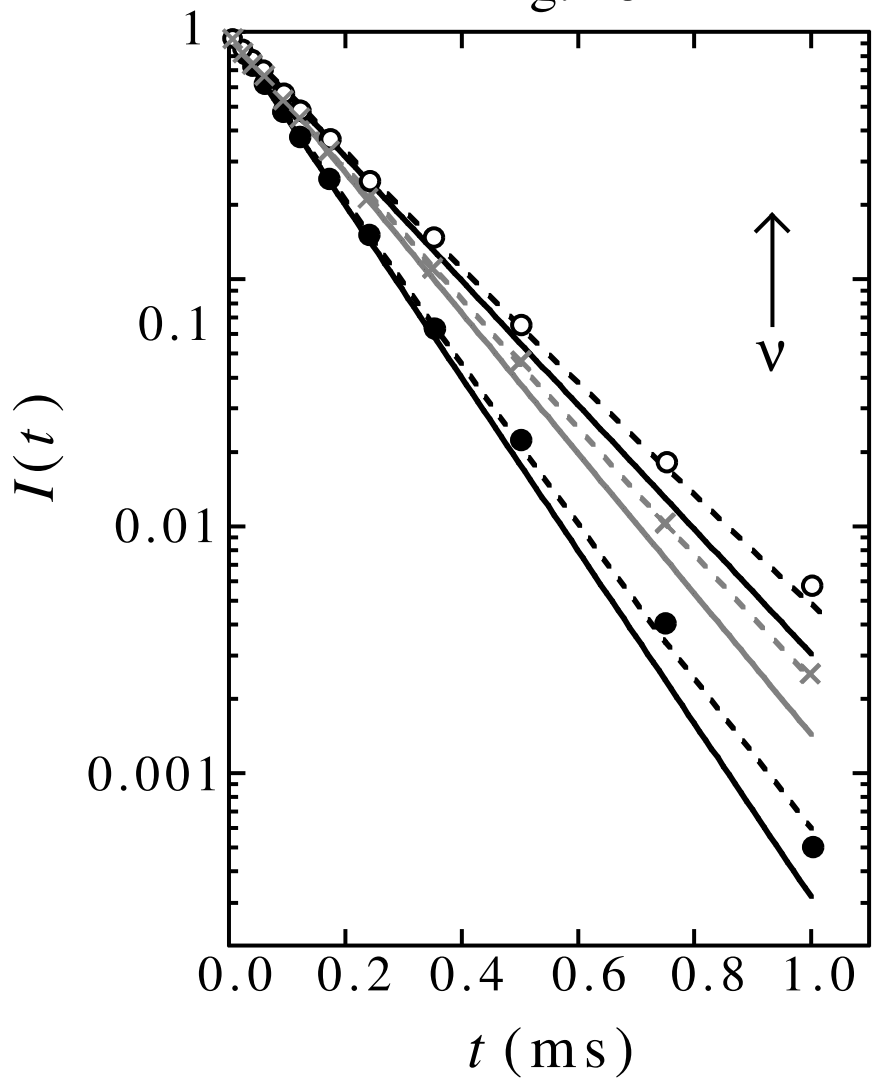
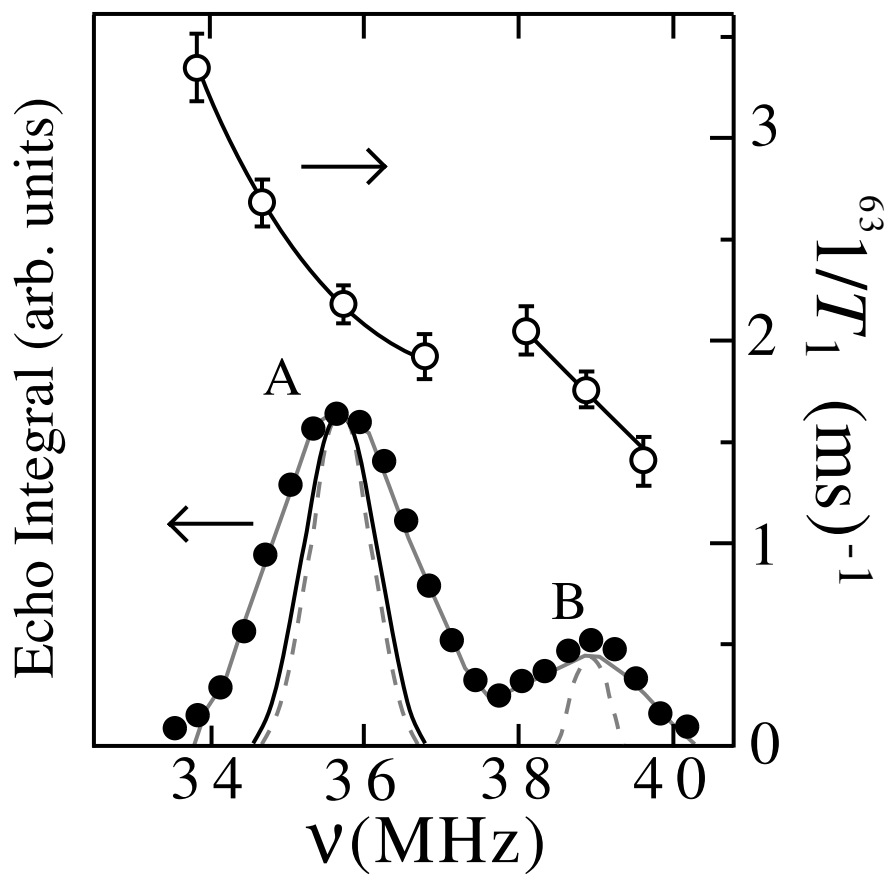
\end{document}